\documentclass{aa}  

\usepackage{graphicx}
\usepackage{xcolor}
\usepackage{gensymb}
\usepackage[colorinlistoftodos]{todonotes}
\usepackage{txfonts}
\usepackage{indentfirst}
\usepackage{ulem}
\usepackage[draft]{hyperref}
\usepackage[version=4]{mhchem}
\usepackage{gensymb}
\usepackage{tikzsymbols}
\usepackage{hypcap}

\newcommand{\karan}[1]{\textcolor{blue}{#1}}

\begin{document}

   \title{Understanding the atmospheric properties and chemical composition of the ultra-hot Jupiter HAT-P-7b}

   \subtitle{I. Cloud and chemistry mapping}

    \titlerunning{The ultra-hot atmosphere of HAT-P-7b: I. Cloud and chemistry mapping}

   \author{Ch. Helling \inst{1,2,3}
          \and
          N. Iro \inst{4}
          \and
          L. Corrales \inst{5}
          \and
          D. Samra \inst{1,2}
          \and
          K. Ohno \inst{6}
          \and
          M.K. Alam \inst{7}
          \and
          M. Steinrueck \inst{8}
          \and
          B. Lew \inst{8}
          \and
          K. Molaverdikhani \inst{9}
          \and
          R.J MacDonald\inst{10}
          \and
          O. Herbort \inst{1,2}
          \and
          P. Woitke \inst{1,2}
          \and
          V. Parmentier\inst{11}
          }

   \institute{Centre for Exoplanet Science, University of St Andrews, Nort Haugh, St Andrews, KY169SS, UK\\
             \email{ch80@st-andrews.ac.uk,dbss3@st-andrews.ac.uk,oh35@st-andrews.ac.uk, pw31@st-andrews.ac.uk}
         \and
             SUPA, School of Physics \& Astronomy, University of St Andrews, North Haugh, St Andrews, KY169SS, UK
         \and
         SRON Netherlands Institute for Space Research, Sorbonnelaan 2, 3584 CA Utrecht, NL
         \and
              Institute for Astronomy (IfA), University of Vienna,
              T\"urkenschanzstrasse 17, A-1180 Vienna\\
              \email{nicolas.iro@univie.ac.at}  
         \and
            Department of Astronomy, University of Michigan, Ann Arbor, MI 48109, USA \\
            \email{liac@umich.edu}
        \and
            Department of Earth and Planetary Sciences, Tokyo Institute of Technology, Meguro, Tokyo, 152–8551\\
            \email{ohno.k.ab@eps.sci.titech.ac.jp}
        \and
            Department of Astronomy, Harvard-Smithsonian Center for Astrophysics, Cambridge, MA 02138, USA \\
            \email{munazza.alam@cfa.harvard.edu}
        \and
            Lunar and Planetary Laboratory, University of Arizona, Tucson, AZ 85721, USA\\
            \email{msteinru@lpl.arizona.edu; wplew@lpl.arizona.edu}  
         \and
            Max Planck Institute for Astronomy, K\"onigstuhl 17, 69117 Heidelberg, Germany\\
            \email{Karan@mpia.de}
        \and
            Institute of Astronomy, University of Cambridge, Madingley Road, Cambridge, CB3 0HA, UK\\
            \email{r.macdonald@ast.cam.ac.uk}
        \and
            Department of Physics, University of Oxford, Parks Rd, Oxford, OX1 3PU, UK\\
            \email{vivien.parmentier@physics.ox.ac.uk}
            }
   \date{Received September 15, 2996; accepted March 16, 2997}

 
  \abstract
{Of the presently known $\approx 3900$ exoplanets, sparse spectral
    observations are available for $\approx 100$.  Ultra-hot Jupiters
    have recently attracted interest from observers and theoreticians
    alike, as they provide observationally accessible test cases.}
   {We aim to study cloud formation on the ultra-hot Jupiter HAT-P-7b,
     the resulting composition of the local gas phase, and how their
     global changes affect wavelength-dependent observations utilised
     to derive fundamental properties of the planet.}  
   {We apply a hierarchical modelling approach as a virtual laboratory     to study
   cloud formation and gas-phase chemistry. We utilise 97 vertical 1D profiles of a 3D GCM for HAT-P-7b to evaluate our kinetic cloud formation model consistently with the local equilibrium gas-phase composition. We use maps and slice views 
   to provide a global understanding of the cloud and gas chemistry.}
  {The day/night temperature difference on HAT-P-7b ($\Delta T\approx
     2500$\;K) causes clouds to form on the nightside (dominated by
     \ce{H2}/He) while the dayside (dominated by H/He) retains
     cloud-free equatorial regions. The cloud particles vary in
     composition and size throughout the vertical extension of the
     cloud, but also
     globally. \ce{TiO2}[s]/\ce{Al2O3}[s]/\ce{CaTiO3}[s]-particles of
     cm-sized radii occur in the higher dayside-latitudes, resulting
     in a dayside dominated by gas-phase opacity. The opacity on the
     nightside, however, is dominated by $0.01\ldots 0.1~\mu$m
     particles made of a material mix dominated by silicates. The gas
     pressure at which the atmosphere becomes optically thick is
     $\sim$ $10^{-4}$ bar in cloudy regions, and $\sim$ 0.1 bar in
     cloud-free regions.}
{HAT-P-7b features strong morning/evening terminator asymmetries,
    providing an example of patchy clouds and
    azimuthally-inhomogeneous chemistry. Variable terminator
    properties may be accessible by ingress/egress transmission
    photometry (e.g., CHEOPS and PLATO) or spectroscopy.  The large
    temperature differences of $\approx$2500\;K result in an
    increasing geometrical extension from the night- to the
    dayside. The {chemcial equilibrium} \ce{H2O} abundance at the terminator changes by $<$ 1
    dex with altitude and $\lesssim$ 0.3 dex (a factor of 2) across
    the terminator for a given pressure, indicating that \ce{H2O}
    abundances derived from transmission spectra can be representative
    of the well-mixed metallicity at $P \gtrsim 10$\,bar.  We suggest
    the atmospheric C/O as an important tool to trace the presence and
    location of clouds in exoplanet atmospheres. The atmospheric C/O
    can be sub- and supersolar due to cloud formation. Phase curve
    variability of HAT-P-7b is unlikely to be caused by dayside
    clouds.  }



  \maketitle 
%

    \section{Introduction}

Ultra-hot Jupiters, the hottest close-in giant planets known to date,
have dayside temperatures $\gtrapprox$ 2200\;K
{(\citealt{Parmentier18,2018AJ....156...17K,2019A&A...625A.136A,2018ApJ...866...27L,2018ApJ...857L..20B})}. These highly irradiated planets are suggested to
display several notable features, including thermal inversions (e.g.,
\citealt{Haynes15,Evans17}), and a lack of strong water absorption at
1.4\;$\mu$m (e.g., \citealt{Arcangeli18,2018AJ....156...10M}).
\cite{2010ApJ...722..871S} suggested that planets like HAT-P-7b would
exhibit a very hot upper atmosphere and a dayside temperature
inversion based on their 1D convective-radiative transfer models in
hydrostatic equilibrium. Their model used a parameterised day-night
heat distribution and an ad hoc opacity source for the upper
atmosphere to model potential temperature inversions.  This new class
of exoplanets is a rare outcome of planet formation \citep{Wright2012}
that provides a unique opportunity for understanding the atmospheric
physics and chemistry of extremely hot giant planets
(e.g. \citealt{lothringer2018extremely,2019arXiv190108640H,
  molaverdikhani2019cold}).

\capstartfalse
\begin{table*}
\caption{System parameters for the HAT-P-7 system}
\label{table:comparison}
\centering                      
\begin{tabular}{c c c }     
\hline\hline                
Orbital Parameters & & \\
\hline\hline 
Period~ $P$	~(days)							 		  & 2.204735471 $\pm$ 0.0000024				&   \citet{2015ApJ...805...28M} 	\\
Eccentricity~ $e$								      &	0.0										& 	\citet{2008ApJ...680.1450P}	    \\
Inclination~ $i$ (degrees)						      & 86.68 $\pm$ 0.14						&	\citet{vaneylen13}   \\
Semi-major axis~ $a$ ~(AU)							  & 0.0379 $\pm$ 0.0004						& 	\citet{2008ApJ...680.1450P}	    \\
\hline\hline 
Stellar Parameters & & \\
\hline\hline  
Stellar mass~ M$_{\star} ~$(M$_{\odot}$)       	  	  & 1.36 $\pm$ 0.02 						&	\citet{vaneylen13}   \\
Stellar radius~ R$_{\star} ~$(R$_{\odot}$)  	 	  	  & 1.90 $\pm$ 0.01  					    &  	\citet{vaneylen13} 	 \\
Surface gravity~ log(g$_{\star}$)\;(cgs)	 	  	  & 4.01 $\pm$ 0.01 						& 	\citet{vaneylen13}   \\
Stellar Temperature~ T\textsubscript{eff}\;(K)     	 	  	  & 6259 $\pm$ 32 							&   \citet{vaneylen13} 	 \\
Metallicity~ [Fe/H]          			     	  	  & $+$0.26 $\pm$ 0.08						& 	\citet{2008ApJ...680.1450P} 	     \\
\hline\hline 
Planetary Parameters & & \\
\hline\hline 
Planetary mass~ $M_{p}$ ($M_{\rm Jup}$) 			  & 1.74 $\pm$ 0.03 	  	    		    & 	\citet{vaneylen13}   \\
Planetary radius~ $R_{p}$ ($R_{\rm Jup}$) 			  & 1.43 $\pm$ 0.01    					    & 	\citet{vaneylen13}	 \\
Equilibrium temperature~ T\textsubscript{eq}\;(K)   		  &	2139 $\pm$ 27     					    &   \citet{heng13}	   \\
\hline\hline                                 
\end{tabular}
\end{table*}

\label{tab:params}
\capstarttrue

The subject of this study is the ultra-hot Jupiter HAT-P-7b
\citep{2008ApJ...680.1450P}. Discovered by the Hungarian Automated
Telescope Network (HATNet) Exoplanet
Survey\footnote{\url{https://hatnet.org/}}, this hot giant orbits a
bright (V = 10.5) F6 star with an orbital period of 2.2 days, a mass
of M$_{\rm }\approx 1.78$ M$_{\rm Jup}$, and a radius of R$_{\rm
}\approx 1.36$ R$_{\rm Jup}$ (see Table \ref{table:comparison}).
\cite{2018AJ....156...10M} confirmed HAT-P-7b as an ultra-hot Jupiter
with a dayside disk-integrated temperature of T $\approx$ 2700\;K and
a geometric albedo of A$_{\rm g}$ = $0.077\pm 0.006$ (Kepler and
\textit{HST}/WFC3 secondary eclipse depth). The 1D retrieval suite
used suggests C/O $<$ 1 and a somewhat sub-solar atmospheric
metallicity of [M/H] $\sim$ $-0.87$. \textit{Spitzer} phase curves at
3.6 and 4.5\;$\mu$m suggest a dayside thermal inversion of the
atmospheric temperature profile and relatively inefficient day-night
circulation (\citealt{2016ApJ...823..122W}) based on 1D retrieval
procedures.  The transit light curve deformation by the stellar
gravity darkening suggests a nearly pole-on host star-planet
configuration with $\psi =101{}\degree \pm 2{}\degree$ or $87{}\degree
\pm 2{}\degree$ \citep{2015ApJ...805...28M}.

\cite{2018AJ....156...10M} also compared cloud-free 3D general
circulation model (GCM) results for the HAT-P-7b system parameters
derived in \cite{2016ApJ...828...22P} with the observed
\textit{HST}/WFC3 thermal emission spectrum. They find that the GCM
needs a very poor redistribution of heat from the day- to the
nightside in order to fit the WFC3 data and argue that it could be
caused by magnetic drag for reasonable planetary magnetic fields.
They further point out that water not being present has a large impact
on the emission spectrum, limiting the range of pressures probed in
the HST/WFC3 bandpass. This effect produces a blackbody-like spectrum,
whereas emission features are expected if water is not dissociated.

Currently, none of the models for HAT-P-7b have included the effect of
cloud formation.  Clouds affect the local gas phase through element
depletion and enrichment, and as strong opacity sources that affect
the local temperature {(\citealt{1986ApJ...310..238L,
  1996A&A...305L...1T,2001ApJ...556..872A,2001ApJ...556..357A,2009A&A...506.1367W,Lee2015,Lee2016,2017A&A...608A..70J,2018A&A...615A..97L})}. Clouds
can also strongly impact thermal emission and reflected light spectra,
transit observations, and phase curves. \citet{2016NatAs...1E...4A}
report that the phase curve of HAT-P-7b in the \textit{Kepler}
bandpass varies significantly with time, with the offset of the phase
curve peak changing over time, occurring before secondary eclipse and
after secondary eclipse. They suggest that this variability might be
caused by a temporally varying degree of cloud coverage on the dayside
due to variations in the strength of the equatorial jet. {Works
  concerned with cloud-cover variability range from modelling include
  turbulence studies
  (\citealt{2001A&A...376..194H,2004A&A...423..657H}), large-scale HD
   simulations (\citealt{2008PhST..133a4005F,Lee2015,2018A&A...615A..97L}) to observational studies (incl. \citealt{2012ApJ...750..105R,2013ApJ...767..173H,2013ApJ...768..121A,2015ApJ...798..127B,2015ApJ...812..161S,2016ApJ...829L..32L,2019MNRAS.483..480V}).} Such a
variation could also come from a thermal emission variation, with the
planet's hottest spot moving from east to west of the substellar point
through oscillation triggered by magnetohydrodynamic
effects~\citep{Rogers2014,Rogers17_magnetic}.  A thorough
understanding of the ion chemistry and cloud formation is needed to
understand this particular behaviour.

In this paper, we study cloud formation and gas-phase chemistry in the atmosphere of HAT-P-7b for the collisionally dominated part of the atmosphere.
We chose a hierarchical approach of modelling environments where we utilise 97 1D profiles extracted from the 3D GCM calculated for the cloud-free HAT-P-7b (\citealt{2018AJ....156...10M}) and apply our kinetic cloud formation model consistently linked to an chemical equilibrium code. {Our study is the continuation of previous investigations of cloud formation in the giant gas planets  HD\,189733b and HD\,209458b (\citealt{Lee2015,2016MNRAS.460..855H,Lee2016,2018A&A...615A..97L}), and the super-hot gas giant WASP-18b (\citealt{2019arXiv190108640H}) in order to enable a comparative overview for these two classes of giant gas planets given that observations mainly provide snapshots of certain aspects (e.g. presence of certain gas species, existence of clouds, a hot spot off-set) of each of these planets so far.}

In Section~\ref{s:ap}, we outline our approach {and our
  assumptions} for modelling the cloud formation and the gas phase chemistry
on HAT-P-7b. Section~\ref{s:cloud} presents our results on cloud
formation on HAT-P-7b, providing details on cloud formation
(Sect.~\ref{ss:how}) before discussing the correlation between the
various cloud properties (e.g., nucleation rate, cloud particles
sizes, material composition). Section~\ref{ss:maps1} presents how the
fundamental cloud properties (e.g., nucleation rate) and derived cloud
properties (e.g., dust-to-gas ratio) vary globally on
HAT-P-7b. Section~\ref{sec:map_comp} focuses on how the material
composition of cloud particles changes globally. The gas-phase
composition of the atmospheric gas is discussed in Sect.~\ref{s:maps},
deriving a distinctly different gas-composition for the day- and the
nightside as a result of the local temperatures and the locally
confined cloud formation on HAT-P-7b. Here we also elude to deriving
water abundances using transmission spectroscopy, and to the resulting
inhomogeneous carbon-to-oxygen ratio in the atmosphere.
Section~\ref{s:opacity} combines the results from all previous
sections to study the opacity profiles of the atmosphere of HAT-P-7b
in terms of p$_{\rm gas}(\tau_{\lambda}=1)$ (gas and cloud) and
optical-depth weighted cloud properties. Section~\ref{s:conclusions}
summarises our conclusions about cloud formation and the collisionally
dominated gas-composition on HAT-P-7b.


\begin{figure*}
\includegraphics[width=\textwidth]{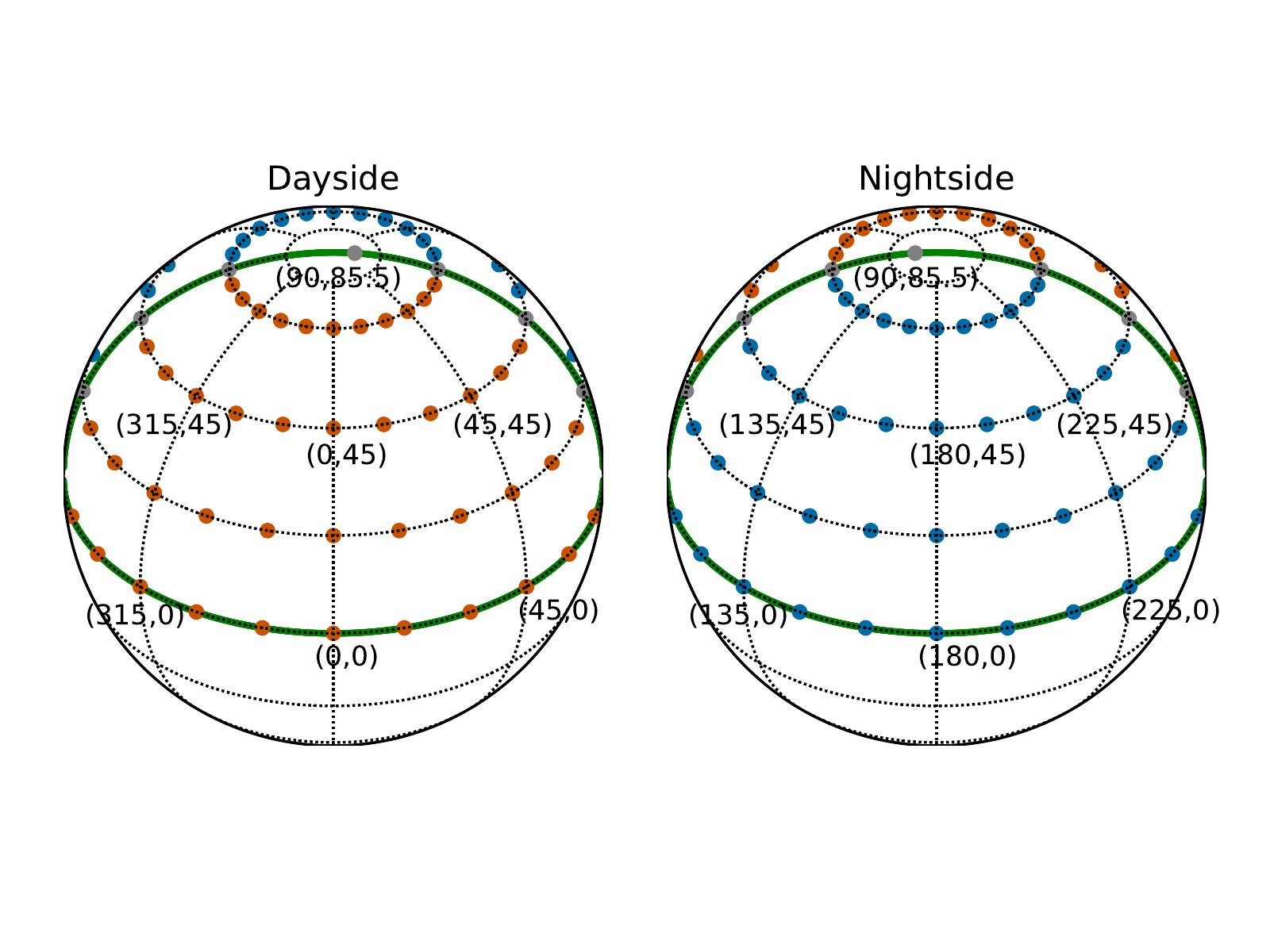}\\*[-3cm]
\caption{Positions of the 1D profiles taken from the GCM as viewed from the dayside (left) and nightside (right) of the planet, with the points $(\phi, \theta)=(0\degree,0\degree)$ and $(\phi, \theta)=(180\degree,0\degree)$ being the sub-stellar and anti-stellar points, respectively ($(\phi, \theta)$ = (longitudes, latitudes)). 1D profiles that sample the dayside (red), nightside (blue), morning  and evening terminator (both grey) are indicated. The green lines show the locations of slices used in subsequent  maps along the equator ($\theta = 0\degree$) and the terminator (longitudes $\phi = 90\degree$ --  evening terminator,  $\phi = 270\degree$ --  morning terminator).
}
\label{fig:Dayside_trajectories}
\end{figure*}

\begin{figure}
    \centering
    \includegraphics[width=0.5\textwidth]{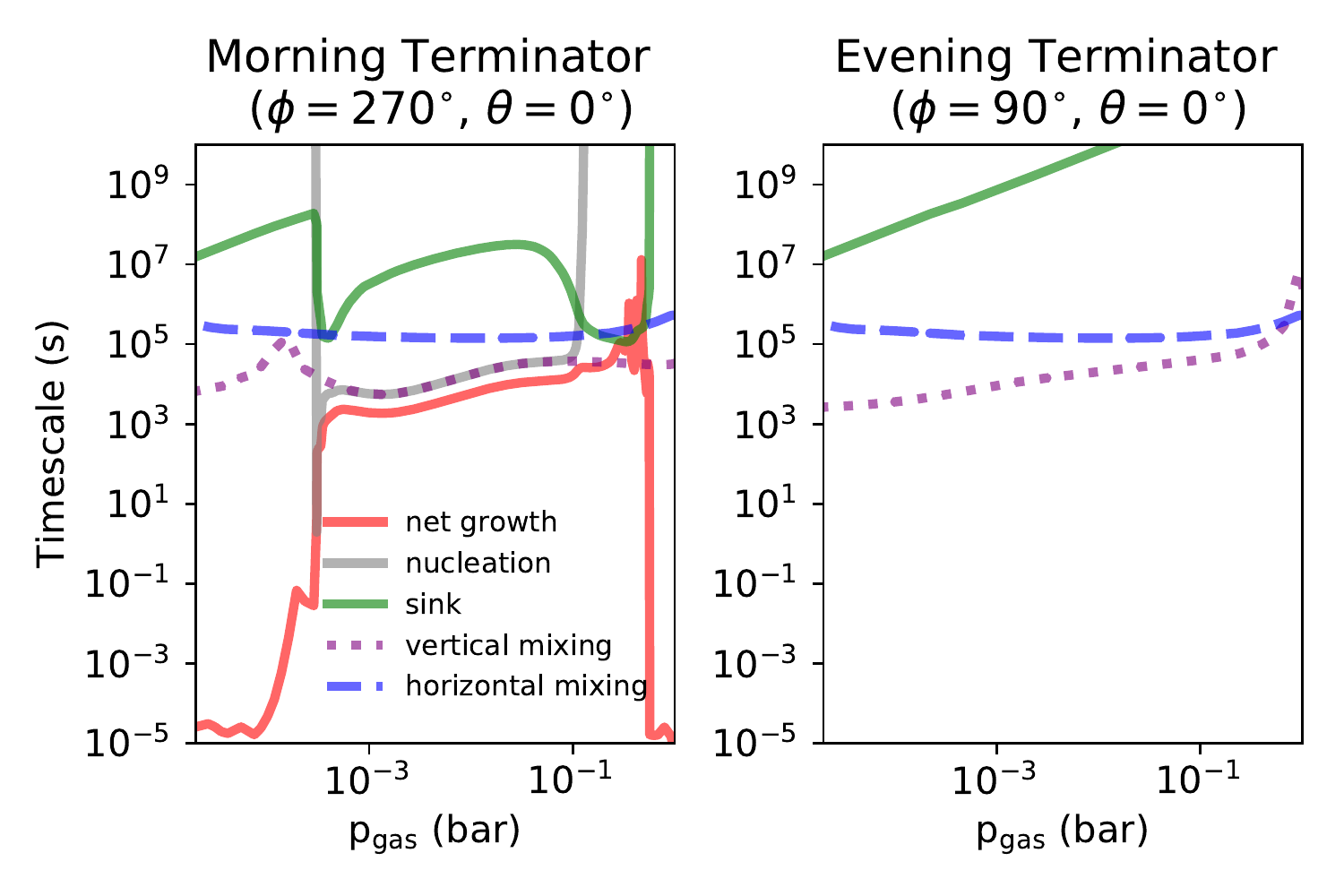}
    \caption{The comparison of different timescales at the morning and evening terminators. The horizontal mixing timescale (blue dotted) is not the dominant (fastest) timescale compared to the vertical mixing (purple dotted) and net grain growth (solid red) timescales.}
    \label{fig:timescales}
\end{figure}

    \section{Approach}\label{s:ap}

 We adopt a hierarchical two-model approach in order to examine the cloud structure on HAT-P-7b similar to works on HD\,189733b and HD\,209458b (\citealt{Lee2015,2016MNRAS.460..855H}), and WASP-18b (\citealt{2019arXiv190108640H}): The first modelling step produces pre-calculated, cloud-free 3D GCM results. These results are used as input for the second modelling step which is a kinetic cloud formation model consistently combined with a equilibrium gas-chemistry calculations. We utilise 97 1D (T$_{\rm
  gas}$(z), p$_{\rm gas}$(z), v$_{\rm z}(x,y,z)$)-profiles for HAT-P-7b as visualised in Fig.~\ref{fig:Dayside_trajectories}: T$_{\rm gas}$(z) is the local gas temperature [K], p$_{\rm gas}$(z) is the local gas pressure [bar], and v$_{\rm z}$(x,y,z) is the local vertical velocity component [cm s$^{-1}$]. We will also use these results to investigate chemical quenching, which will be presented in a subsequent paper (Molaverdikhani et al.: Paper II). Implications for observations will be considered in another forthcoming paper (MacDonald et al.: Paper III). Below, we describe our approach and will refer to previous papers for more details.

The hierarchical two-model approach has the
  limitation of not explicitly taking into account the potential effect of
  horizontal winds on cloud formation {and the radiative feedback of the cloud particles.  The effect of horizontal winds on cloud formation only enters implicitly through the temperature structure while horizontal advection of cloud particles themselves is neglected.} However, processes governing the formation of clouds are determined by local thermodynamic
  properties which are the result of 3D dynamic atmosphere  simulations. Horizontal transport can affect both the cloud formation and the cloud properties. As an order-of-magnitude estimation, we approximate the horizontal transport timescale by dividing the planet's circumference over the horizontal wind velocity. As shown in Fig.~\ref{fig:timescales} the horizontal transport timescale is longer than both the nucleation and the growth timescale, indicating that the cloud should not be strongly affected by the horizontal transport. Horizontal advection is, however, faster than the settling timescale of the cloud particles. As a consequence, the cloud particle properties such as particle size or particle composition should be smeared out in longitude compared to the results shown here. That being said, and as shown by comparing~\citet{Lee2015} (without horizontal advection) and ~\citet{Lee2016} (including horizontal advection), the non-coupled problem is both more computationally feasible, easier to interpret and provides reasonable first order insights into the expected mean cloud particle size and composition, cloud distribution and gas phase depletion in exoplanet atmospheres. {As we will demonstrate in our results, HAT-P-7b dayside clouds are made of big particles which, hence, will have little effect on the local temperature. However, the clouds on the nightside are expected to increase the temperature everywhere in the planet by $\approx$ 100K as demonstrated in the discussion of cloudy models in \citealt{2016ApJ...828...22P}). This will not change the global cloud structure drastically.}


\subsection{Kinetic cloud formation}\label{ss:cm}
Cloud particles form through a sequence of processes (see  \cite{2018arXiv181203793H} for a review). First, condensation seeds form from the atmospheric gas (nucleation). Once these condensation seeds have formed, other materials grow a substantial mantle on top of the seed particles through surface reactions. Only materials that are thermally stable will undergo this surface condensation process which produces the bulk of the cloud particle. Many materials are thermally stable in a very narrow temperature interval in oxygen-rich planetary atmospheres, resulting in the growth of a mix of materials. Cloud particles interact with the atmospheric gas through elemental depletion and through collisions (friction, radiation). Nucleation and surface growth deplete the atmospheric gas by those elements now captured in the cloud particle's lattice structure. Each cloud particle also interacts mechanically with the surrounding gas through friction and gravity. The cloud particle's fall is determined by the equilibrium between friction and gravity which is established very quickly \citep{Woitke2003}. This gravitational settling can be disturbed by horizontal winds if these gas densities are high enough and the particles small enough to achieve frictional coupling. \cite{2016ApJ...823..122W} suggest that the day/night circulation is rather inefficient on HAT-P-7b. A stationary cloud structure in the vertical direction is, hence,  a reasonable assumption for HAT-P-7b. This is further corroborated by comparing horizontal and vertical mixing timescales from the GCM with the timescales of cloud formation processes (Sect.~\ref{ss:ib}).

\smallskip
\noindent
    {\it Nucleation (seed formation):} We calculate the homogeneous nucleation rates (\citealt{helling2013RSPTA,2018A&A...614A.126L}) of \ce{TiO2}, SiO and carbon \citep{2015A&A...575A..11L,2017A&A...603A.123H}.  The effective
    nucleation rate, $J_* = \sum_i J_{\rm i=\ce{TiO2}, SiO, C}$ [cm$^{-3}$
      s$^{-1}$], determines the number of cloud particles, n$_{\rm d}$
    [cm$^{-3}$], and hence, the total cloud surface (as a sum of the
    surface of the cloud particles). 

\smallskip
\noindent
    {\it Bulk growth/evaporation:} The seed forming species also need to be considered as surface growth material, since both processes (nucleation and growth) compete for the participating elements (Ti, Si, O, and C in this work). We consider the formation of 15
    bulk materials (s=\ce{TiO2}[s], \ce{Mg2SiO4}[s], \ce{MgSiO3}[s],
    MgO[s], SiO[s], \ce{SiO2}[s], Fe[s], FeO[s], FeS[s], \ce{Fe2O3}[s],
    \ce{Fe2SiO4}[s], \ce{Al2O3}[s], \ce{CaTiO3}[s], \ce{CaSiO3}[s],
    C[s]) that form from 9 elements (Mg, Si, Ti, O, Fe, Al, Ca, S, and C)
    by 126 surface reactions (Tables B.1 and B.2 in \citealt{2019arXiv190108640H}). We solve
    moment equations for the cloud particle size distribution function
    that consider nucleation, growth/evaporation, gravitational
    settling, and mixing
    (\citealt{Woitke2003,Helling2006,2008A&A...485..547H,
      helling2013RSPTA}). 
In addition to the local element abundances, {$\varepsilon(z)$}, the local thermodynamic properties T$_{\rm gas}(z)$ and
$\rho_{\rm gas}(z)$ (local gas density, [g cm$^{-3}$])  determine if atmospheric clouds can form, to which sizes the cloud particles grow, and
of which material mix, $s$ (e.g. \ce{TiO2}[s], \ce{Mg2SiO4}[s], \ce{MgSiO3}[s]), they will be composed of. The gravitational
settling velocity ($v_{\rm drift}$ [cm s$^{-1}$]) is determined
by the local gas density, $\rho_{\rm gas}(z)$, and the cloud particle
size (where $\langle a \rangle$
  [$\mu$m] is mean cloud particle size).

\smallskip
\noindent
{\it Element conservation:} An set of equations for all
involved elements is solved with source/sink terms for
nucleation, surface growth/evaporation, and gravitational settling {(\citealt{2008A&A...485..547H})}.

\smallskip
\noindent
{\it Element replenishment:} Cloud particle formation depletes the
local gas phase, and gravitational settling causes these elements to
be deposited, for example, in the inner (high pressure) atmosphere where the cloud
particles evaporate. For a stationary cloud to form, element
replenishment needs to be modelled. We apply the approach outline in
\cite{Lee2015} using the local vertical velocity to
calculate a mixing time scale, $\tau_{\rm mix}\sim v_{\rm
  z}(r)^{-1}$. 


\subsection{Gas phase chemistry equilibrium  modelling} 
We apply chemical equilibrium  to
calculate the gas composition (number densities n$_{x}$ [cm$^{-3}$]) of the atmosphere as part
of our cloud formation approach.  We use the 1D (T$_{\rm gas}(z)$,
p$_{\rm gas}(z)$) profiles and element abundances $\varepsilon_{\rm
  i}(z)$ (i=O, Ca, S, Al, Fe, Si, Mg, Ti, C) depleted by cloud
formation processes.  All other elements are assumed to be of solar
abundance. 
A combination
of $156$ gas-phase molecules, $16$ atoms, and various ionic species
are included.  
High velocities
and/or strong radiation may cause departure from
LTE. \cite{2006ApJ...648.1181V,2010ApJ...716.1060V,2009ApJ...701L..20Z,line2010high,2012ApJ...745...77K,2011ApJ...737...15M,venot2018better}
have suggested that in warm exoplanet atmospheres gases thermodynamic equilibrium hold for T$_{\rm gas}>1200$\;K assuming a species-independent, constant diffusion constant. We will evaluate this assumption in Paper II.

No condensates are part of the chemical equilibrium calculations, in contrast to equilibrium condensation models. The influence of cloud
formation on the gas phase composition results from the reduced or
enriched element abundances due to cloud formation and the impact of cloud
opacity on the radiation field, and hence, on the local gas
temperature and gas pressure.  The element depletion or enrichment due
to cloud formation is therefore directly coupled with the gas-phase
chemistry calculation.

\begin{figure*}
    \vspace*{-0.5cm}
    \includegraphics[width=8.5cm]{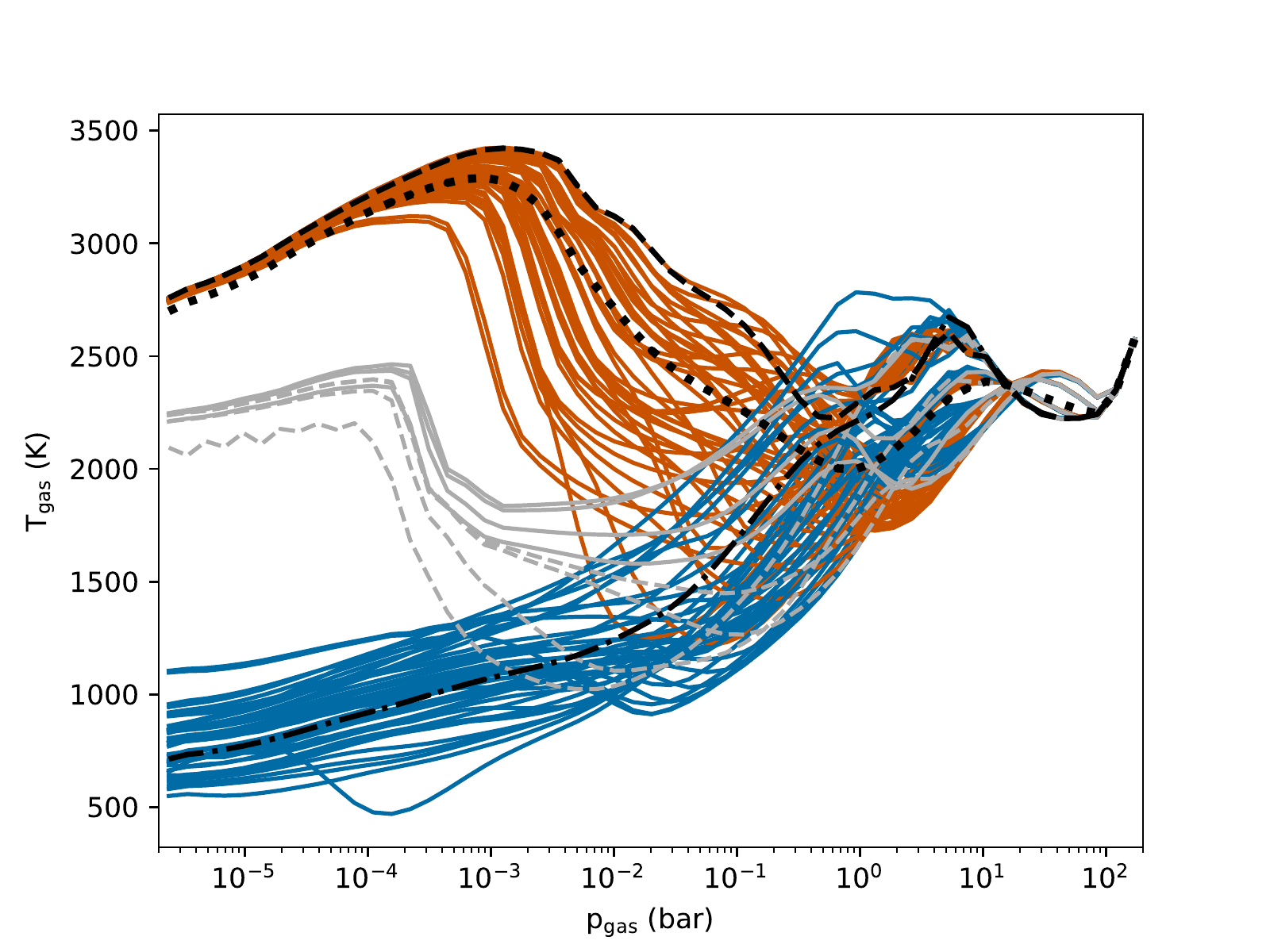}    
    \includegraphics[width=8.7cm]{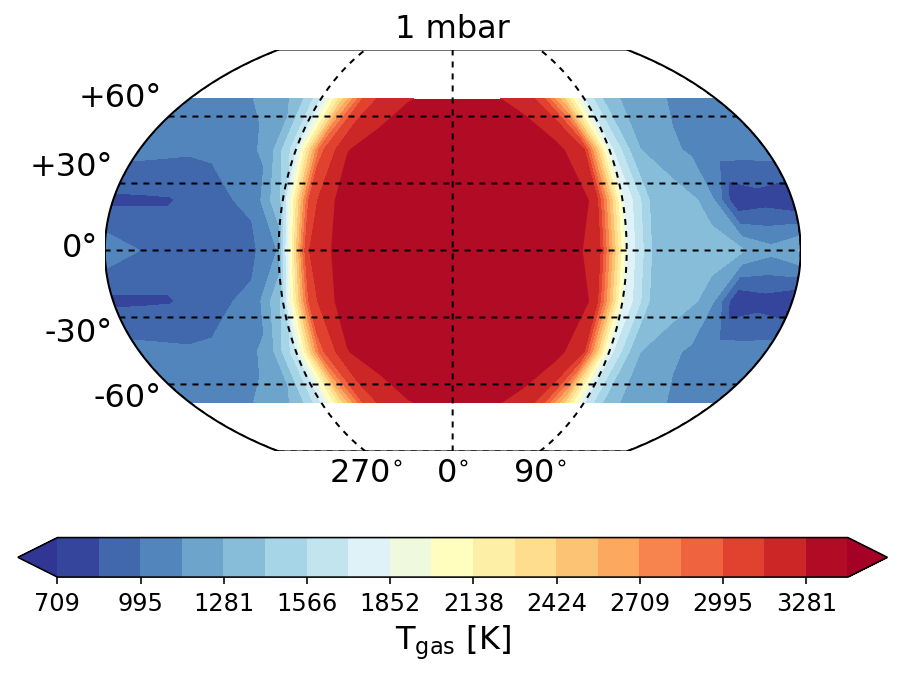}\\ 
    \includegraphics[width=8.7cm]{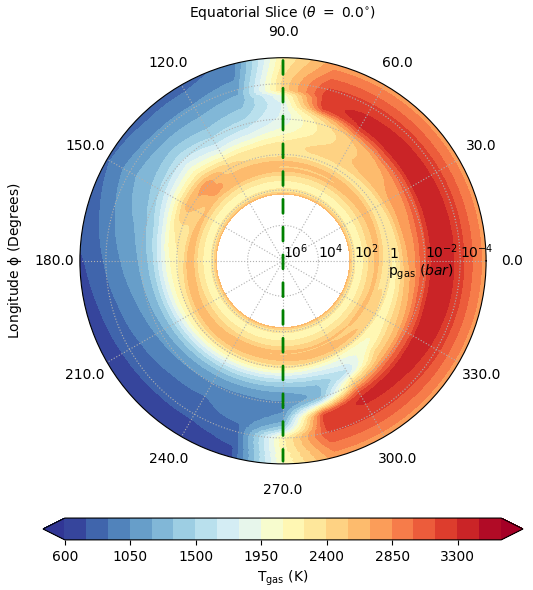}
        \includegraphics[width=8.7cm]{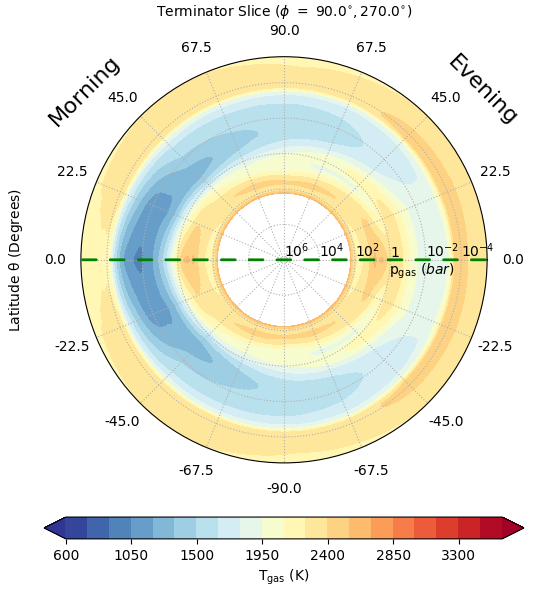}
    \caption{The global gas temperature, T$_{\rm gas}$ [K], distribution on  HAT-P-7b. 
    {\bf Top left:} 
    1D structures as function of p$_{\rm gas}$\;[bar] of the probed 97 profiles (red -- dayside ($\phi=60\degree\,\ldots\,0\degree\ldots\,300\degree$),  blue --  nightside ($\phi=120\degree\,\ldots\, 240\degree$), grey dashed -- evening terminator $(\phi=90\degree)$, grey solid  -- morning terminator $(\phi=270\degree)$,  black dashed --  substellar point  $(\phi=0\degree)$, black dot-dashed -- antistellar point  $(\phi=180\degree)$).  The dotted black line represents a dayside-average profile (using an average weighted by the projected area that is observed in secondary eclipse). {\bf Top right:} 
    Global 2D maps at p$_{\rm gas}=10^{-3}$\;bar. The 2D temperature maps visualise the strong day-night differences and also the somewhat inhomogeneous nightside ($(\theta,\phi)=(\pm 22.5\degree, 165\degree), (\pm 22.5\degree, 180\degree))$. These local temperature minima ($\Delta$T$_{\rm gas}\approx 200$\;K to surrounding) will have an effect on our subsequent analysis results.
     {\bf Bottom left:} 2D cut through the equatorial plane. {\bf Bottom right:}  2D cut along the terminator.  The dashed green lines indicate where these two slice plots overlap. For the viewing geometry of the slice plots in the bottom row, please see Figs.~\ref{fig:1DTp_1},~\ref{fig:vgeo}.}
    \label{fig:1DTp_1}
\end{figure*}

\subsection{Cloud and gas opacity calculation}\label{sec:cloud_opac}

\paragraph{Cloud opacity and optical depth:}
We calculate the cloud opacity following the same approach described in \citep{2008A&A...485..547H,Lee2015}.
To calculate the opacity of cloud particles consisting of various materials,  we apply the effective medium theory with the Br\"uggeman mixing rule \citep{1935AnP...416..636B}.
According to Br\"uggeman's rule, the effective refractive index of a solid or liquid particle can be derived from the effective dielectric function $\epsilon_{\rm eff}$ given by
\begin{equation}
    \Sigma_s \left(\frac{V_{\rm s}}{V_{\rm tot}}\right)\frac{\epsilon_{\rm s}-\epsilon_{\rm eff}}{\epsilon_{\rm s}+2\epsilon_{\rm eff}}=0,
    \label{eq:tauweighted}
\end{equation}
where $V_{\rm s}/V_{\rm tot}$ is {the volume ratio between the volume of the condensed species s, $V_{\rm s}$,  and the total condensed volume $V_{\rm tot}$  and calculated in the formalism summarised in} Sect.~\ref{ss:cm} and $\epsilon_{\rm s}$ is the dielectric function of a material $s$. 
The refractive index of the cloud materials from the literature are listed in Table~\ref{table:opt}.
The effective refractive index $\epsilon_{\rm eff}$ is found by solving Eq. \eqref{eq:tauweighted} with the Newton-Raphson method.
With the derived effective refractive index $\epsilon_{\rm eff}$, we calculate extinction efficiencies, $Q_{\rm ext}$,   of cloud particles at each atmospheric layer using Mie theory and assuming  spherical cloud particles \citep[e.g.,][]{1983asls.book.....B}.
The vertical cloud optical depth is then calculated as 
\begin{equation}
    \tau_{\rm cloud}(\lambda) = \int Q_{\rm ext}(\lambda,a)\pi a^2n_{\rm d}dz,
    \label{eq:taud}
\end{equation}
where $a$ and $n_{\rm d}$ are the radius and number density of cloud particles.
In this study, we use the mean particle radius $\langle a \rangle$ for the opacity calculations.

\paragraph{Gas opacity and optical depth:}
The wavelength-dependent extinction coefficient for gaseous species $x$ is given by
\begin{equation}
\kappa_x(\lambda) = n_x \, \sigma_{x}(\lambda),
\end{equation}
where $\sigma_{x}(\lambda)$ is the absorption cross section of species $x$ at wavelength $\lambda$. For pair processes, such as collisionally induced absorption (CIA) and free-free absorption, the extinction coefficient is instead given by 
\begin{equation}
\kappa_{x_1-x_2}(\lambda) = n_{x_1} n_{x_2} \, \alpha_{x_1-x_2}(\lambda),
\end{equation}
where $\alpha_{x_1-x_2}(\lambda)$ is the binary absorption coefficient between species $x_1$ and $x_2$ at wavelength $\lambda$. The wavelength-dependent optical depth due to either a species or a pair is then
\begin{equation}
\tau_{x / x_1-x_2}(\lambda) =\int \kappa_{x / x_1-x_2}(\lambda) \, dz,
\label{eq:taug}
\end{equation}
where the integral is evaluated from the top of the atmosphere vertically downwards along the z-direction (similar to Eq.~\ref{eq:taud}). The absorption cross sections, $\sigma_{x}(\lambda)$, for various gas-phase atoms and molecules were obtained from the opacity database introduced in \citet{MacDonald2019} (and subsequently expanded). Namely, we consider opacities due to the following species: \ce{H2O, C2H2, CH4, CO2}, CO, CH, OH, HCN, NH, NO, \ce{NH3, O3}, SiH, SiO, AlH, AlO, MgH, Cs, FeH, \ce{H2S, SO2, H2, N2, O2}, TiO, CaO, CaH, TiH, LiH, VO, Na, K, Li, Fe, Ti, and H-. Binary absorption cross sections, $\alpha_{x_1-x_2}(\lambda)$, are obtained from \citet{Richad2012} for \ce{H2-H2 and H2-He} CIA and \citet{Gray2008} for H- free-free absorption. In Sect.~\ref{ss:tau1mol_abovecloud}, these extinction coefficients will be employed, in tandem with Eqs.~\ref{eq:taud} and ~\ref{eq:taug}, to determine the gas pressure where the optical depth reaches one for each particular species, i.e. $p_{\rm gas}(\tau_{x}(\lambda)=1)$.

\subsection{Input and Boundary conditions}\label{ss:ib}

{\it Element abundances:} We assume that HAT-P-7b has an oxygen-rich
atmosphere of approximately solar elemental composition. We use the
solar element abundances from \citet{2009ARA&A..47..481A} as the non-depleted values for the cloud formation simulation and outside the cloud forming domains. The carbon-to-oxygen ratio (C/O) is 0.53 for an oxygen-rich, undepleted gas.

\noindent
{\it Input profiles from a 3D atmosphere simulation for HAT-P-7b:}\\
We use 1D pressure-temperature and wind profiles calculated from the 3D global circulation model SPARC/MITgcm global circulation model~\citep{Showman2009} adapted to the specific case of HAT-P-7b. The hydrodynamic model solves the primitive equations on a cube sphere and is coupled to a radiative transfer scheme~\citep{Marley1999}. The model includes the radiative effects of the main gaseous absorbers expected in this temperature regime, including TiO/VO and H$^-$~\citep{Parmentier2018}. The model is the same as the solar compositon, dragless model described in~\citet{2018AJ....156...10M}. The model was run for 300 earth days, with the last 100 days used to calculate time averaged quantities. The pressure in the model ranges from $2\cdot 10^{-6}\,\ldots\,200$ bars. 53 levels are used in the vertical direction, equally separated in log pressure. {Our cloud formation code will increase the pressure resolution as result of the implicit method used to solve the system of  ODEs}.  The cubed-sphere grid has a horizontal resolution of C32, equivalent to 128 cells in longitude and 64 in latitude. Cloud opacities were not taken into account in the global circulation model. Although we will show that clouds form on the planet's nightside, they are not expected to drastically  change the energy budget of the atmosphere. As a consequence, features such as the dayside temperature profiles or the strong day/night temperature contrast are robust even in the presence of cloud opacities~\citep[e.g.][]{parmentier16_phase-curve,Lee2016}. 
The atmospheric circulation in the global circulation model is characteristic for hot Jupiters, with a superrotating equatorial driven by the day-night heating contrast dominating the circulation deep in the atmosphere \citep{ShowmanPolvani2011}. At lower pressures, day- to night flow becomes more important. The vertical velocity field is characterized by large-scale upwelling on the dayside and downwelling on the nightside. In addition, there are localized regions of strong updrafts and downdrafts associated with convergence and divergence in the equatorial jet region near the terminators (Fig. \ref{fig:orgData}). 

Given the north/south symmetry of the global circulation outputs, we use our results calculated for the 24 profiles at the equator and the 72 ones of the northern hemisphere and mirror these results to the southern hemisphere across the equator which provides us with 168 (24 + 72 + 72) 1D profiles equally spaced in latitude and longitude. These 168  profiles are then used as the base for our 2D maps.

Figure~\ref{fig:1DTp_1} (top left) summarises the input profiles that
we use to study cloud formation at the day- (e.g., longitude $\phi=
0\degree$, $45\degree$, $-45\degree$) and nightside (e.g., longitude
$\phi=180\degree$, $135\degree$, $-135\degree$) of HAT-P-7b as well as
at the evening/morning terminators (longitude $\phi=90\degree$ and
$\phi=270\degree$, respectively).  The differences in the
thermodynamic structures are very large. Dayside and nightside
temperature structures differ by {\it more than 2400\;K}. The dayside
profiles can be as hot as 3500\;K at a relatively low pressure of
p$_{\rm gas} = 10^{-3}$\;bar. The terminator regions show strong
temperature inversions that cause a steep local (inward) drop in
temperature of up to 1000\;K
($\phi=-90\degree$). Figure~\ref{fig:1DTp_1} (top left) also
demonstrates that the dayside-averaged profile (dotted black line)
does not approach an isothermal situation, therefore suggesting that
the assumption of isothermal temperature profiles for other close-in
giant gas planets is unjustified (e.g.,
\citealt{2019arXiv190207944M}).

Figure~\ref{fig:1DTp_1} (top right) visualises the global T$_{\rm
  gas}$ distribution at p$_{\rm gas}=10^{-3}$\;bar representing the
pressure regime that appears accessible by transmission
spectroscopy. While the temperature difference between day- and
nightside decrease for p$_{\rm gas}<10^{-3}$bar
(Fig.~\ref{fig:1DTp_1}, top left) to $\Delta T_{\rm gas}\approx
2000$K, p$_{\rm gas}\approx 10^{-3}$\;bar represents the atmospheric
regions with the strongest temperature difference for HAT-P-7b of
$\Delta T_{\rm gas}\approx 2500$K. The dayside is almost uniformly at
$T_{\rm gas}\approx 3000\pm 200$K at p$_{\rm gas}\approx
10^{-3}$bar. The slice-plots in Fig.~\ref{fig:1DTp_1} (bottom)
underpin this finding and show that the global temperature differences
are smaller in the terminator plane (right) compared to the equatorial
plane (left). The largest temperature differences between the two
terminators occur at p$_{\rm gas}\approx 10^{-3}$bar, too, and amount
to $\Delta T_{\rm gas}\approx 500$K. The dashed green lines indicate
where these two slice plots overlap.

\begin{figure*}
    \includegraphics[width=\textwidth]{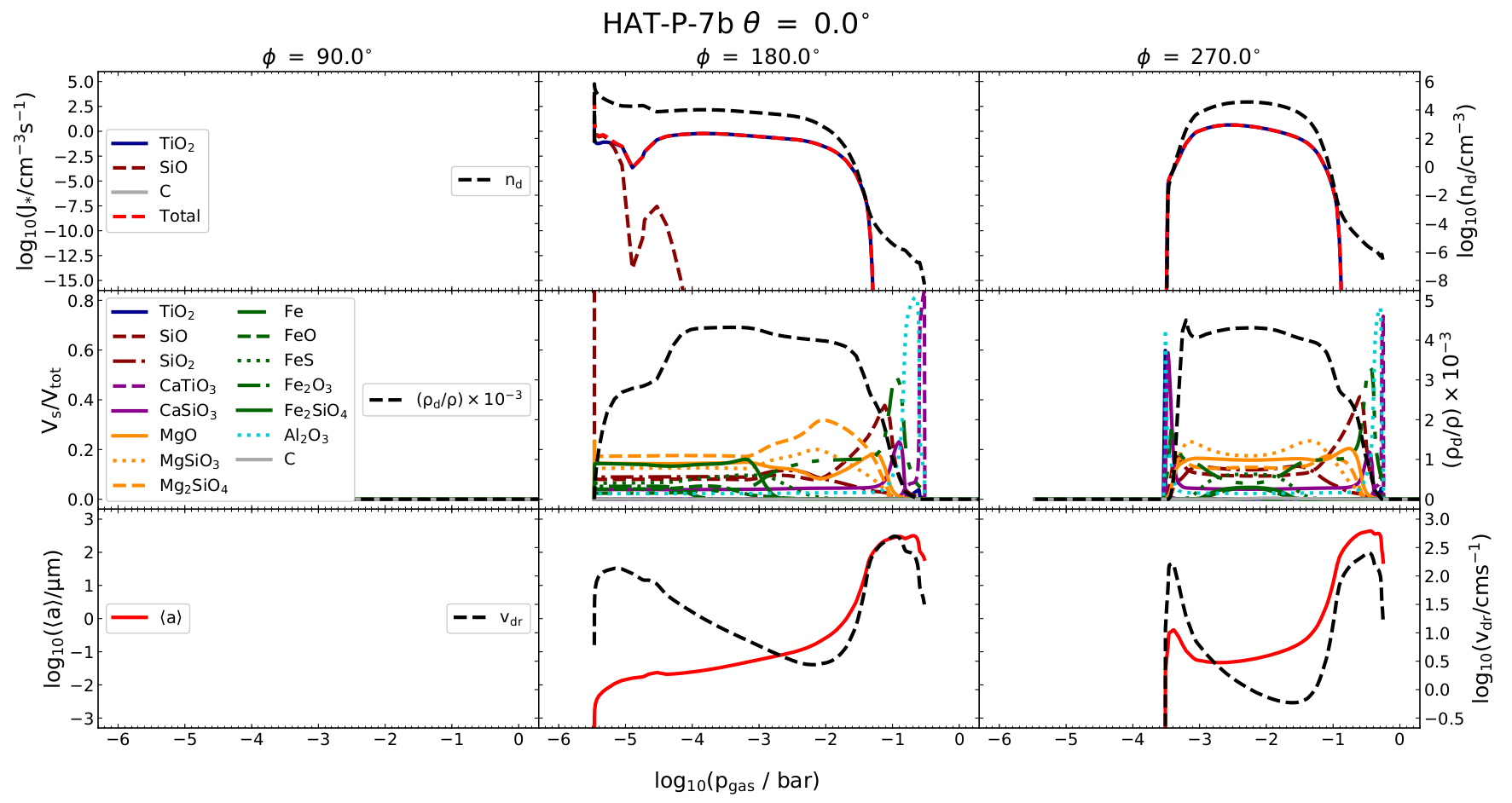}
    \caption{
    The cloud properties at the evening terminator, anti-stellar point, and morning terminator $(\phi\ =\ 90\degree,180\degree, 270\degree)$.
    \textbf{Top:} Left axis shows the individual, $J_{\rm i}$ (i=\ce{TiO2}, \ce{SiO}, \ce{C}),  and the total nucleation rates $J_* = \sum_i J_{\rm i}= $ [cm$^{-3}$
      s$^{-1}$] (cut-off at $10^{-15}$). Right axis (black dashed line) is the number density of cloud particles, $n_{\rm d}$ [cm$^{-3}$].
    \textbf{Middle:} Volume fractions, $V_{\rm s}/V_{\rm tot}$ (s=\ce{TiO2}[s], \ce{Mg2SiO4}[s], \ce{MgSiO3}[s],
    MgO[s], SiO[s], \ce{SiO2}[s], Fe[s], FeO[s], FeS[s], \ce{Fe2O3}[s],
    \ce{Fe2SiO4}[s], \ce{Al2O3}[s], \ce{CaTiO3}[s], \ce{CaSiO3}[s],
    C[s]),  of the materials that grow the bulk of the cloud particles.  Right axis (black dashed line) is the local  dust-to-gas ratio, $\rho_{\rm d}/\rho$ scales to $10^{-3}$.
    \textbf{Bottom:} Left axis (red solid line) is the average grain size, $\langle a\rangle$ [$\mu$m]. Right axis (black dashed line) is the vertical drift velocity,  $v_{\rm dr}$ [cm s$^{-1}$]  of the cloud particle of size  $\langle a\rangle$.     \label{fig:cloud_prop_grid}}
\end{figure*}

\section{Cloud formation on HAT-P-7b}\label{s:cloud}

First, we examine how and which kind of clouds form in the atmosphere of HAT-P-7b. 
In this section, we endeavour to provide insight into the global and local cloud structure for the ultra-hot Jupiter HAT-P-7b anticipated to display a vast asymmetry between its dayside and nightside.

{\it Our first result} is that no clouds form 
in the dayside equatorial regions. Some day-side clouds  form at higher and lower latitude outside the equatorial jets (Fig.~\ref{fig:1DJ*}), hence forming some patchy, but very optically thin clouds. A non-depleted, very warm gas-phase chemistry emerges inside the cloud-free equatorial jets, and a low-depletion gas in the higher latitudes on the dayside. The majority of in-situ cloud formation takes place  on the nightside of HAT-P-7b.  
At the northern evening terminator profile ($\theta=45\degree$, $\phi=90\degree$), seed particles form, but no condensation seeds can form at the equatorial evening terminator profile ($\theta=0\degree$, $\phi=90\degree$; 'empty' panels in Fig.~\ref{fig:cloud_prop_grid}), which therefore remains cloud-free.  Atmospheric superrotation affects the equatorial temperature such that the mid-latitudes are colder than the equator (compare also Figs.~\ref{fig:1DTp_1},~\ref{fig:orgData}) on the dayside. As a result of the high equatorial temperatures the gas is undersaturated such that  cloud particle growth in the dayside equator regions is not possible even if condensations seeds would be swept along with the winds from the nightside. Therefore, the different thermodynamic conditions in the two terminator regions cause the equatorial evening terminator ($\phi=90\degree$) to be cloud free and the  equatorial  morning terminator ($\phi=270\degree$) to form clouds. On the nightside, a vast amount of cloud particles form leading to a dust-to-gas-ratio of $\geq10^{-2.5}$ and an increase in the atmospheric C/O values to $\sim 0.7$.  


The combination of atmospheric regions with and without clouds on any given hemisphere can have profound implications for spectral observations. In particular, transmission spectra of terminators with such `patchy clouds' exhibit broader molecular features than either cloud-free or uniformly cloudy models predict \citep{Line2016,MacDonald2017}. We predict HAT-P-7b realises such inhomogeneous terminator clouds. However, the large thermodynamical differences between the day- and nightside, resulting in hemispheres with different geometrical extensions, poses an additional complication to the interpretation of transmission spectra. Conceptually, we can imagine the paths of stellar rays during transit traversing a greater distance 
through the larger, extended, largely cloud-free dayside, before impinging on the cloud-dominated nightside. This geometrical difference could potentially introduce biases in transmission spectra, arising from varying cloud properties along the width of the terminator.

The impact of 3D terminator differences on transmission spectra has recently been examined by \cite{Caldas2019}. Though they did not examine the impact of biases from rays crossing day / night cloud transitions, they noted, for the test case of GJ~1214b, that spectral features may be altered when a given chemical species undergoes a large abundance change between the day- and nightside hemispheres, producing an opacity discontinuity. In the case of HAT-P-7b, which is warmer than GJ 1214b by $\sim$ 2000K (GJ~1214b: $T_{\mathrm{eq}}\approx$500\;K; \citealt{Charbonneau09}), the atmospheric chemistry changes more continuously between the day- and nightside hemispheres (see Sect.~\ref{s:opacity}). However, one may imagine a similar opacity discontinuity arising from day / night cloud asymmetries along a tangent ray. Here, we will briefly examine whether this effect merits further consideration by estimating the width of HAT-P-7b's terminator.

The width of a terminator can be conceptually expressed as a simple function controlled by the ratio of the atmospheric scale height to planetary radius \citep{Caldas2019}. We first evaluate H/$R_p$ for the sub-stellar $(\phi, \theta)=(0\degree, 0\degree)$ and anti-stellar $(\phi, \theta)=(180\degree, 0\degree)$ points (i.e. the centre of the day- and nightsides). We follow the definition of an exponential scale height according to  Equ.~A.1 in \cite{Caldas2019}, such that 
\begin{equation}
    H = \frac{z_{\mathrm{max}}}{\mathrm{ln}(n(z_{\mathrm{max}})/n_{\rm tot})}. 
\end{equation}
Here, we use the total gas number density, $n_{\rm tot}$, and the maximal vertical extension of the atmosphere, $z_{\rm max}$, from our model for the sub-stellar and anti-stellar points in turn. $n_{\rm tot}$ is dominated by atomic hydrogen on the dayside and by molecular hydrogen on the nightside. We find H/$R_p = 2.8 \times 10^{-3}$ and $4.6 \times 10^{-3}$ at the anti-stellar and sub-stellar points, respectively. Inserting the average of these values into Eq. 3 from \cite{Caldas2019}, and taking $10^{-5}$ and $10^{-2}$~bar for the terminator photosphere pressure range (as in \citealt{Caldas2019}), we estimate the angular width of the terminator to be $\approx$ 18\degree. This suggests that HAT-P-7b will experience a less-pronounced bias due to hemispheric asymmetry than \cite{Caldas2019} found for GJ~1214b. We leave a full quantitative assessment of such 3D biases on observable spectra to the forthcoming Paper III.

\smallskip

In what follows, we give a general overview of our results to build an understanding of the cloud properties of HAT-P-7b.  Section~\ref{ss:how} concentrates on the cloud formation properties on the evening terminator ($\phi=90\degree$), anti-stellar point ($\phi=180\degree$), and morning terminator ($\phi=270\degree$), utilising the respective equator profiles. Section~\ref{ss:maps1} combines all 168 profiles (72 + 97) into maps of cloud properties at a certain pressure level to demonstrate their distribution across the globe. In order to capture the pressure (hence, atmospheric height) dependence, we use slice-plots which represent slices through the globe at a certain $\phi$ (longitude) or $\theta$ (latitude). The location of the slice-plots on the planet's globe  are visualised as green solid contours in Fig.~\ref{fig:Dayside_trajectories}.

\subsection{How clouds form on HAT-P-7b}\label{ss:how}
Cloud formation is triggered by the local thermodynamic conditions in
the atmosphere, which determine if sufficient gas species are
available to form condensation seeds and subsequently grow to
macroscopic particles.  Figure~\ref{fig:cloud_prop_grid} shows the
details of the cloud formation on HAT-P-7b for the anti-stellar point
$(\phi\ =\ 180\degree)$ and the morning terminator
$(\phi\ =\ 270\degree)$ at the equator ($\theta=0\degree)$. Both
regions form clouds which differ with respect to their local
details. Since the atmospheric gas is being transported from the hot
dayside to the nightside, no clouds form at the equatorial region of
the evening terminator ($(\phi, \theta)= (90\degree,
0\degree)$). These results are supplemented by the gas-phase chemistry
results in Sect.~\ref{s:maps} (Fig.~\ref{fig:Molecs}).

\begin{figure*}
    \vspace*{-0.5cm}
    \includegraphics[width=9.0cm]{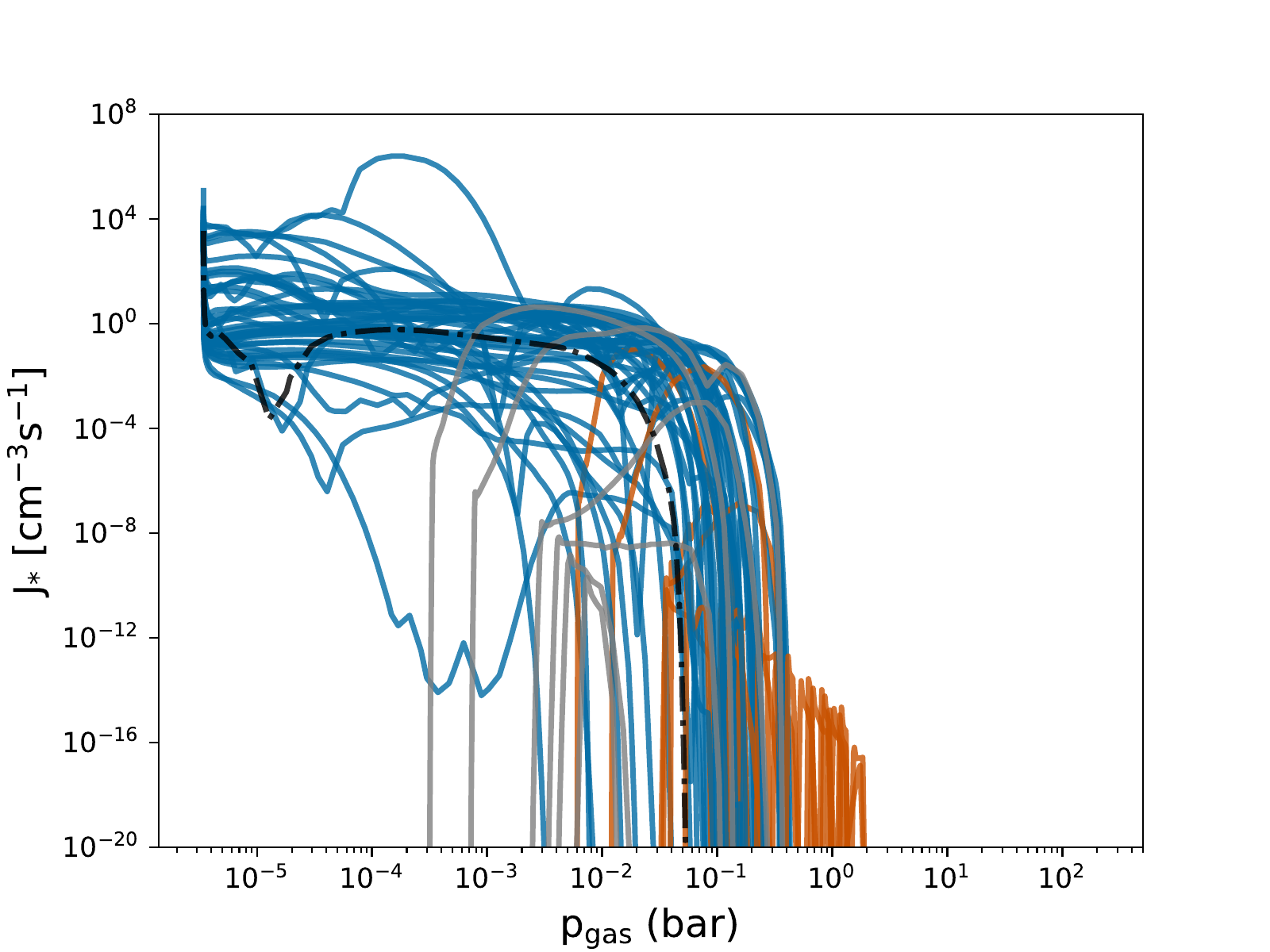}
   \includegraphics[width=8.6cm]{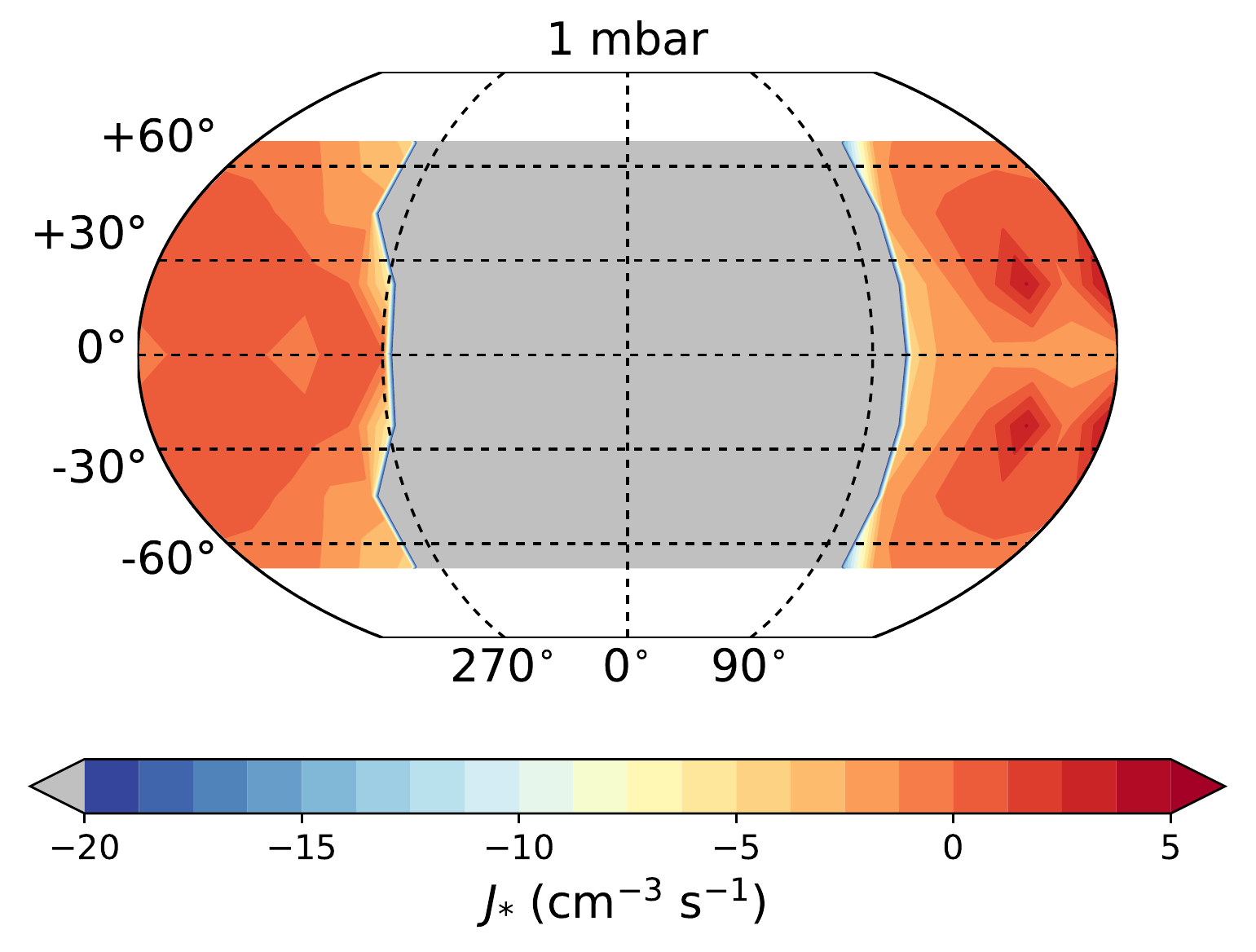}\\
    \includegraphics[width=8.7cm]{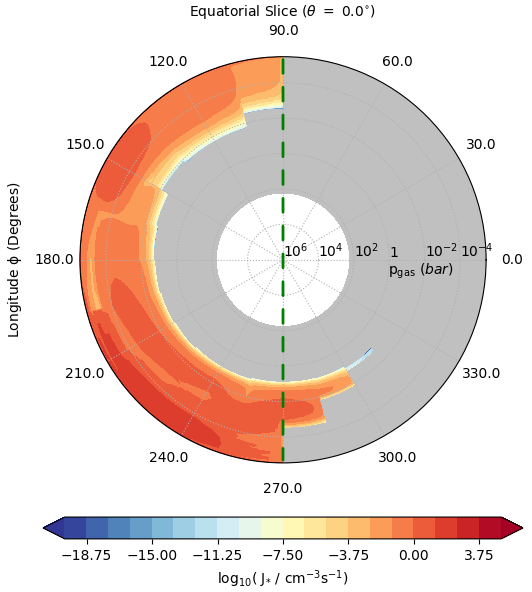}
    \includegraphics[width=8.7cm]{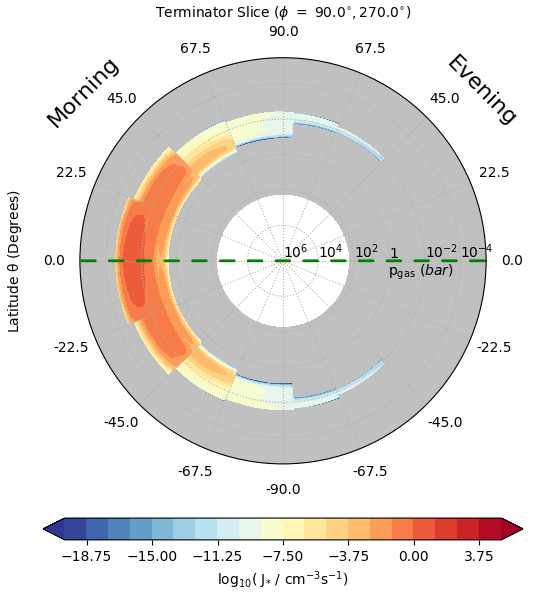}\\
    \caption{The global distribution of the  total nucleation rate, J$_*$ [cm$^{-3}$\,s$^{-1}$] ($J_*=\Sigma J_{\rm i}$, i=\ce{TiO2}, SiO, C) that triggers cloud formation on HAT-P-7b. 
    {\bf Top left:} 1D J$_*$ structure as function of  p$_{\rm gas}$ [bar] for  the probed 97 profiles (red -- dayside, blue --  nightside, grey dashed -- evening terminator $(\phi=90\degree)$, grey solid  -- morning terminator $(\phi=270\degree)$, black dot-dashed -- antistellar point  $(\phi=180\degree)$)
    {\bf Top right:}  Global 2D maps at p$_{\rm gas}=10^{-2}$\;bar.
    {\bf Bottom:} 2D cut through the equatorial plane. The smallest value shown in the slice plots is $10^{-20}$ [cm$^{-3}$\,s$^{-1}$].  The dashed green lines indicate where these two slice plots overlap. For the viewing geometry of the slice plots in the bottom row, please see Fig.~\ref{fig:vgeo}.
    }
    \label{fig:1DJ*}
\end{figure*}

The {\it cloud formation on HAT-P-7b is triggered} mainly by the formation of \ce{TiO2} condensation seeds (solid blue line, top row Fig.~\ref{fig:cloud_prop_grid}), with SiO  nucleation (dashed dark red line, top row Fig.~\ref{fig:cloud_prop_grid})  dominating at high altitudes on the nightside. Of the three nucleation species considered, $J_i$ [cm$^{-3}$ s$^{-1}$] ($i$=\ce{TiO2}[s], SiO[s], C[s]), carbon does not nucleate on HAT-P-7b. Nucleation at the equator only occurs up to $\phi\ =\ 315\degree$ into the dayside (Fig.~\ref{fig:1DJ*}, bottom left) due to favourable thermal conditions.  

\begin{figure*}
     \vspace*{-0.5cm}
    \includegraphics[width=8.7cm]{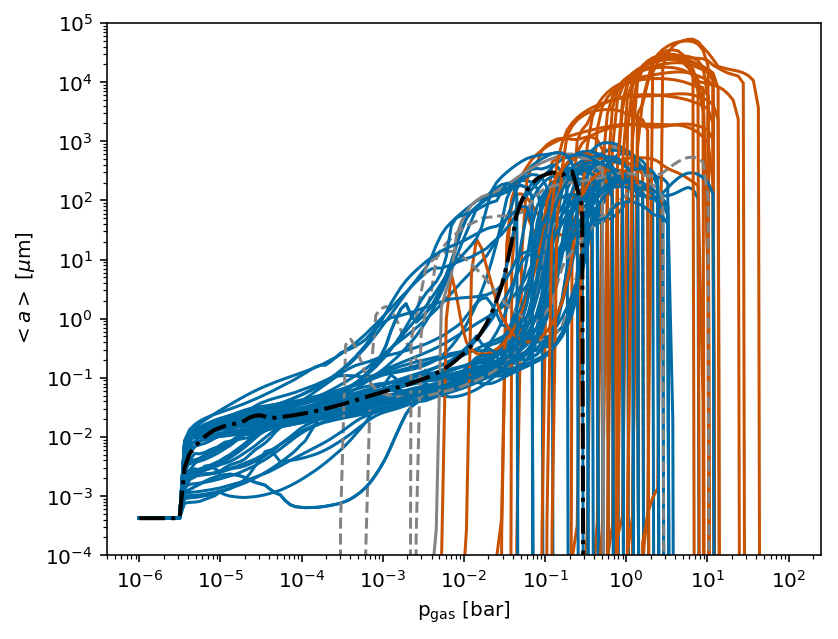}
     \includegraphics[width=8.7cm]{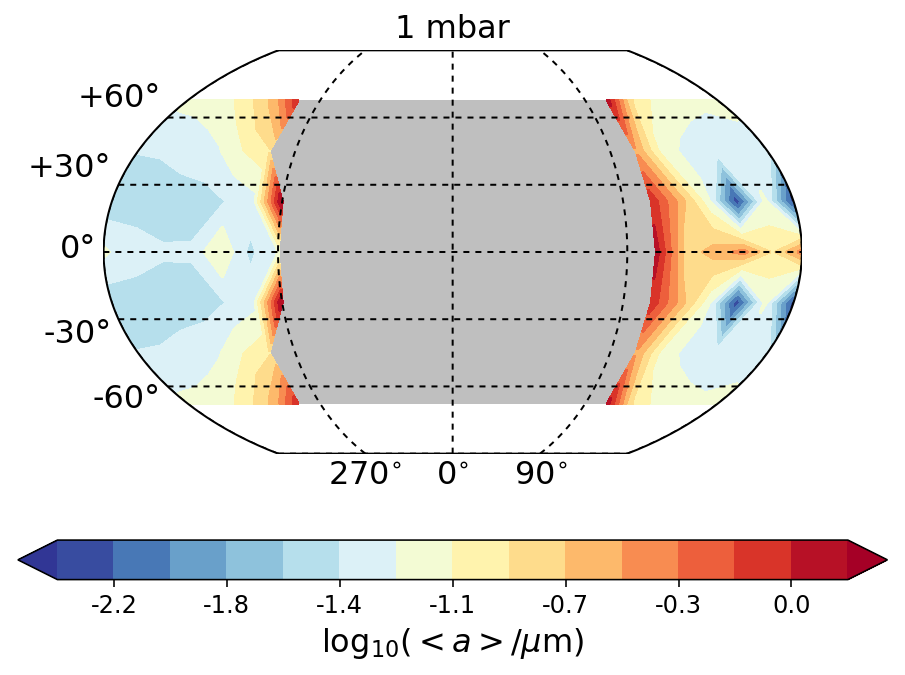}\\    
     \includegraphics[width=8.7cm]{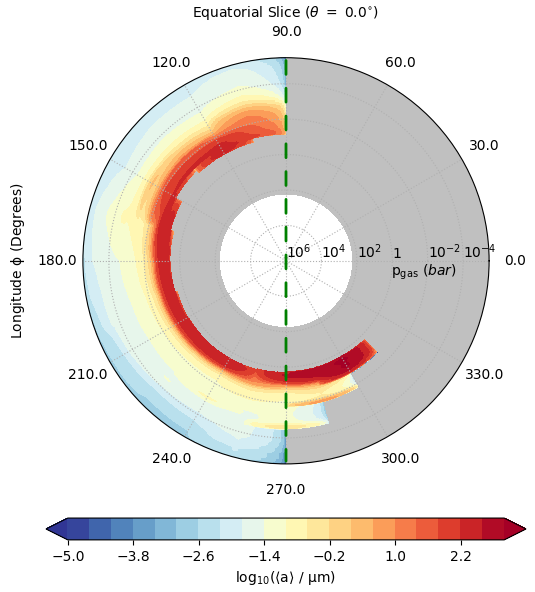}
\includegraphics[width=8.7cm]{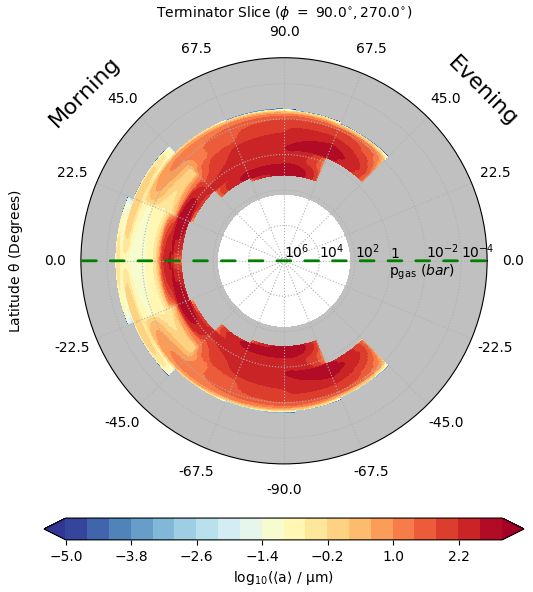}
    \caption{The global distribution of mean cloud particle sizes  $\langle a \rangle$ in microns on HAT-P-7b. The color coding is the same as in Fig.~\ref{fig:1DTp_1}: The day-side profiles are shown in red, the night-side profiles in blue, the black dot-dashed line is the anti-stellar point ($\phi, \theta)=(180\degree, 0\degree)$, the terminator profiles in grey.  {\bf Top Left:} 1D structures as function of p$_{\rm gas}$) [bar] of the probed 97 profiles. {\bf Top Right:} Global 2D map of mean particle size shows that sub-micron cloud particle dominates the night-side atmosphere at p$_{\rm gas}=10^{-2}$\;bar. {\bf Bottom left} 2D cut through the equatorial plane, {\bf Bottom right:}  2D cut along the terminator. The slice plots' cut of is at $\langle a\rangle =10^{-5}$\;[$\mu$m].  The dashed green lines indicate where these two slice plots overlap. For the viewing geometry of the slice plots in the bottom row, please see Fig.~\ref{fig:vgeo}.
 }
    \label{fig:1Damean_1}
\end{figure*}

\begin{figure*}
     \vspace*{-0.5cm}
    \includegraphics[width=8.7cm]{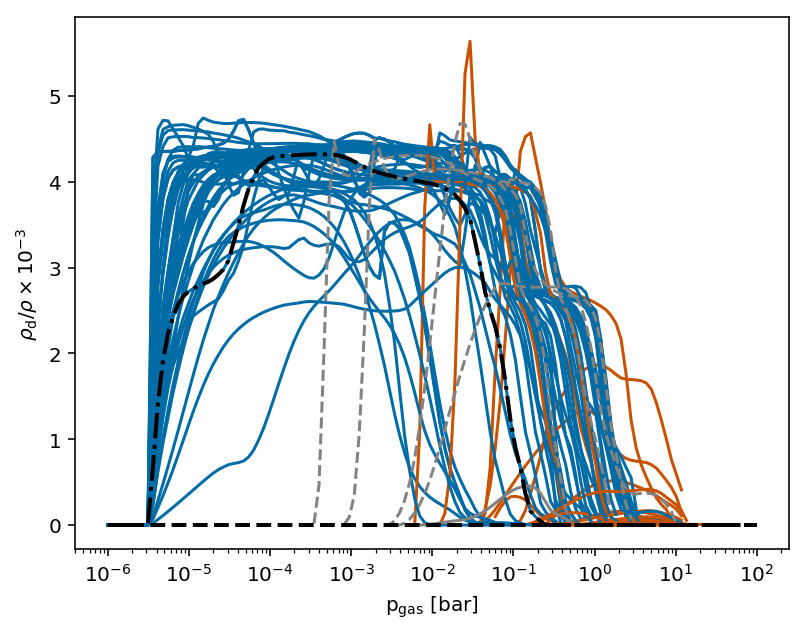}
     \includegraphics[width=8.7cm]{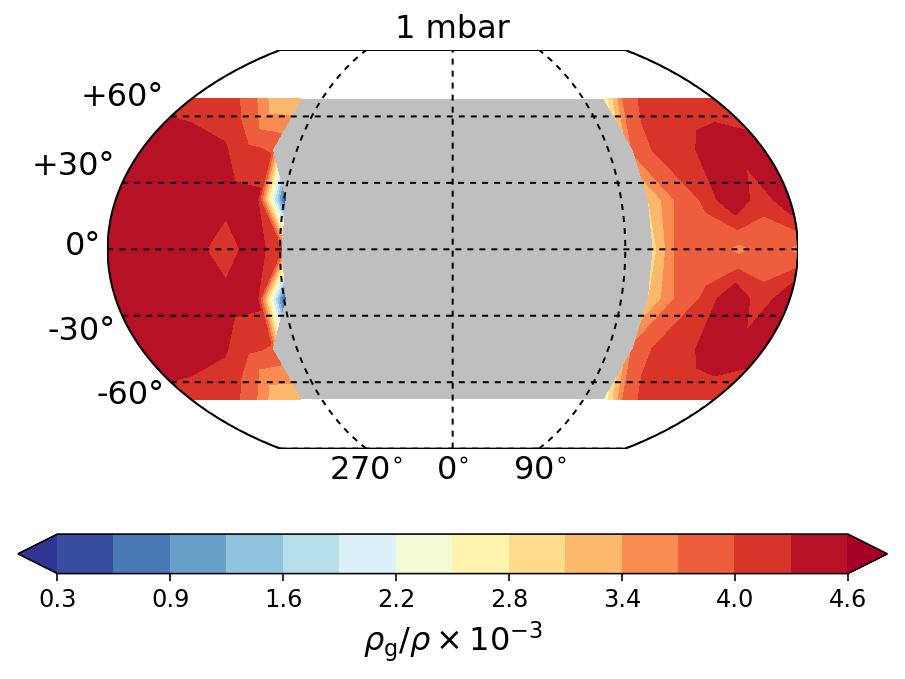}\\     \includegraphics[width=8.7cm]{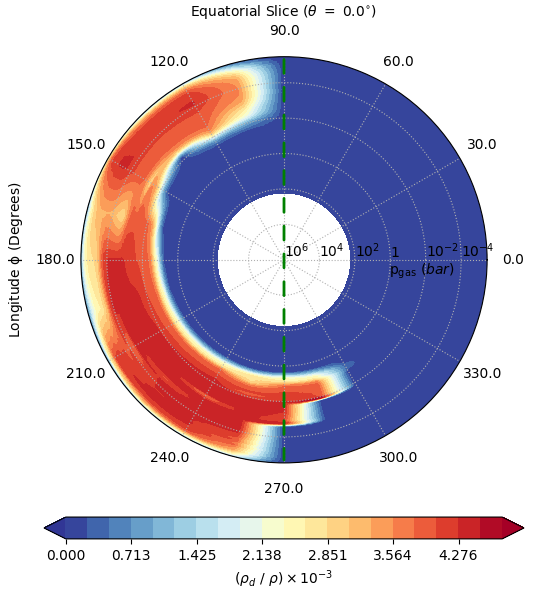}
     \includegraphics[width=8.7cm]{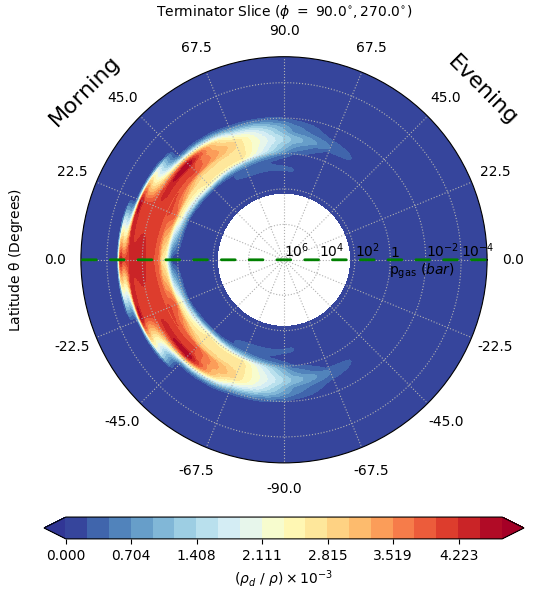}
    \caption{The global distribution of the cloud particle load in term of the dust-to-gas mass ratio  $\rho_{\rm d}/\rho$ on HAT-P-7b. The color coding is the same as Fig.~\ref{fig:1DTp_1}: The day-side profiled are shown in red, the night-side profiles in blue, the black dashed-dot line is the anti-stellar point $(\phi, \theta)=(180\degree, 0\degree)$, the grey lines are the terminator profiles (dashed -- evening, $\phi=90\degree$; solid -- morning  $\phi=270\degree$). {\bf Top right:} Dust-to-gas mass ratio as a function of p$_{\rm gas}$ [bar] of the probed 97 profiles. {\bf Top left:} global maps at p$_{\rm gas}=10^{-2}$\;bar. {\bf Bottom left} 2D cut through the equatorial plane, {\bf Bottom right:}  2D cut along the terminator. The slice plot cut-off is at $\rho_{\rm d}/\rho=10^{-3}$. The dashed green lines indicate where these two slice plots overlap. For the viewing geometry of the slice plots in the bottom row, please see Fig.~\ref{fig:vgeo}.}     \label{fig:rhodrhog}
\end{figure*}

\begin{figure*}
\centering
     \includegraphics[width=7.5cm]{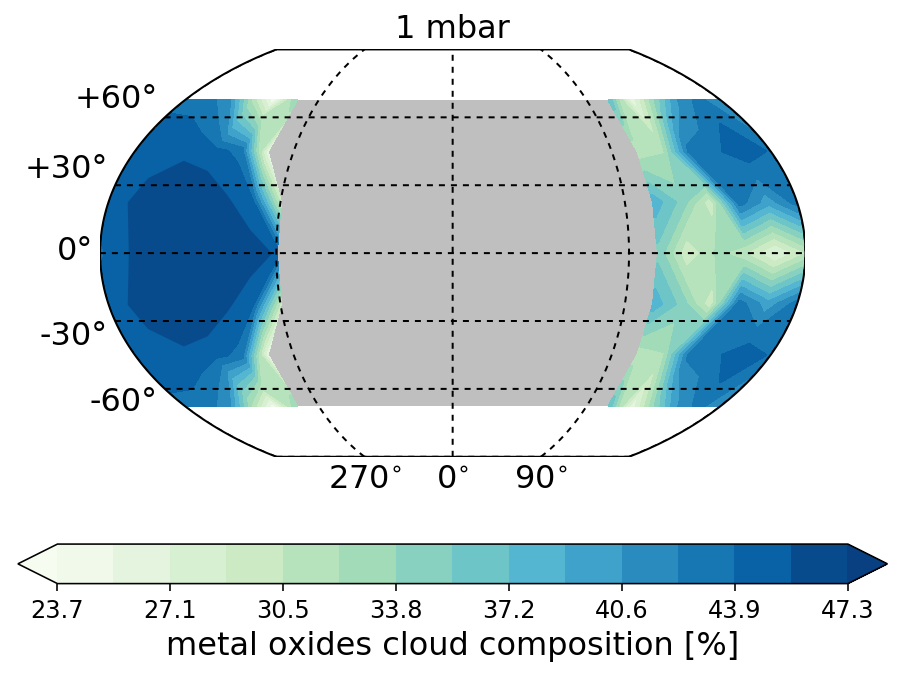}
     \includegraphics[width=7.5cm]{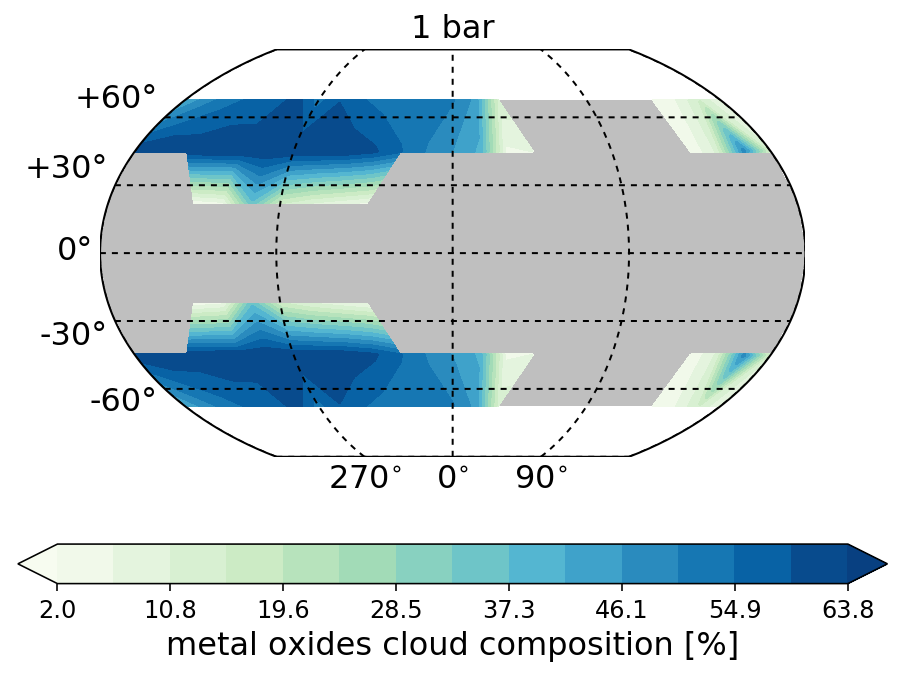}\\
     \includegraphics[width=7.5cm]{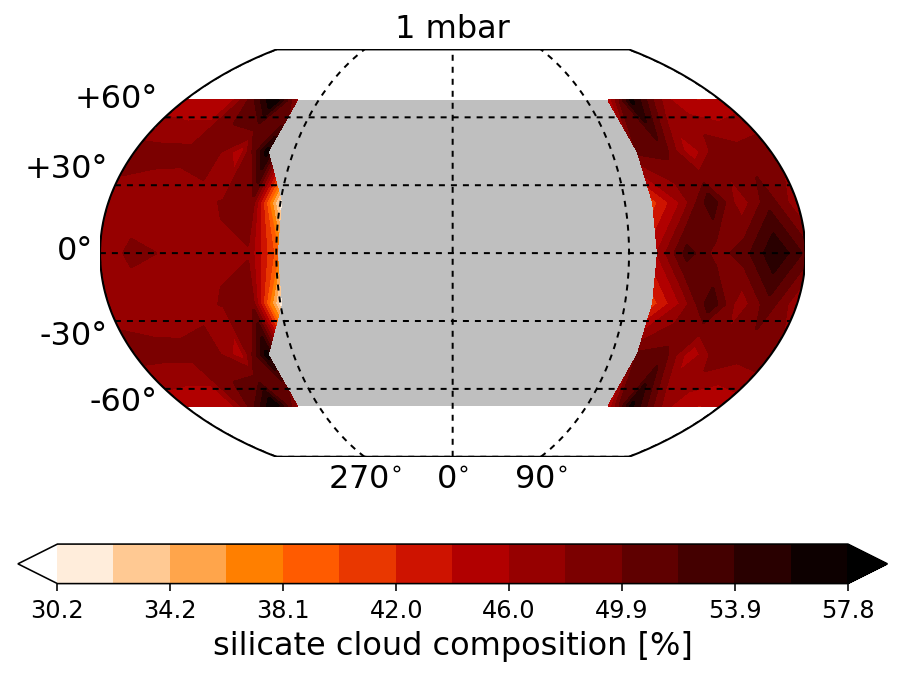}
     \includegraphics[width=7.5cm]{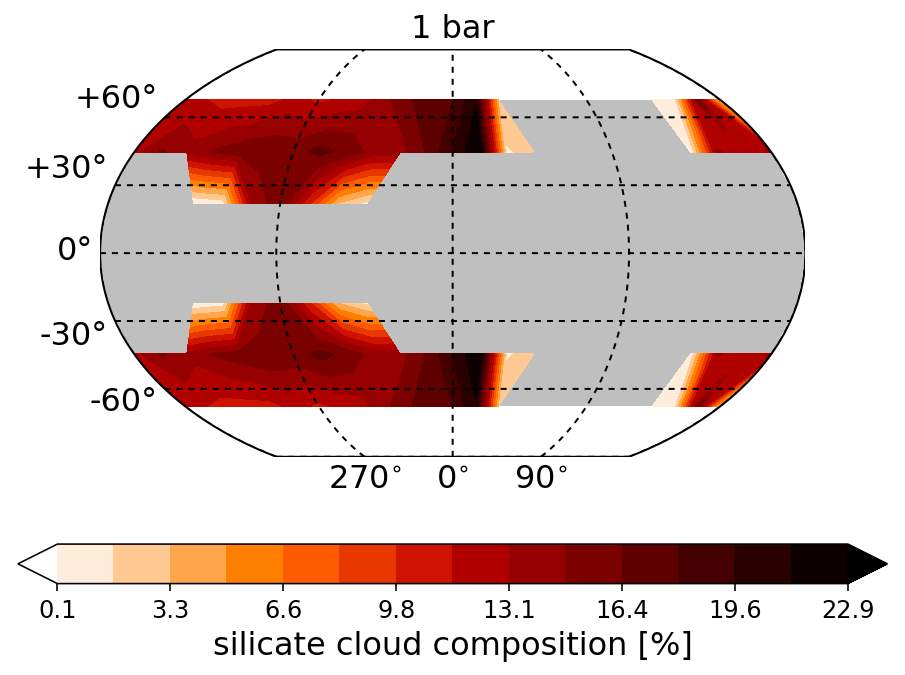}\\
     \includegraphics[width=7.5cm]{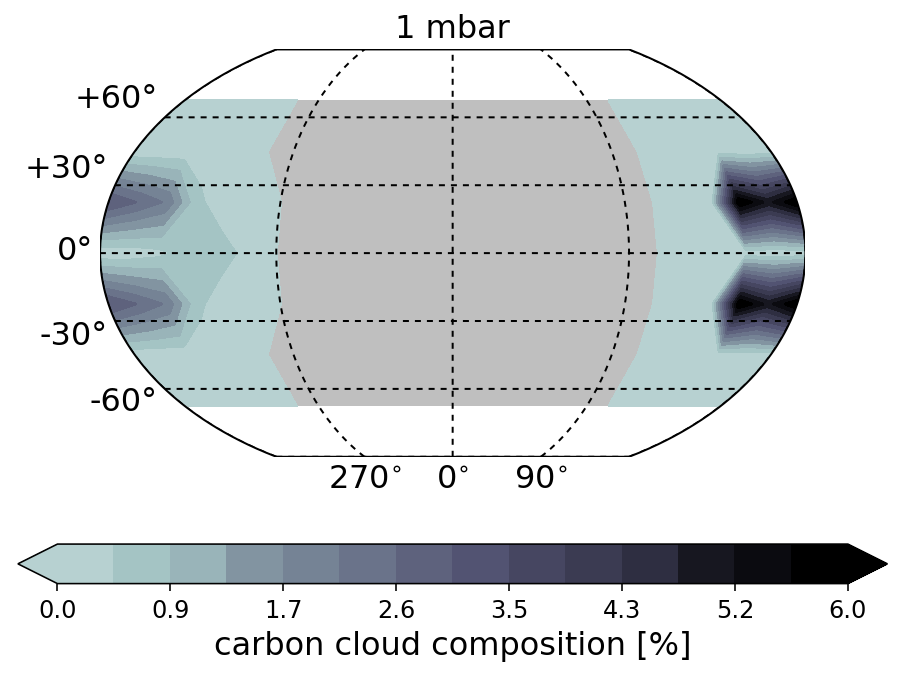}
     \includegraphics[width=7.5cm]{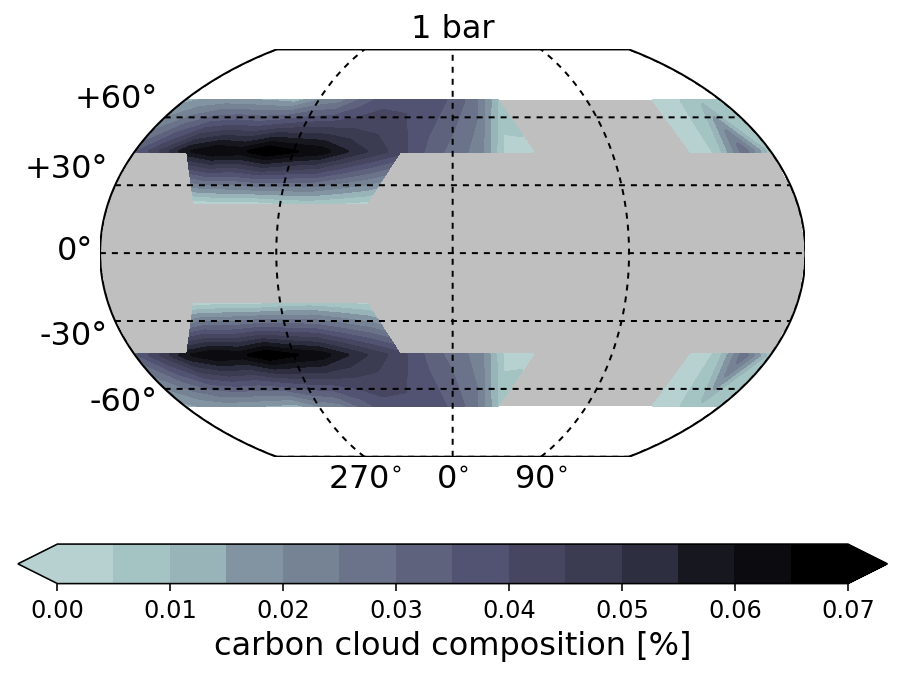}\\
    \includegraphics[width=7.5cm]{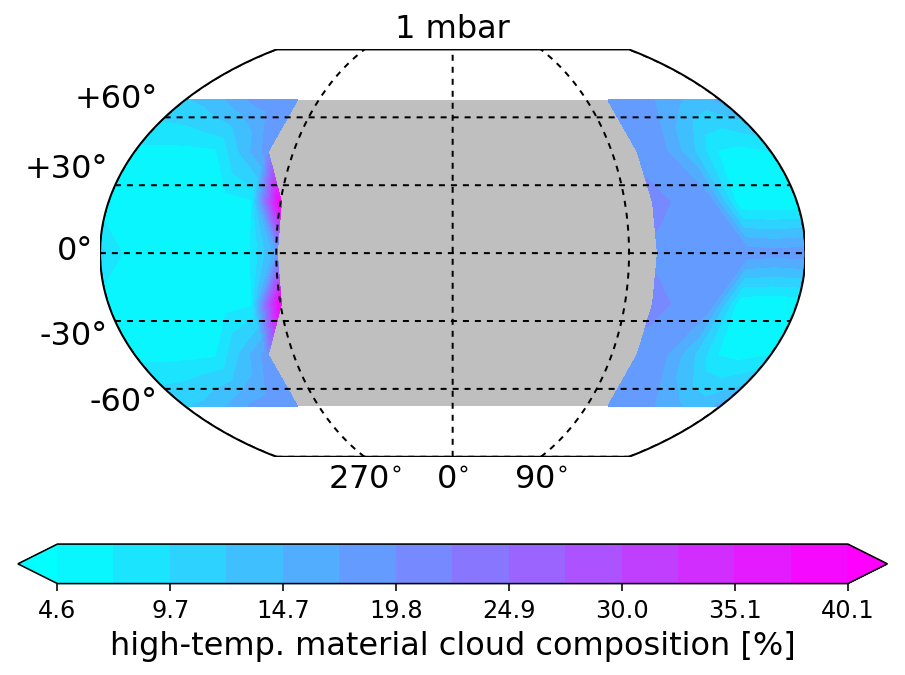}
    \includegraphics[width=7.5cm]{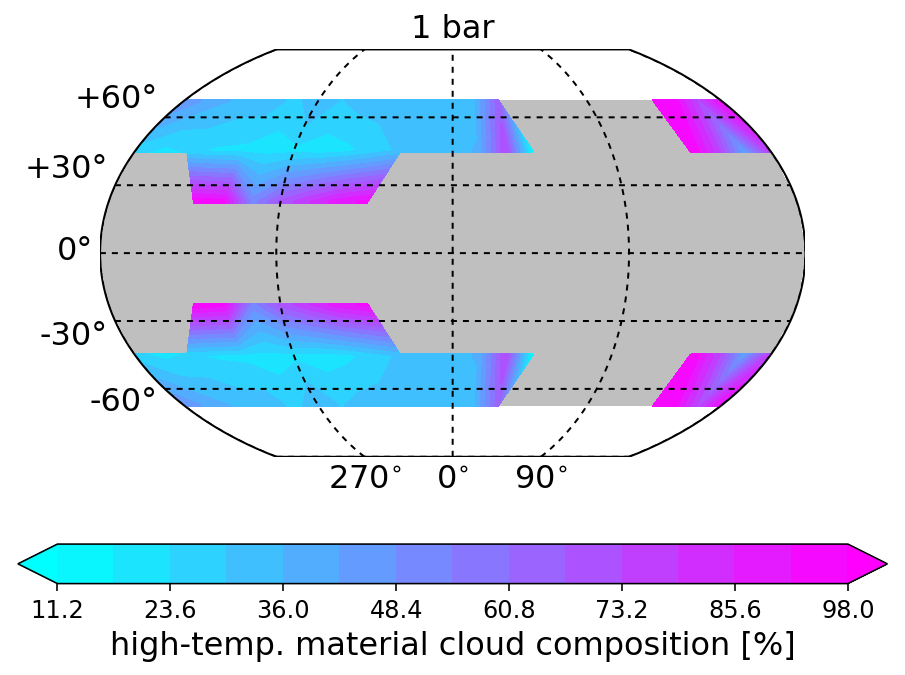}\\
  \caption{The cloud material composition in volume fractions, $\Sigma_{\rm s_i}V_{\rm s_i}/V_{\rm tot}$,  categorized into material groups: {\bf 1st row}: metal oxides ($s_i=\,$SiO[s], \ce{SiO2}[s], MgO[s], FeO[s], \ce{Fe2O3}[s]); {\bf 2nd row}: silicates ($s_i=\,$\ce{MgSiO3}[s], \ce{Mg2SiO4}[s], \ce{CaSiO3}[s],\ce{Fe2SiO4}[s]); {\bf 3rd row}: carbon material ($s_i=\,$C[s]); {\bf 4th row}: high temperature condensates ($s_i=\,$\ce{TiO2}[2], Fe[s], \ce{Al2O3}[s], \ce{CaTiO3}[s],FeS[s]). The left column shows the 1\;mbar level, the right column shows the 1\;bar level.}
      \label{fig:cloud_map_1bar_2}
\end{figure*}

The {\it number density} of cloud particles, $n_{\rm d}$ [cm$^{-3}$], (dashed black line, top row, Fig.~\ref{fig:cloud_prop_grid}) mostly follows the total nucleation rate, $J_*$, with a tail expanding into higher pressures regions inside the atmosphere due to gravitational settling of cloud particles that form higher in the atmosphere. Gravitational settling becomes the main cause of the presence of cloud particles at deep atmospheric layers reaching pressures  $\sim 10^{-1.3}$\;bar for the anti-stellar point, and at $\sim 10^{-1}$\;bar for the morning terminator.
 
The {\it gravitational settling} is characterised by the height-dependent drift velocity, $v_{dr}$ [cm\,s$^{-1}$] (black dashed line, 3rd row,  Fig.~\ref{fig:cloud_prop_grid}) which depends on the cloud particle sizes ($\langle a\rangle$ [$\mu$m]; red line, 3rd row, Fig.~\ref{fig:cloud_prop_grid}), its bulk mass density ($\rho_{d}$ [g cm$^{-3}$]) and the density of the surrounding gas phase ($\rho_{\rm gas}$ [g cm$^{-3}$]).  It is defined as the relative velocity between the cloud particles and the gas flow $v_{dr}\ =\  v_{d}-v_{gas}$ \citep{Woitke2003}.  The mean equilibrium drift velocity for large Knutsen numbers is 
$    \langle v_{dr}\rangle  = -(\sqrt{\pi}/2)\cdot(g\rho_{d}\langle a\rangle)/(\rho_{gas}c_T),$
where $c_T$ is the speed of sound in the gas which is $\sim\sqrt{T_{\rm gas}}$.  From $\langle v_{dr}\rangle \sim  \langle a\rangle/\rho_{gas}$ follows that the $1/\rho_{gas}$ dominates in the low-density regions where cloud particles are small and  $\langle v_{dr}\rangle \sim  \langle a\rangle$ dominates in the inner atmospheric regions where cloud particles are big. Here nucleation has almost completely stopped, and cloud particles are present due to gravitational settling from higher layers. These cloud particles continue to grow as they fall and, hence,  the mean particle size increases  the drift velocity to values of $10^3$ cm s$^{-1}$. 
The drift velocity gradual rises towards higher altitudes owing to a decreasing atmospheric density ($    \langle v_{dr}\rangle \sim 1/\rho_{\rm gas}$) and thus less resistance to gravitational settling by frictional forces. This continues until the point where the vast majority of dust particles are recently nucleated clusters that have experienced minimal growth, and hence are very small and buoyant even in a rarefied environment. The peak in the mean particle size appears only at the morning terminator (right most panels of Fig.~\ref{fig:cloud_prop_grid}) and is explained by the shift to lower altitudes as discussed previously, this atmosphere is much more suitable to growth, and thus particles rapidly grow after nucleation to a mean size of $\sim 1\;\mu$m.


We measure the {\it cloud particle load} of the atmosphere of HAT-P-7b by the dust-to-gas mass (density) ratio, $\rho_{\rm d}/\rho$ (black dashed line in Fig.~\ref{fig:cloud_prop_grid}) which traces the location of the cloud very well. $\rho_{\rm d}/\rho$ shows that the top part of the clouds on the morning terminator ($\phi=270\degree$) have a thin layer with a high cloud particle load (peaking at $\rho_{\rm d}/\rho_{\rm gas}\approx 4.5\cdot10^{-3}$) and the top layers of the nightside cloud at the anti-stellar point ($\phi=180\degree$) are less mass-loaden ($\rho_{\rm d}/\rho_{\rm gas}\approx 2.5\cdot10^{-3}$).

The {\it cloud particle composition} varies dramatically both throughout atmospheric pressure levels and at different points around the globe. This is typified by what we see at the anti-stellar point and the morning terminator in Fig.~\ref{fig:cloud_prop_grid}. At the anti-stellar point, in regions where nucleation is the dominant driver of the cloud particle number density (i.e., at high altitudes) the composition is mostly SiO[s] which is the main nucleating species at this profile and at this altitude. \ce{TiO2}[s]  and other high-temperature condensates make up the upper-most cloud layer at the morning terminator ($\phi=270\degree$). Once seed particles are present, many materials can grow almost simultaneously on these surfaces as surface growth is energetically easier than the gas-solid phase transition that forms the condensation seeds. Therefore, 
the cloud particles are made of all materials that are thermally stable, 
leading to so-called 'mixed grains', up until pressures above $\sim$ $10^{-1.5}\;$bar which are dominated by silicates such as \ce{Mg2SiO4}[s], \ce{MgSiO3}[s], MgO[s], and \ce{Fe2SiO4}[s] which together contribute $\approx 60\%$ of the cloud particle volume. In atmospheric regions below this,  a chemical transition regions emerges where the complex Mg/Si/Fe/O-materials have evaporated but smaller metal oxides like MgO[s], SiO[s] and FeO[s] are still thermally stable.  The high temperature species (\ce{Al2O3}[s] and \ce{CaTiO3}[s]) dominate the innermost cloud layers, as  all other species have evaporated at these temperatures. Using these detailed results, we will define material groups in Sect.~\ref{sec:map_comp} for our study of global distributions of cloud properties on HAT-P-7b.

The nightside and the morning terminator have very comparable cloud characteristics with respect to the mean particle sizes, nucleation rates, dust-to-gas ratios and even the cross-material composition in the centre part of the cloud. Differences, however, are apparent, too, which are linked to the locally different temperatures and pressures. Therefore,  different materials are the most abundant at specific atmospheric locations and the mean particle sizes being lower by ca. 2 order of magnitudes in the upper regions of the night-side cloud compared to the morning terminator.   Section~\ref{ss:maps1} will discuss such global changes of  the cloud properties in HAT-P-7b's atmosphere. For now, we point out that the clouds forming on the morning terminator are bracketed by $\approx 1\mu$m \ce{Al2O3}[s]/ \ce{CaTiO3}[s]\ce{TiO2}[s] cloud particles at the top and  $>\,100\mu$m \ce{Al2O3}[s]/ \ce{CaTiO3}[s]\ce{TiO2}[s] cloud particles at the cloud's inner rim.


The {\it cloud extension} at a given location in an exoplanet atmosphere is controlled by two main factors, thermal stability of the cloud particles in the inner atmosphere and seed formation in the upper atmosphere. At high pressures, all 1D temperature profiles converge for HAT-P-7b (Fig.~\ref{fig:1DTp_1}), meaning that in cloud forming regions (nightside and morning terminator) the lower cloud boundary is roughly uniform. However, the morning terminator, at low pressures, has a temperature inversion (at $\sim 5\times 10^{-4}\;$bar). This is unlike the nightside profiles, and limits the altitude at which condensation seeds can form. This leads to an overall lower cloud top and a higher cloud-top pressure, which has implications for atmospheric opacity (see Sect.~\ref{s:opacity}). We also note that the geometrical extension of the cloud will depend on the geometrical extension of the atmosphere which is different for the day- and the nightside as pointed out previously.
\subsection{Mapping seed formation, 
mean grain sizes and dust-to-gas ratios}\label{ss:maps1}

Having discussed how clouds form and the causal dependence of their characteristic properties in Sect.~\ref{ss:how}, we now study the cloud formation on HAT-P-7b on global scales. We aim to demonstrate if and how the cloud properties change from the day- to the nightside and provide an understanding of the underlying cloud formation processes. We utilize 2D maps for specific pressure levels ($10^{-3}$ bar) and 2D slices through the equatorial plane and along the terminator (green circumferences in Fig.~\ref{fig:Dayside_trajectories}). Grey areas indicate regions where no cloud particles are present, white areas indicate regions that are outside our computational domain. The physical quantities are colour coded.

We begin our discussion again with the nucleation rate as it determines how many particles are formed in a planetary atmosphere (assuming no other mechanisms like volcanoes or wild fires are available). Figure~\ref{fig:1DJ*} demonstrates  how the nucleation rate changes globally and the comparison with Fig.~\ref{fig:1DTp_1} shows that the nucleation rate is directly linked to the atmospheric temperature distribution: Seed formation occurs most efficiently on the nightside, hence, the cloud particle number density is highest where the nucleation is most efficient in our model.

Figure~\ref{fig:1DJ*} shows that seed formation is confined to a
pressure interval of up to 6 orders of magnitude
($10^{-6}\,\ldots\,0.5$\;bar) and that the inner pressure boundary for
nucleation varies for all the profiles tested here. This pressure
range is confined at the top by the upper pressure boundary of the
GCM, hence, could extend to lower pressures and temperature.  The seed
formation can prevail at higher temperatures if the pressure is very
high because thermal stability increases with increasing pressure (see
night/day terminator at $\phi=-90\degree$ in 2D cut in
Fig.~\ref{fig:1DJ*}). Similar observations have been made for the
inner boundaries of the clouds in the atmosphere of the cooler giant
gas planet HD\,189733b \citep{Lee2016}. The resulting cloud particle
number density, $n_{\rm d}$ [cm$^{-3}$], therefore changes rapidly
from the night- to dayside (not shown as map) where no or only very
little seed formation occurs.  The highest cloud particle number
density correlates with where the seed formation is most efficient as
long as the horizontal mixing time scale is longer than the nucleation
time scale.

Comparing the nucleation rate (Fig.~\ref{fig:1DJ*}) to the mean cloud particle size (Fig.~\ref{fig:1Damean_1}) shows that regions with efficient seed formation contain on average the smallest particles, $\sim 0.01-0.03$\;$\mu$m. This observation holds for the vertical profiles as well as for the global distribution of the cloud particles sizes at different pressure levels (see also Fig.~\ref{fig:cloud_map_1bar}).
The mean grain size varies throughout the vertical extension of the cloud and changes from the size of a seed particles to a maximum size of  $\langle a\rangle_{\rm max}\approx 0.1\mu$m$\,\ldots\,5$\;cm.
The largest cloud particles  
are found in the hotter part of the atmosphere where the very high pressure support their thermal stability (left top panel in Fig.~\ref{fig:1Damean_1}) and were the rate of growth is much higher due to  higher gas-number densities.  Cloud particle sizes are bigger at higher pressures and in regions where the nucleation rate is very low (Fig.~\ref{fig:cloud_prop_grid}).  Low seed formation rates imply a low number of cloud particles  which therefore can grow bigger for any given amount of condensing gas. Particular low (but non-zero) seed formation occurs in the terminator regions and at higher pressure in the higher latitudes 
where the cloud particles can, hence,  grow to substantial sizes of $>10^3\mu$m  (compare also middle plot in Fig.~\ref{fig:cloud_map_1bar}). A comparison with $\rho_{\rm d}/\rho$ shows that the cloud particle load remains very low in these regions. The global and local size differences of the cloud particles are furthermore correlated to the material composition of the cloud particles as both are determined by the local gas temperature and gas pressure (Sect.~\ref{sec:map_comp})

Our results show that the cloud particle sizes vary strongly globally at the nightside and across the terminator regions but also vertically throughout the atmosphere, hence, no one cloud particle size will be sufficient to characterise the clouds that form on HAT-P-7b.  The size of the cloud particles are  also of interest for later opacity calculations, but do not necessarily provide a good measure for where most of the cloud mass is concentrated compared to the atmospheric gas.  It is therefore instructive to measure the cloud particle mass load of the atmosphere in terms of the dust-to-gas ratio, $\rho_{\rm d}/\rho$ (where $\rho_{\rm d}$
is the cloud particle mass density and $\rho$ is the gas mass density, both in [g cm$^{-3}$]).  $\rho_{\rm d}/\rho$  is used to measure the enrichment of a gaseous medium with solid particles, for example in planet-forming discs, cometary tails, the ISM, in winds of AGB stars ($\rho_{\rm d}/\rho\approx 10^{-3}$ \citealt{2000A&A...361..641W}) but also as as indicator for dust evolution  and metallicity in cosmological simulations  ($\rho_{\rm d}/\rho \approx 10^{-1.5}\,\dots\,10^{-5}$ \citealt{2019MNRAS.485.1727H}) and super novae ejecta \citep{2019MNRAS.485..440P}. The dust-to-gas ratio changes with temperature for systems in thermal equilibrium and the maximum values in thermal equilibrium are $(\rho_{\rm d}/\rho)_{\rm ThE} \approx 0.004 \,\ldots\, 0.0052$ for a gas of solar element abundances \citep{2018A&A...614A...1W}.

Hence, mapping $\rho_{\rm d}/\rho$ in 2D can be used as indicator for
the spatial distribution of the cloud particle load in a planetary
atmospheres. In the equatorial plane, the cloud particle mass load in
HAT-P-7b ranges from zero (in cloud-free regions) to values of
$(\rho_{\rm d}/\rho)_{\rm ThE}\approx 0.0043$ (bottom left of
Fig.~\ref{fig:rhodrhog}).  The shift of the maximum $\rho_{\rm
  d}/\rho$ towards lower pressure (higher altitudes) compared to the
maximum in cloud particles sizes indicates a higher number of cloud
particles that contribute to the atmospheric cloud particles
load. This is consistent with our findings in Sect.~\ref{ss:how} and
our previous argument that bigger cloud particles are an indication
for low cloud particles number densities (unless found in very-high
pressure regions where their growth is super-supported). We also
identify one profile with $\rho_{\rm d}/\rho > (\rho_{\rm
  d}/\rho)_{\rm ThE}$ which is linked to non-solar element abundances
due to depletion/enrichment by cloud formation, and the kinetic
character of the cloud formation process in contrast to a
phase-equilibrium approach.
Furthermore,   $\rho_{\rm d}/\rho$ varies throughout the cloud and also globally. This is a result of the different thermodynamic conditions that occur in planetary atmospheres, particularly the large day and night temperature differences on ultra-hot planets like HAT-P-7b. 

\begin{figure*}
     \vspace*{-0.5cm}
     \centering
    \includegraphics[width=7.2cm]{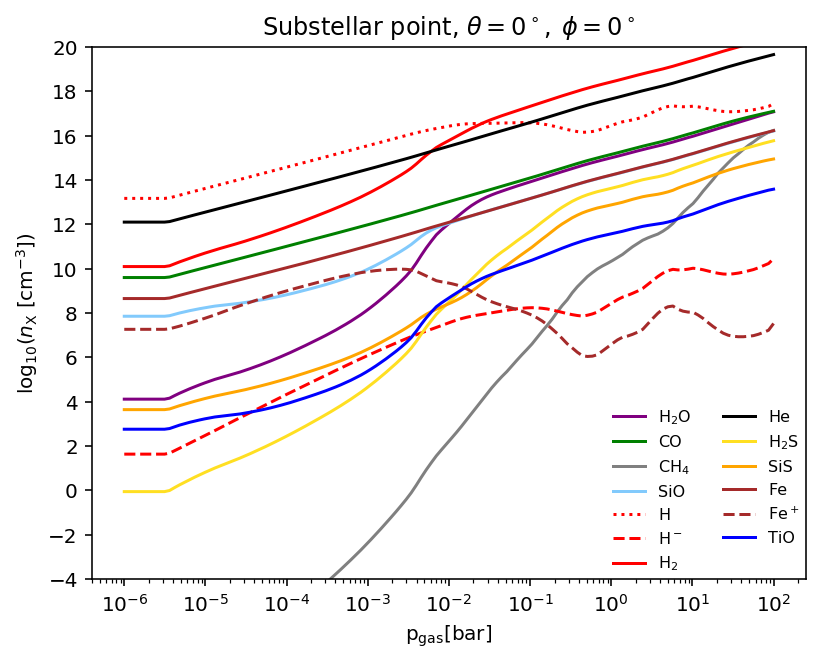}
    \includegraphics[width=7.2cm]{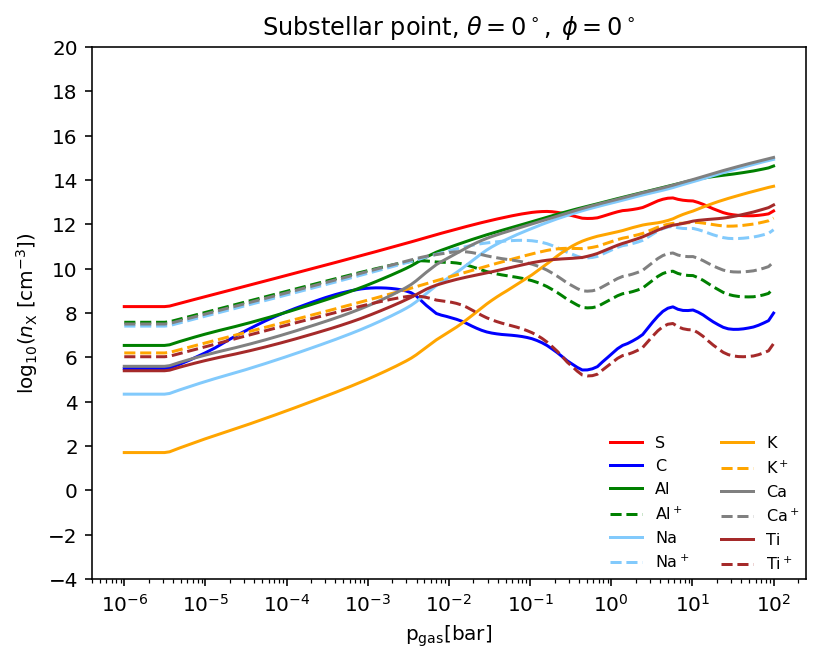}\\
    \includegraphics[width=7.2cm]{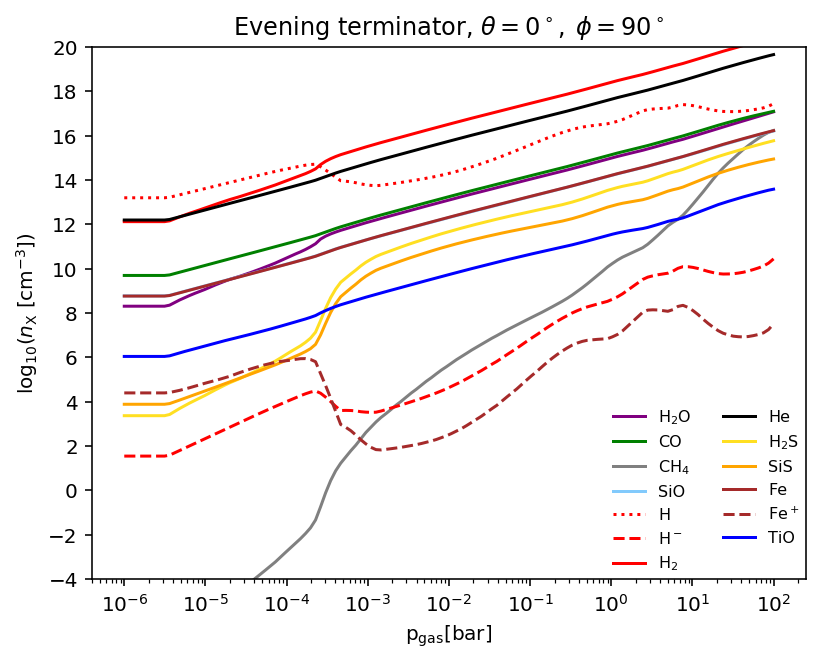}
    \includegraphics[width=7.2cm]{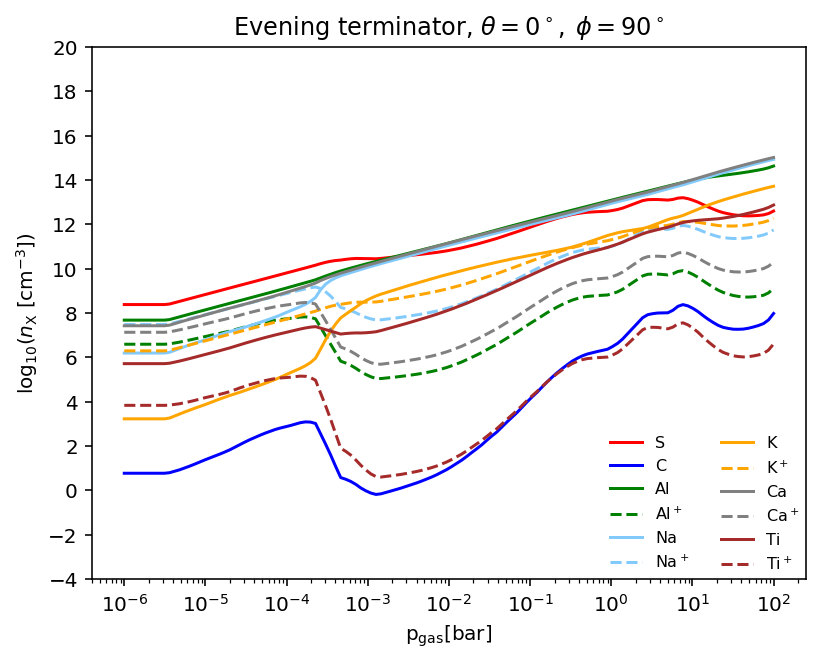}\\
    \includegraphics[width=7.2cm]{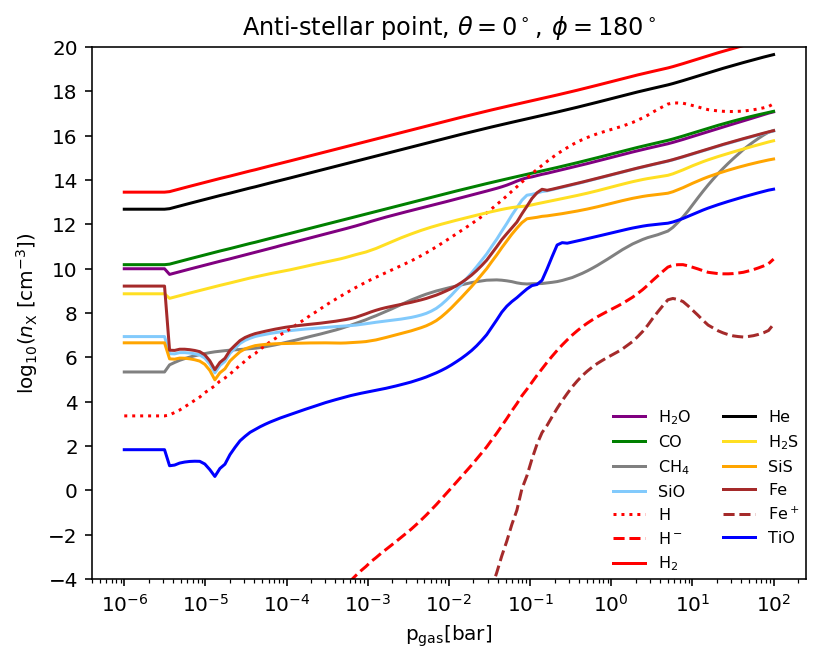}
    \includegraphics[width=7.2cm]{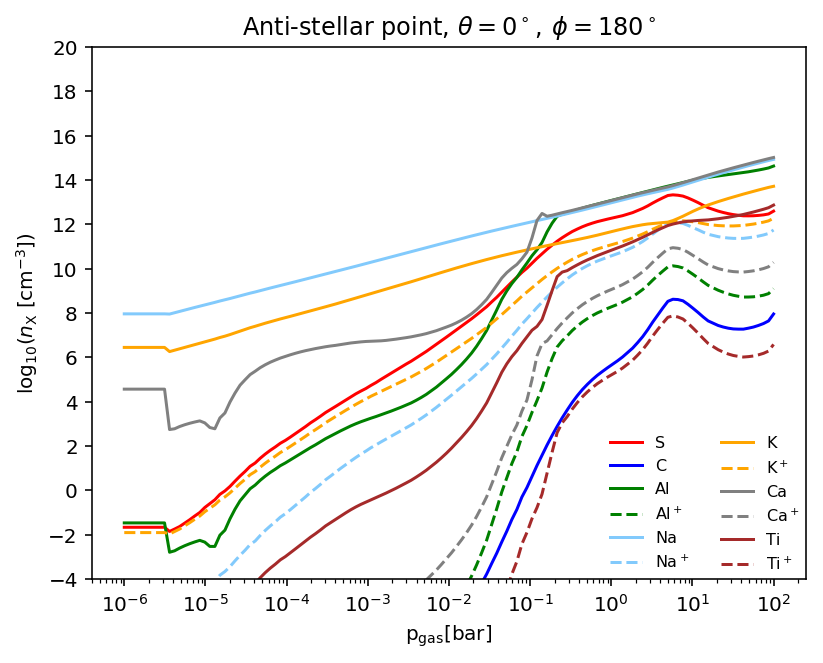}\\
    \includegraphics[width=7.2cm]{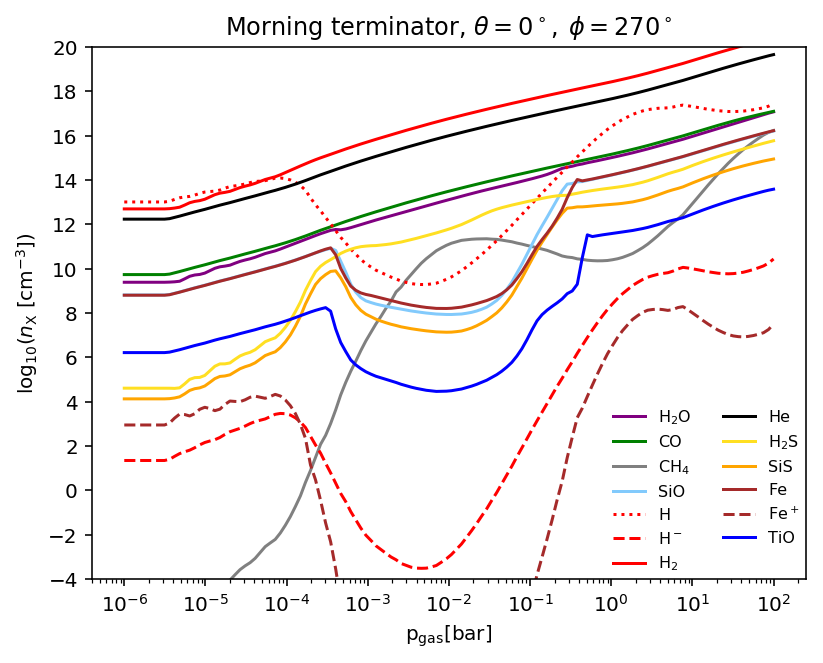}
    \includegraphics[width=7.2cm]{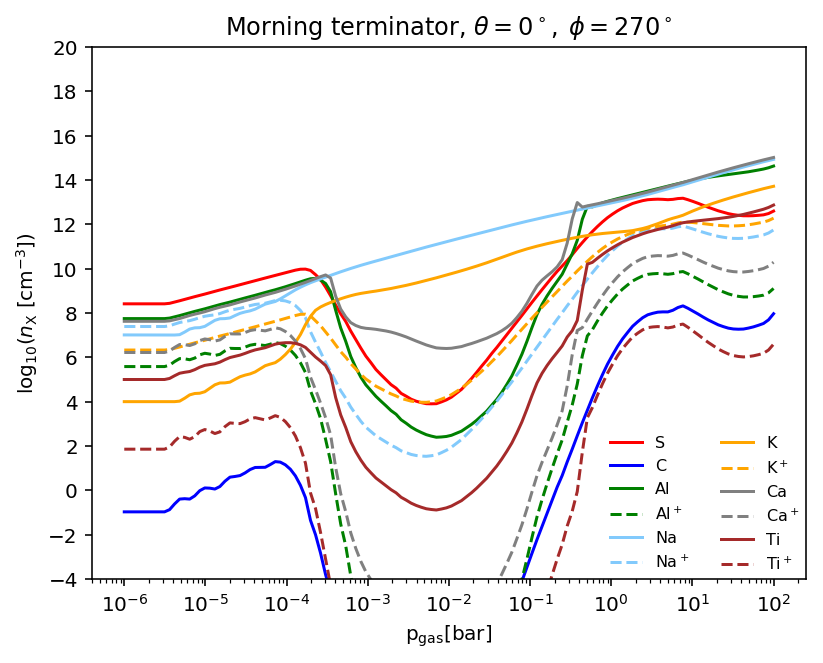}

    \caption{The number densities of essential gas components for the four most distinct equatorial profiles of our study. {\bf First row:} substellar point ($\theta,\phi)=(0\degree, 0\degree)$, {\bf Second row:} evening terminator ($\theta,\phi)=(0\degree, 90\degree)$, {\bf Third row:} anti-stellar point ($\theta,\phi)=(0\degree, 180\degree)$, {\bf Fourth row:} morning terminator ($\theta,\phi)=(0\degree, 270\degree)$. The changing regimes of \ce{H2}/H on the nightside, dayside and at the terminators.}
    \label{fig:Molecs}
\end{figure*}

\subsection{Mapping the cloud particle material composition}\label{sec:map_comp}

We have detailed the principle results for where and why clouds form
on HAT-P-7b in Sect.~\ref{ss:how} and demonstrated in
Fig.~\ref{fig:cloud_prop_grid} how nucleation rate, mean particle,
size and drift velocity relate to the condensation of 15 individual
cloud formation materials. Here, we concentrate on how the material
composition of the cloud particles on HAT-P-7b changes globally which
completes our discussion of the global cloud properties for
HAT-P-7b. The material composition of the cloud particles is important
for deriving optical properties and chemical feedback on the gas-phase
due to element depletion or enrichment. In order to see how the
material composition changes globally, and consequently affect the
spectral appearance, we define material groups by making use of the
detailed results from
Sect.~\ref{ss:how}. Figure~\ref{fig:cloud_prop_grid} shows that the
centre part of the cloud appears dominated by silicates
(e.g. \ce{MgSiO3}[s], \ce{Mg2SiO4}[s], \ce{Fe2SiO4}[s]) and metal
oxides (e.g.  SiO[s]). The inner cloud boundary is dominated by the
thermally most stable materials, i.e. high-temperature condensates
like \ce{TiO2}[s], Fe[s], \ce{Al2O3}[s]. We therefore define the
following four groups of materials for which we calculate the volume
contribution, $\Sigma_{\rm s_i}V_{\rm s_i}$ (${\rm s_i}=$ numbers of
materials in group), relative to the total cloud volume, $V_{\rm
  tot}$:
\begin{itemize}
     \item {\it Metal oxides:}\\ SiO[s], \ce{SiO2}[s], MgO[s], FeO[s], \ce{Fe2O3}[s];
     \item {\it Silicates:}\\ \ce{MgSiO3}[s], \ce{Mg2SiO4}[s], \ce{CaSiO3}[s], \ce{Fe2SiO4}[s];
     \item {\it Carbon material:} C[s];
     \item {\it High-temperature condensates:}\\ \ce{TiO2}[s], Fe[s], \ce{Al2O3}[s], \ce{CaTiO3}[s], FeS[s].
\end{itemize}

Figure~\ref{fig:cloud_prop_grid} demonstrates how cloud particles
change in composition (depicted as volume fractions of single
materials $s$, $V_{\rm s}/V_{\rm tot}$) as they move vertically
through the atmosphere.  Figure~\ref{fig:cloud_map_1bar_2}
demonstrates the contribution of the material groups to the total
cloud volume (depicted as volume fractions of the sum over all
materials per group $\Sigma_{\rm s_i} V_{\rm s_i}/V_{\rm tot}$) for
two selected pressure levels, 1bar and 1mbar.  The chemical
composition is also globally very well correlated with the atmospheric
temperature. Comparing the p$_{\rm gas}=1$~mbar maps in
Fig.~\ref{fig:cloud_map_1bar_2} (left column) to the temperature map
in Fig.~\ref{fig:1DTp_1} (top right), it becomes apparent that metal
oxides (first row in Fig.~\ref{fig:cloud_map_1bar_2}) make up as much
as $\sim$50~\% of the total cloud volume in the coldest regions which
is the nightside at 1mbar but at higher latitude at 1 bar were cloud
formation becomes inefficient in the equatorial regions of the
nightside (compare with Figs.~\ref{fig:1DJ*}
and~\ref{fig:cloud_map_1bar}).  This value decreases as the local
temperature increase. The high temperature condensates follow an
opposite trend. They make up as much as $\sim$~40~\% of the cloud
material volume at 1~mbar and up to $\sim$~90~\% at the 1 bar level.
Silicates (second row of Fig.~\ref{fig:cloud_map_1bar_2}) follow the
same trend as the high temperature species, showing an increasing
fraction with temperature. Contrary to the high temperature
condensates, the silicate fraction decreases with increasing pressure
which is related to an increase of local temperature leading to
evaporation of silicate materials (compare
Fig.~\ref{fig:cloud_prop_grid}, middle panels).  The carbon material
(third row of Fig.~\ref{fig:cloud_map_1bar_2}) condenses in the
cooler, low-pressure regions of the atmosphere of HAT-P-7b. Given the
assumed oxygen-overabundance of the atmospheric gas, the carbon volume
fraction reaches a maximum of only 6\% inside the coldest spots on the
nightside. Its contribution becomes negligible in higher pressure
regions. The carbon contribution is largest in atmospheric regions
that are oxygen-depleted by cloud
formation. Figure~\ref{fig:cloud_map_1bar_2} shows that almost all
cloud particles will be composed of a mix of materials and that the
details of these mixes change globally.

\begin{figure*}
    \includegraphics[width=0.89\columnwidth]{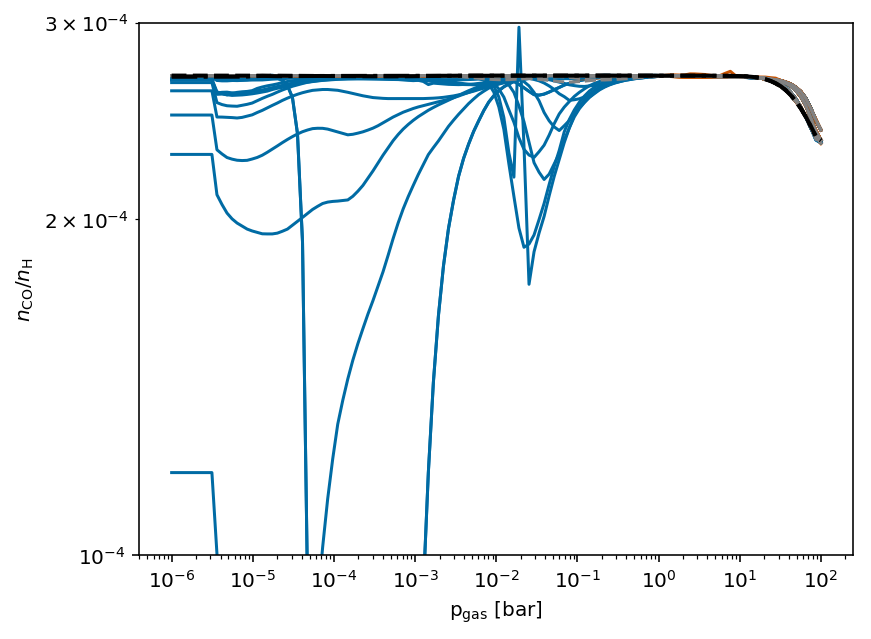}
    \includegraphics[width=0.89\columnwidth]{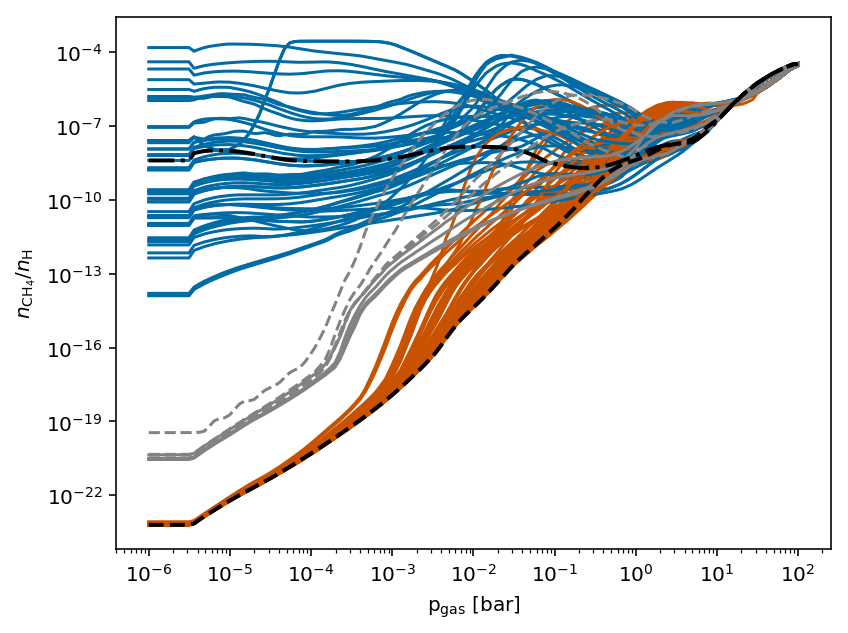}\\
    \includegraphics[width=0.89\columnwidth]{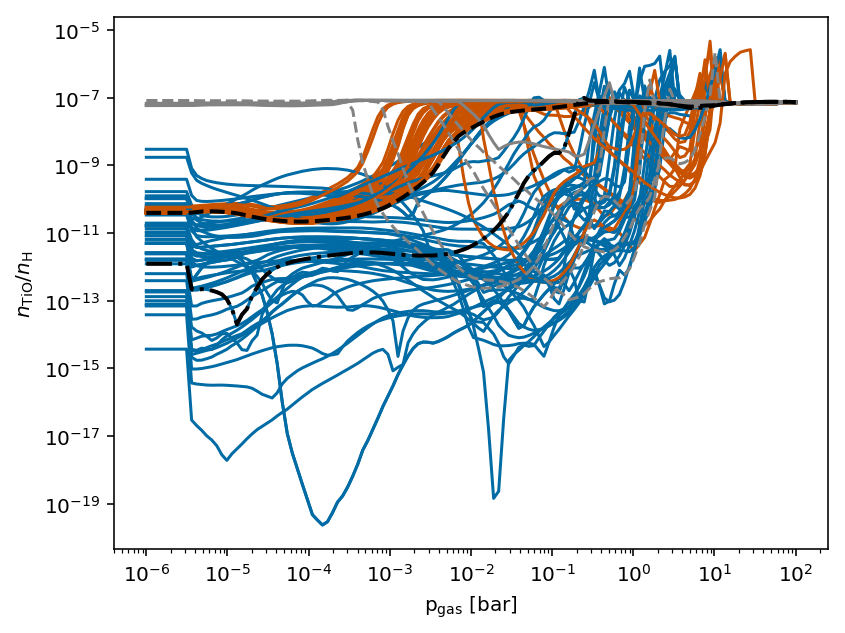}
    \includegraphics[width=0.89\columnwidth]{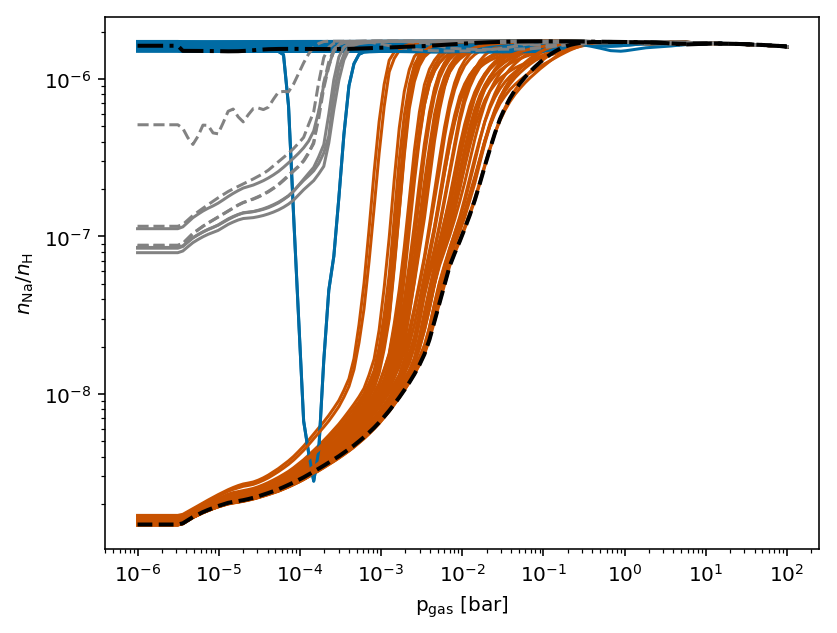}
  \caption{The gas-phase concentrations $q(x)=n_{\rm x}/n_{\rm <H>}$ (with $n_{\rm x}$ [cm$^{-3}$]) for CO, \ce{CH4}, TiO  and Na. We follow the same colour code like in Fig.~\ref{fig:1DTp_1}: blue -- nightside ($\phi=120\degree\,\ldots\, 240\degree$) , orange -- dayside ($\phi=60\degree\,\ldots\,0\degree\ldots\,300\degree$), black-dashed -- substellar point ($\phi=0\degree$), black-dot-dashed -- anti-stellar point ($\phi=180\degree$), grey profiles -- terminators (dashed -- evening $\phi=90\degree$; solid -- morning  $\phi=270\degree$). The largest differences between day- and nightside appear for the molecular species depicted.}
      \label{fig:mixing_ratios}
\end{figure*}

\begin{figure*}
    \includegraphics[width=8.7cm]{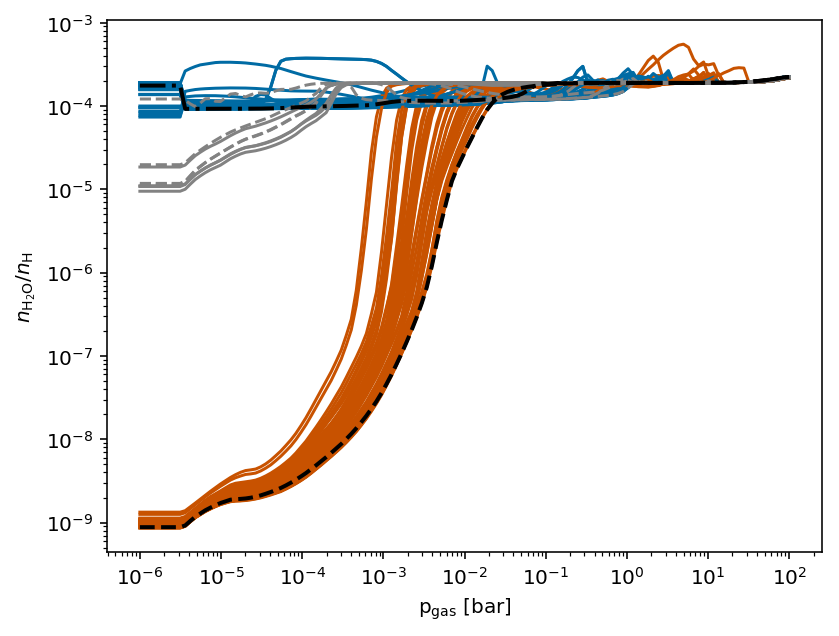}
    \includegraphics[width=8.7cm]{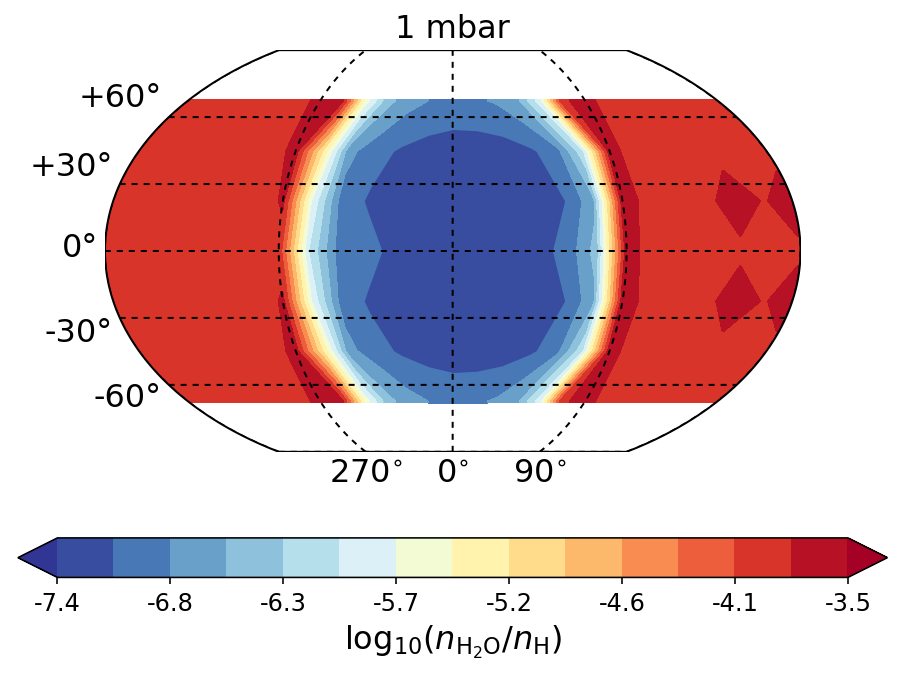}\\
    \includegraphics[width=8.7cm]{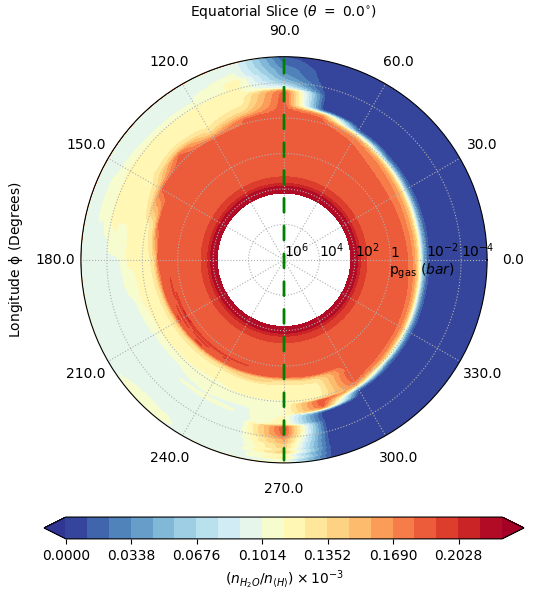}
    \includegraphics[width=8.7cm]{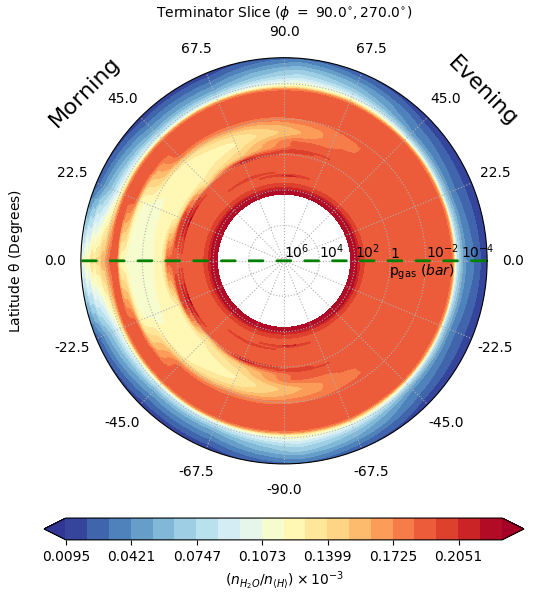}\\
    \caption{The gas-phase concentrations $q(x)=n_{\rm H2O}/n_{\rm <H>}$ (with $n_{\rm x}$ [cm$^{-3}$]), also called mixing ratio.
    {\bf Top left:} The 1D structures of the \ce{H2O} distribution on HAT-P-7b.  We follow the same colour code like in Fig.~\ref{fig:1DTp_1}: blue -- nightside, orange -- dayside; light blue -- morning terminator ($\phi=270\degree$), light orange -- evening terminator ($\phi=90\degree$), black-dashed -- substellar point ($\phi=0\degree$), black-dot-dashed -- anti-stellar point ($\phi=180\degree$), grey profiles -- terminators (dashed -- evening $\phi=90\degree$; solid -- morning  $\phi=270\degree$).  {\bf Top right:} Global \ce{H2O} map at p$_{\rm gas}=10^{-3}$\;bar. {\bf Bottom left:} 2D cut through the equatorial plane. {\bf Bottom right:}  2D cut along the terminator.  The dashed green lines indicate where these two slice plots overlap. For the viewing geometry of the slice plots in the bottom row, please see Fig.~\ref{fig:vgeo}.
    }
    \label{fig:H2O}
\end{figure*}
 
\begin{figure}
    \includegraphics[width=\columnwidth]{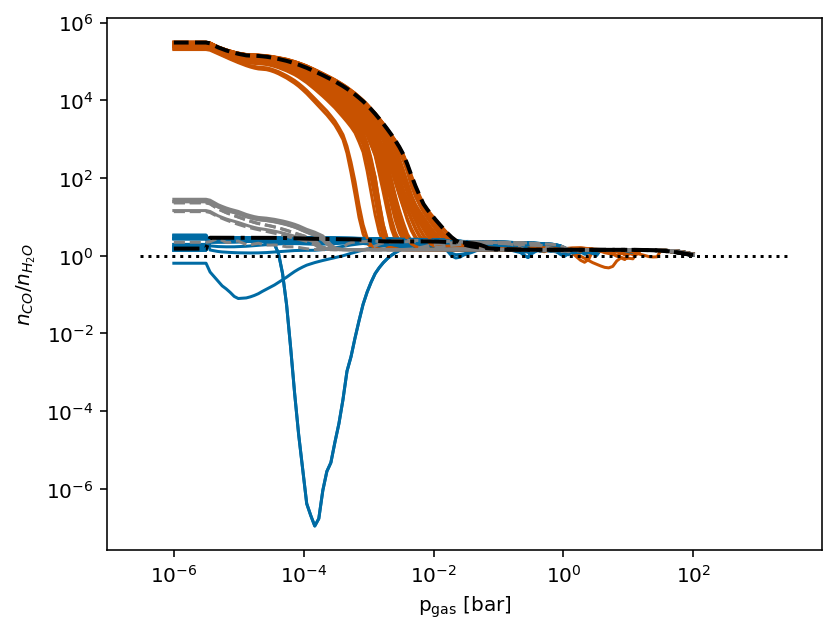}
    \includegraphics[width=\columnwidth]{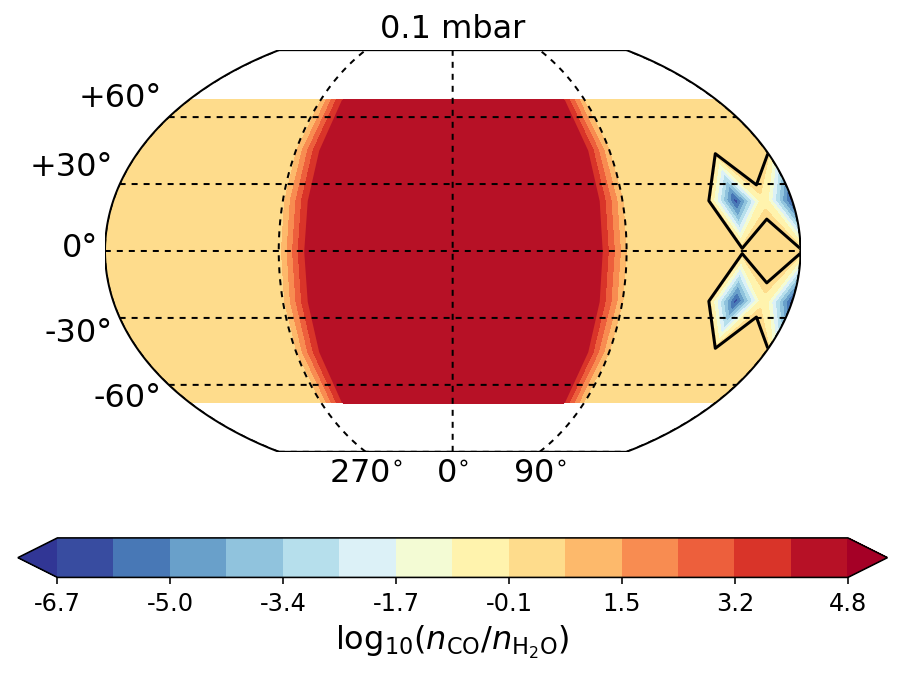}
 \includegraphics[width=\columnwidth]{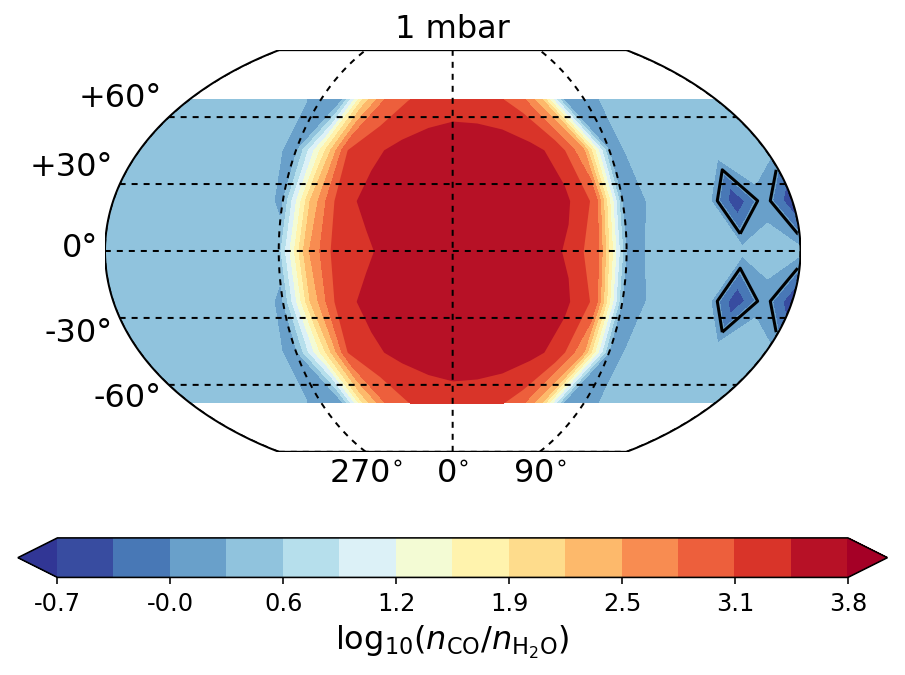}
  \caption{The molecular ratio n(CO)/n(\ce{H2O}). 
    {\bf Top:} The 1D structures of the  n(CO)/n(\ce{H2O}) distribution on HAT-P-7b.  We follow the same colour code like in Fig.~\ref{fig:1DTp_1}: blue -- nightside, orange -- dayside; day-night terminator: light blue and light orange, black-dashed: sub-stellar and anti-stellar points, grey profiles: terminator profiles (dashed -- evening, $\phi=90\degree$; solid -- morning  $\phi=270\degree$) . {\bf Middle:} Global n(CO)/n(\ce{H2O}) map at p$_{\rm gas}=10^{-3}$\;bar representing the possibly optically thin part of the atmosphere. {\bf Bottom:}  Global n(CO)/n(\ce{H2O}) map at p$_{\rm gas}=10^{-2}$bar showing the transitional region from warm to cold (Fig.~\ref{fig:1DTp_1}). The black contour lines in the 2D maps show where n(CO)=n(\ce{H2O}). As seen in 1D plot (top) for p$_{\rm gas}>$10\;mbar, only 2 profiles have n(CO)<n(\ce{H2O}) on the "far" nightside.}
      \label{fig:CoH2O}
\end{figure}

\begin{figure}[h!]
    \includegraphics[width=\columnwidth]{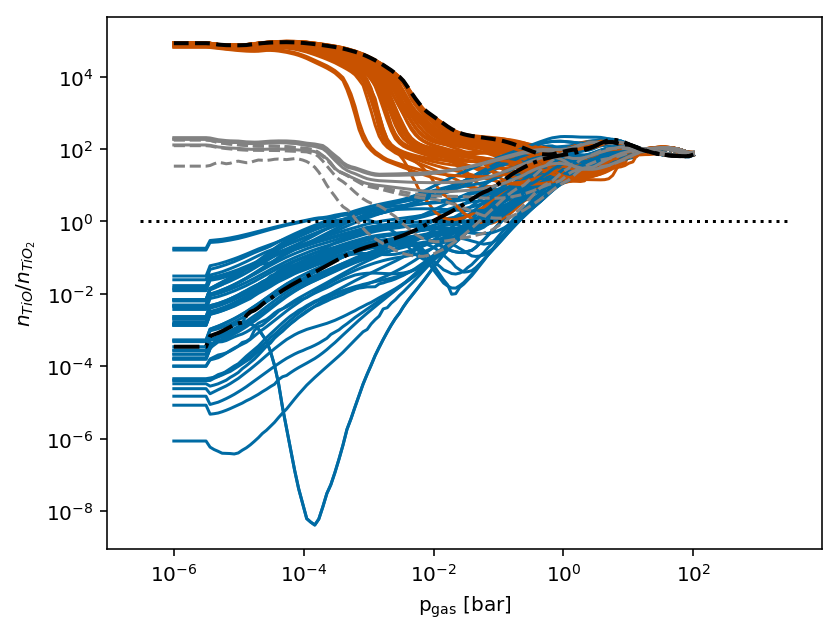}
    \includegraphics[width=\columnwidth]{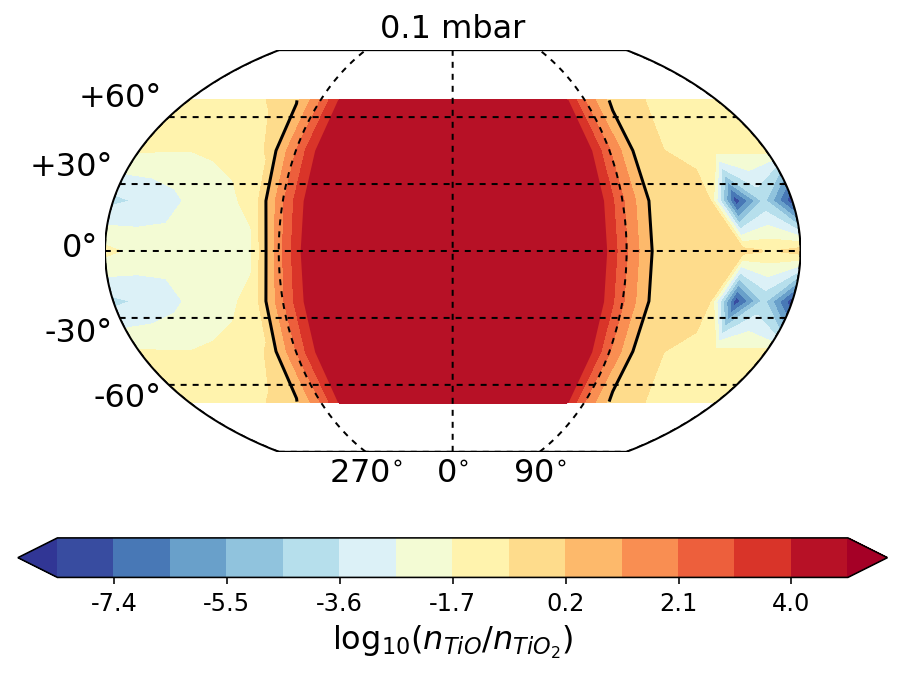}
 \includegraphics[width=\columnwidth]{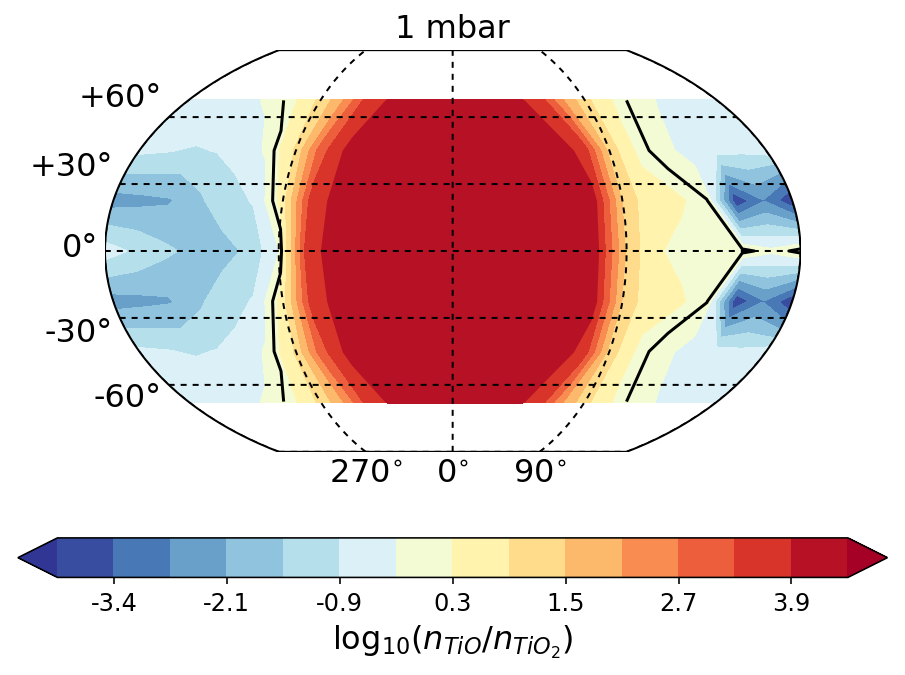}
  \caption{The molecular ratio n(\ce{TiO})/n(\ce{TiO2}). The figure follows the same structure and colour code like Fig.~\ref{fig:CoH2O}. In contrast to n(CO)/n(\ce{H2O}), many more trajectories exhibit n(TiO)<n(\ce{TiO2}) on the nightside (blue line in top panel).}
      \label{fig:TiOTiO2}
\end{figure}

\begin{figure}[h!]
    \includegraphics[width=\columnwidth]{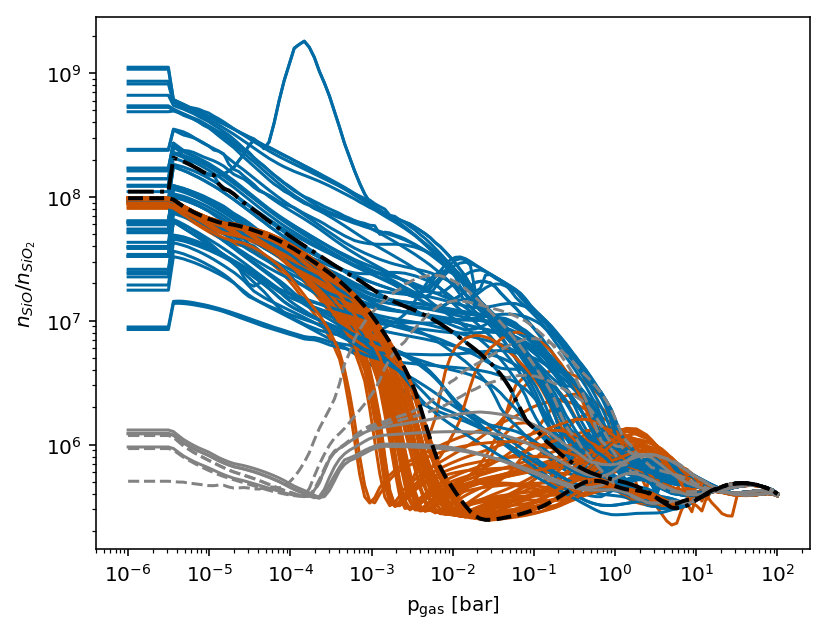}
    \includegraphics[width=\columnwidth]{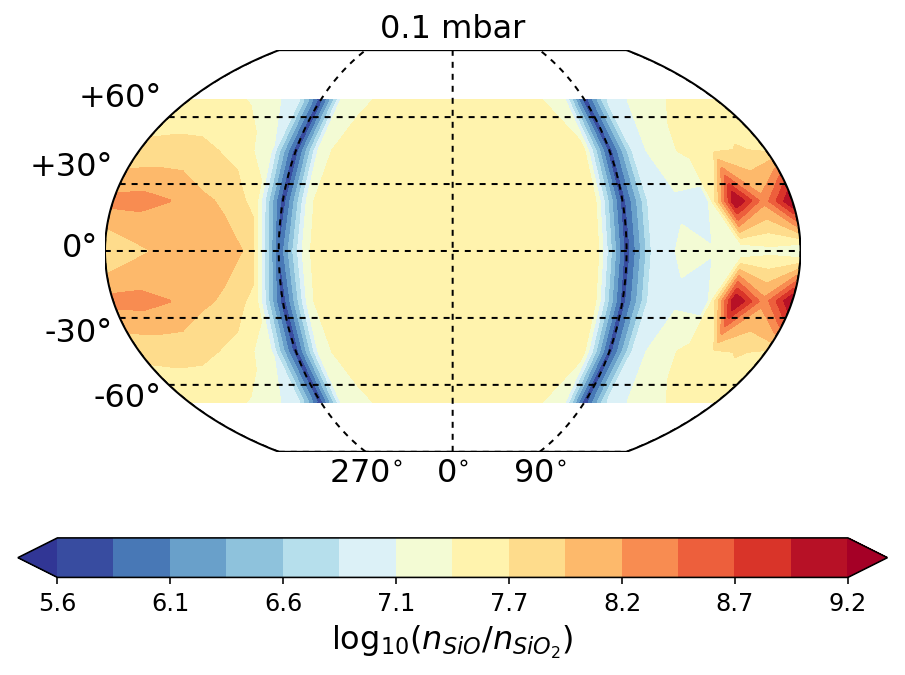}
 \includegraphics[width=\columnwidth]{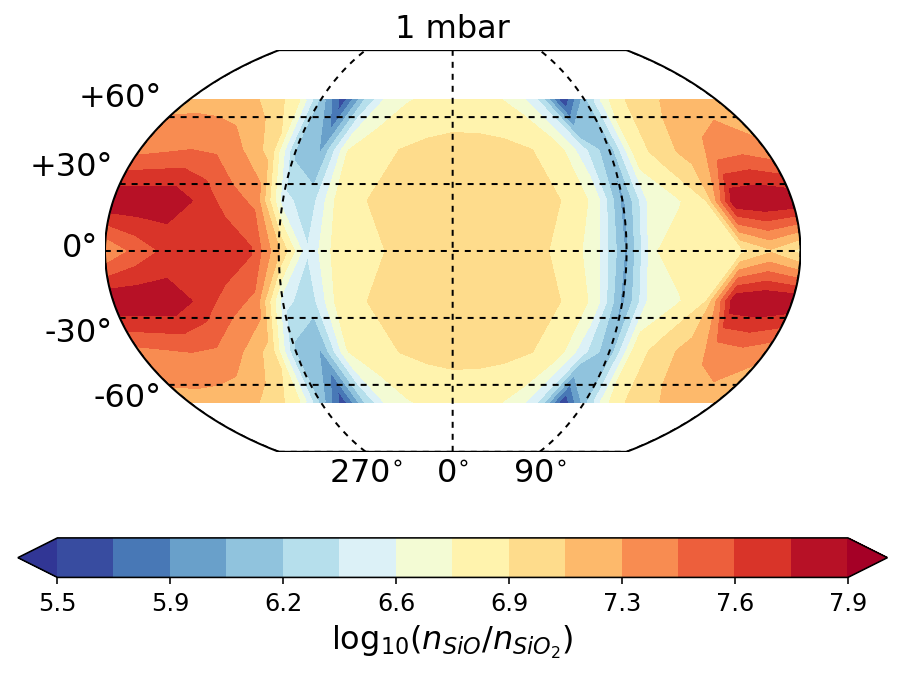}
  \caption{The molecular ratio n(SiO)/n(\ce{SiO2}). The figure follows the same structure and colour code like Fig.~\ref{fig:CoH2O}. In contrast to n(CO)/n(\ce{H2O}) and n(TiO)/n(\ce{TiO2}), n(SiO)>n(\ce{SiO2}) for the whole atmosphere with the lowest ratio in the terminator regions (compare light blue/orange 1D profiles in top panel). 
  }
      \label{fig:SiOSiO2}
\end{figure}

\begin{figure}[h!]
    \includegraphics[width=\columnwidth]{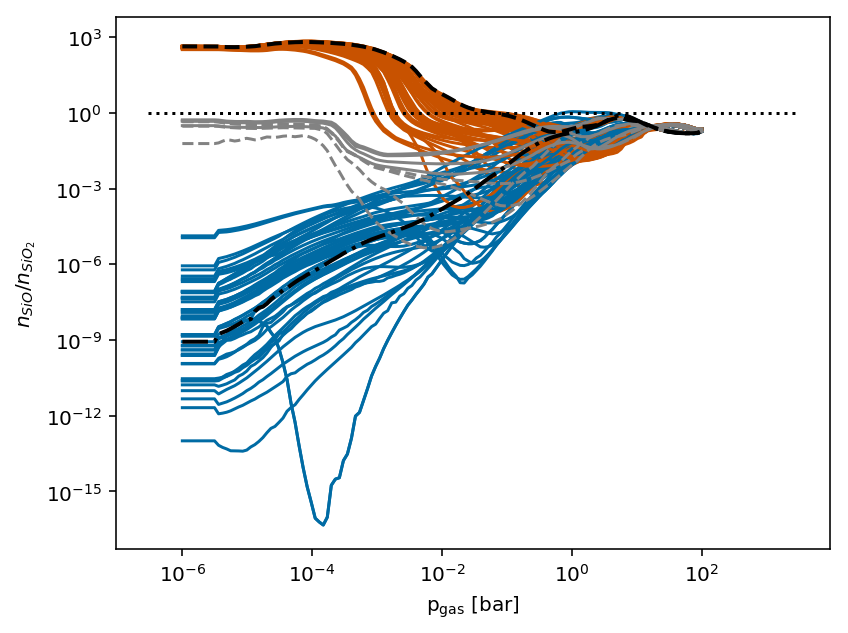}\\
    \includegraphics[width=\columnwidth]{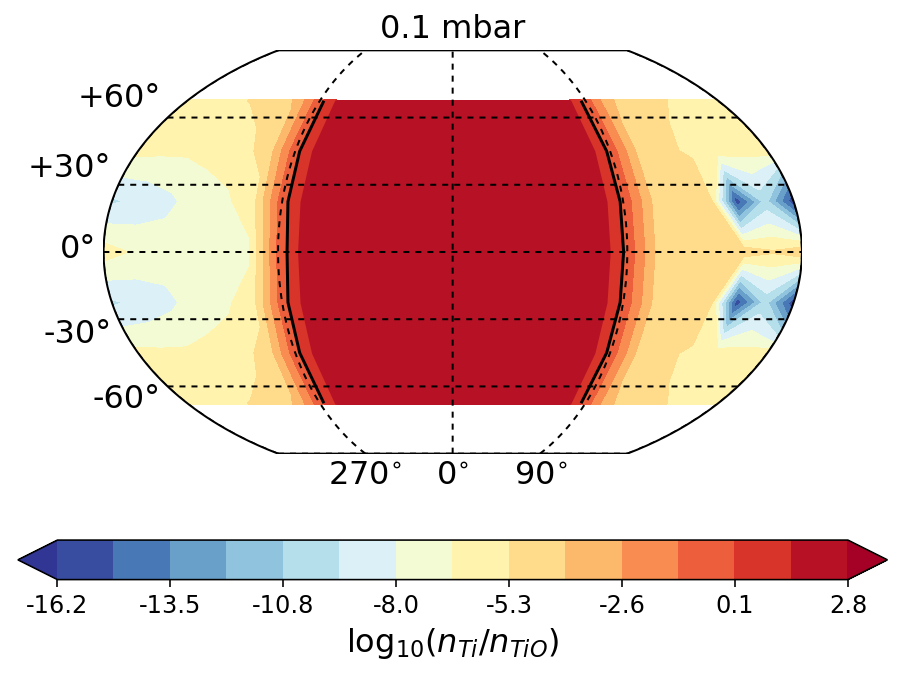}\\
 \includegraphics[width=\columnwidth]{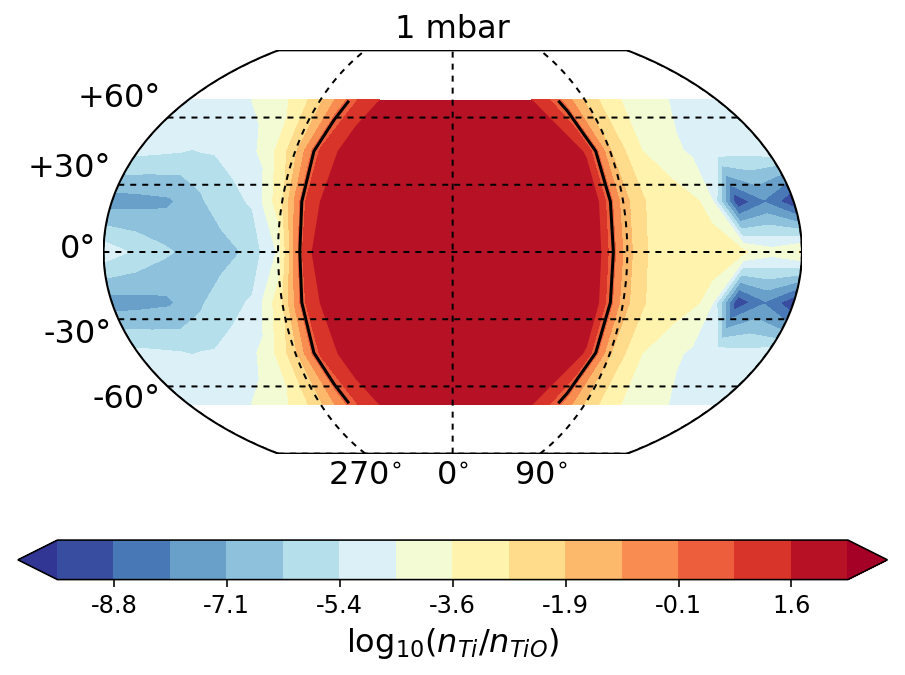}
  \caption{The atom/molecule ratio n(Ti)/n(TiO). The figure follows the same structure and colour code like Fig.~\ref{fig:CoH2O}. The solid black lines indicates where n(Ti)=n(TiO) which is off-set from where n(TiO)=n(\ce{TiO2}). 
  The 1D profile (top panel) peaking at $10^{-3.8}\,$bar  is the one with the lowest temperature in our sample on the nightside.}
      \label{fig:TiTiO}
\end{figure}

\begin{figure}[h!]
    \includegraphics[width=\columnwidth]{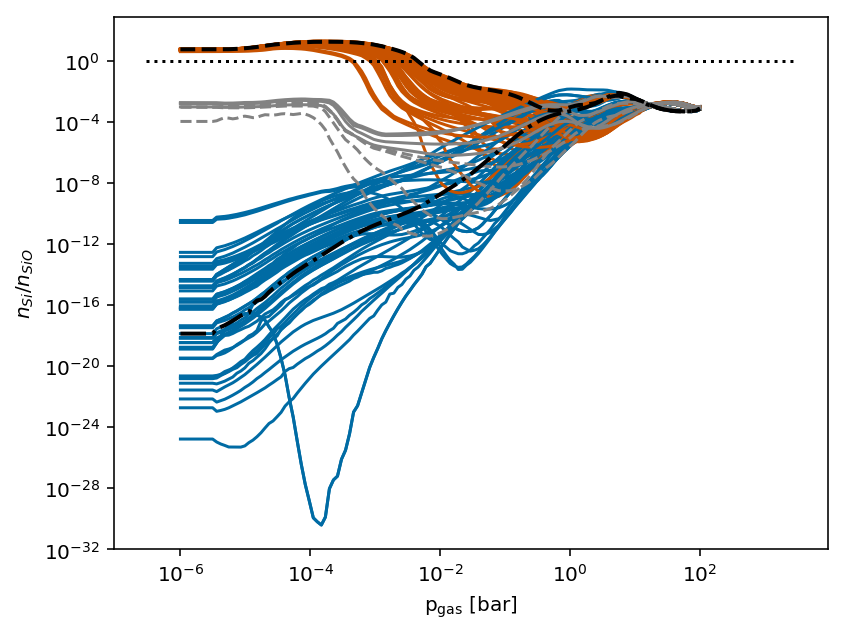}
    \includegraphics[width=\columnwidth]{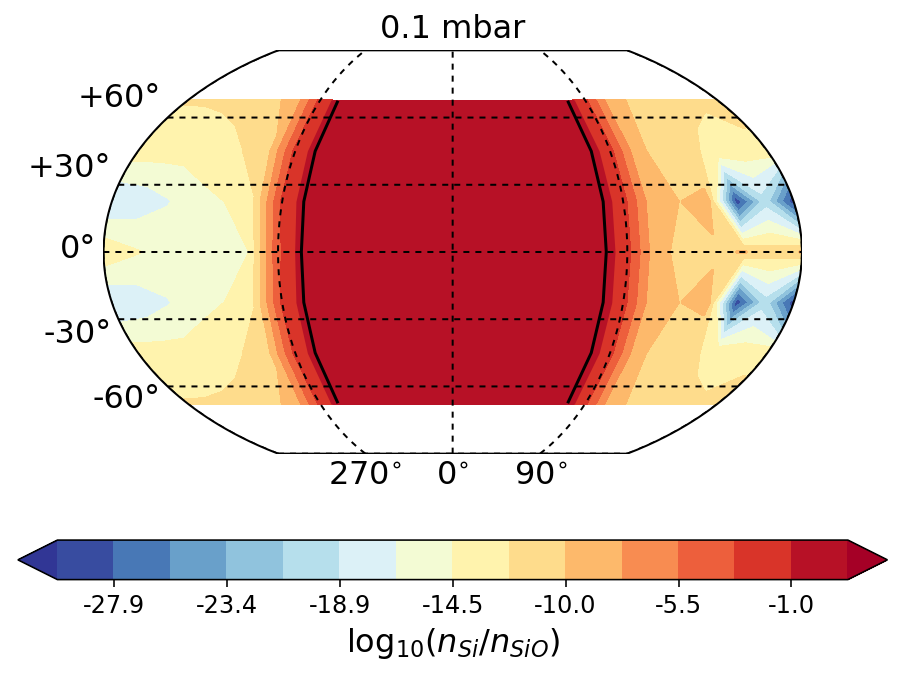}
 \includegraphics[width=\columnwidth]{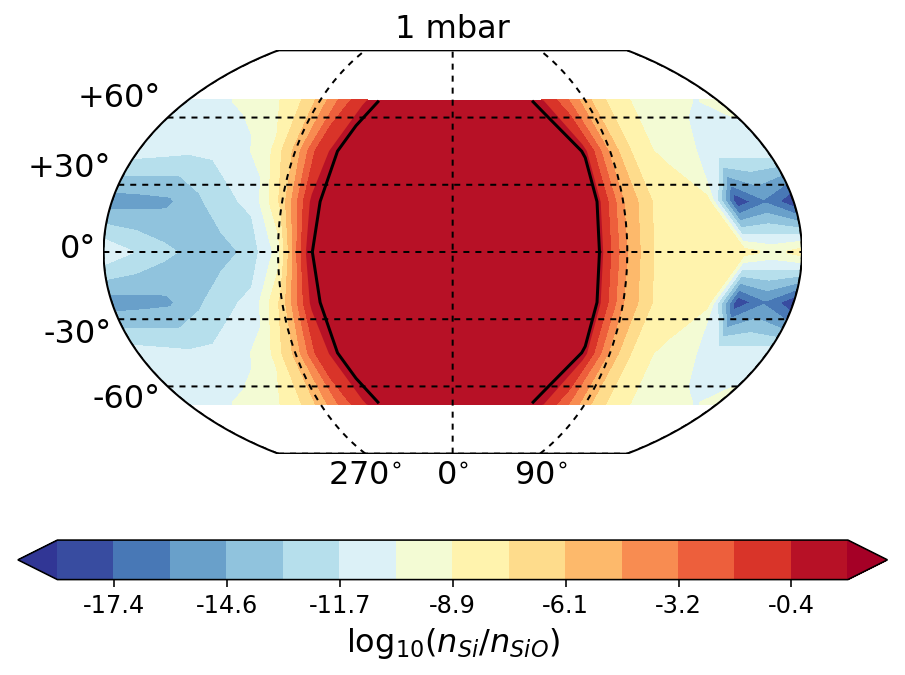}
  \caption{The atom/molecule ratio n(Si)/n(SiO). The figure follows the same structure and colour code like Fig.~\ref{fig:CoH2O}. The solid black lines indicates where n(Si)=n(SiO) which is off-set from where n(SiO)=n(\ce{SiO2}). Similar to Fig.~\ref{fig:TiTiO},
  The 1D profile (top panel) peaking at $10^{-3.8}\,$bar  is the one with the lowest temperature in our sample on the nightside.}
      \label{fig:SiSiO}
\end{figure}

\begin{figure}
    \includegraphics[width=\columnwidth]{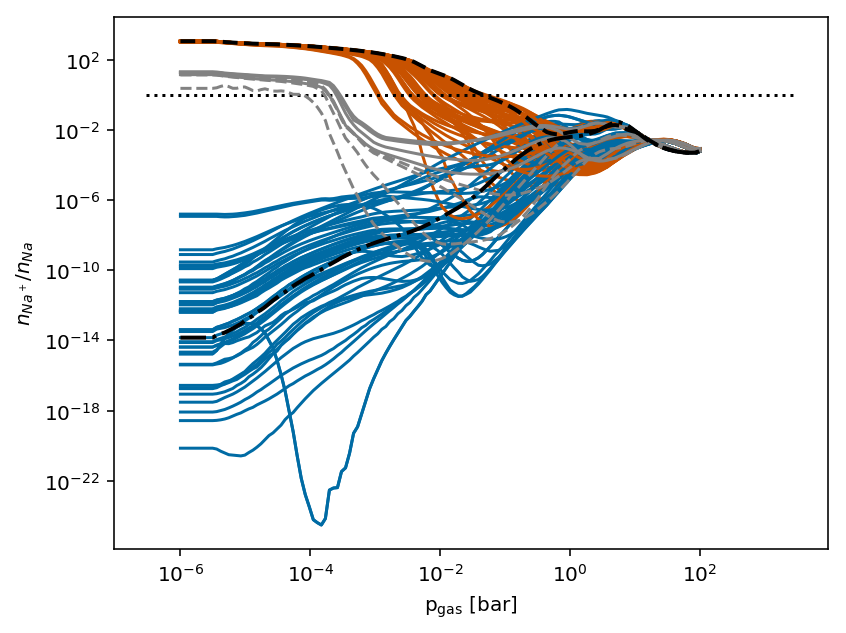}
    \includegraphics[width=\columnwidth]{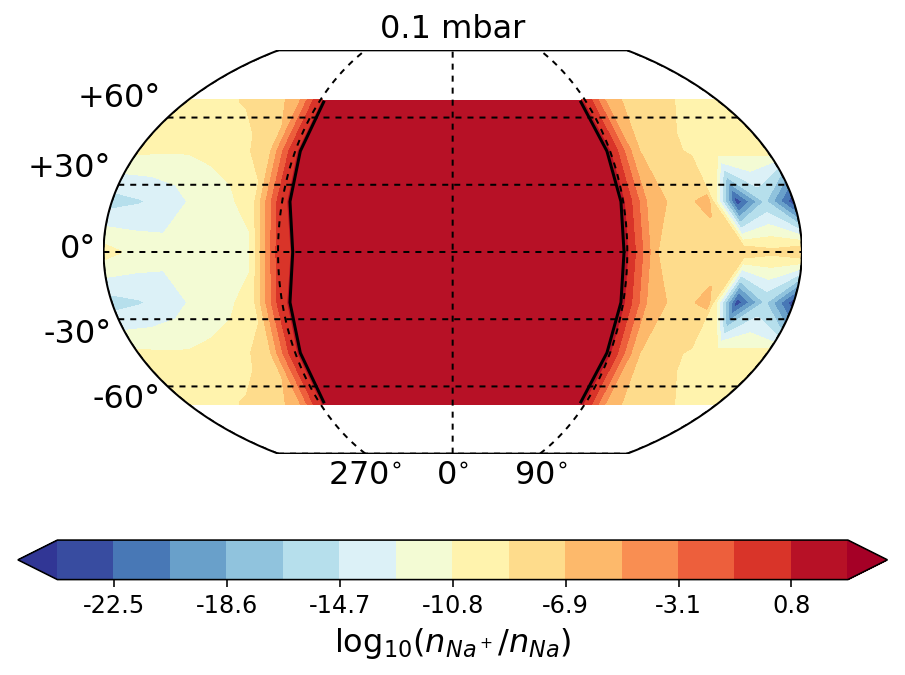}
        \includegraphics[width=\columnwidth]{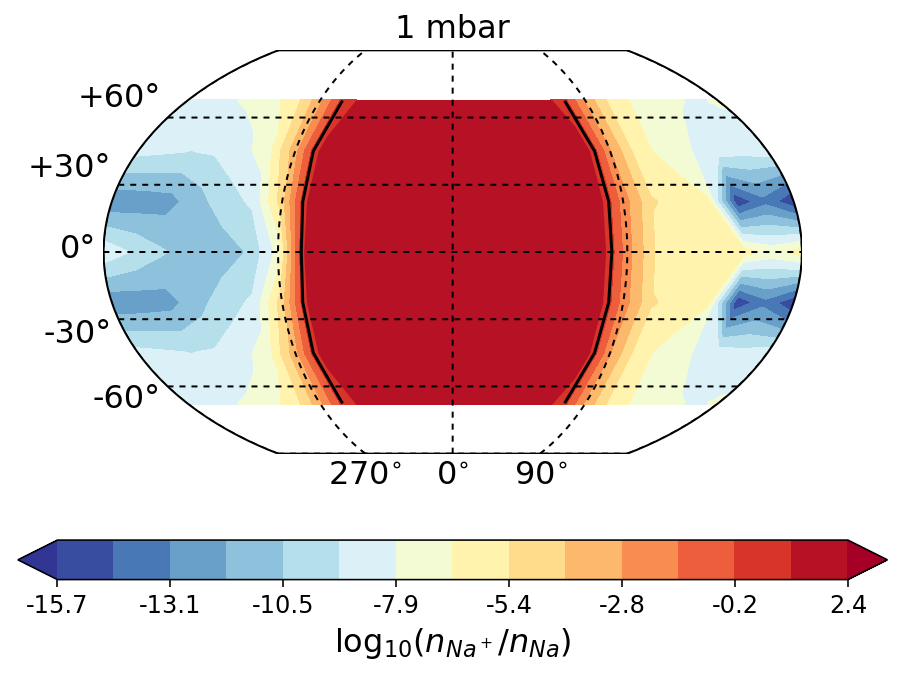}

  \caption{The ion/atom  ratio n(\ce{Na+})/n(Na). The figure follows the same structure and colour code like Fig.~\ref{fig:CoH2O}. The black contour lines shows where n(\ce{Na+})/n(Na) =1. }
      \label{fig:rATIOS_MIXING}
\end{figure}

\begin{figure}
\vspace{-0.5cm}
      \includegraphics[width=\columnwidth]{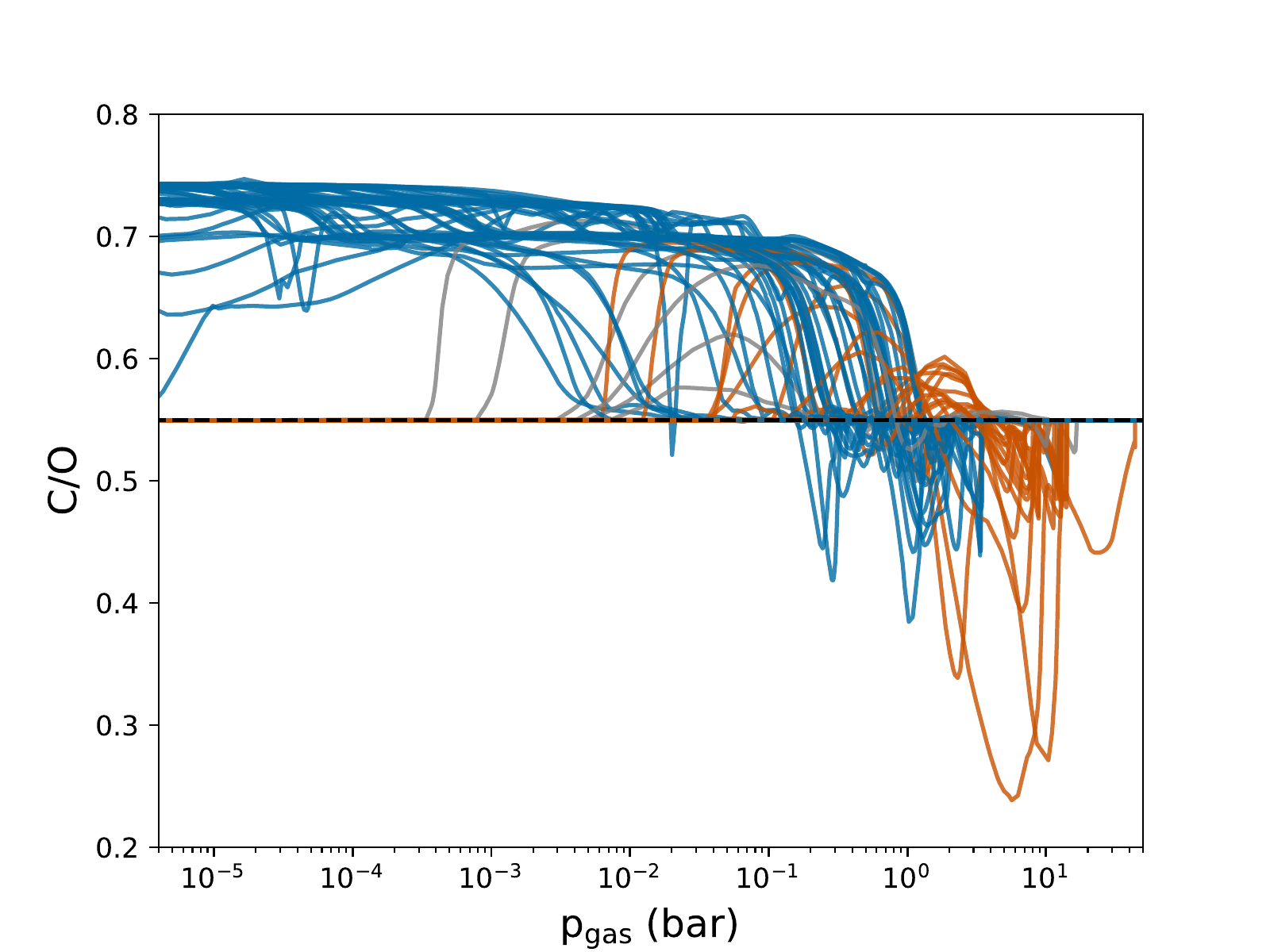}\\       
      \includegraphics[width=\columnwidth]{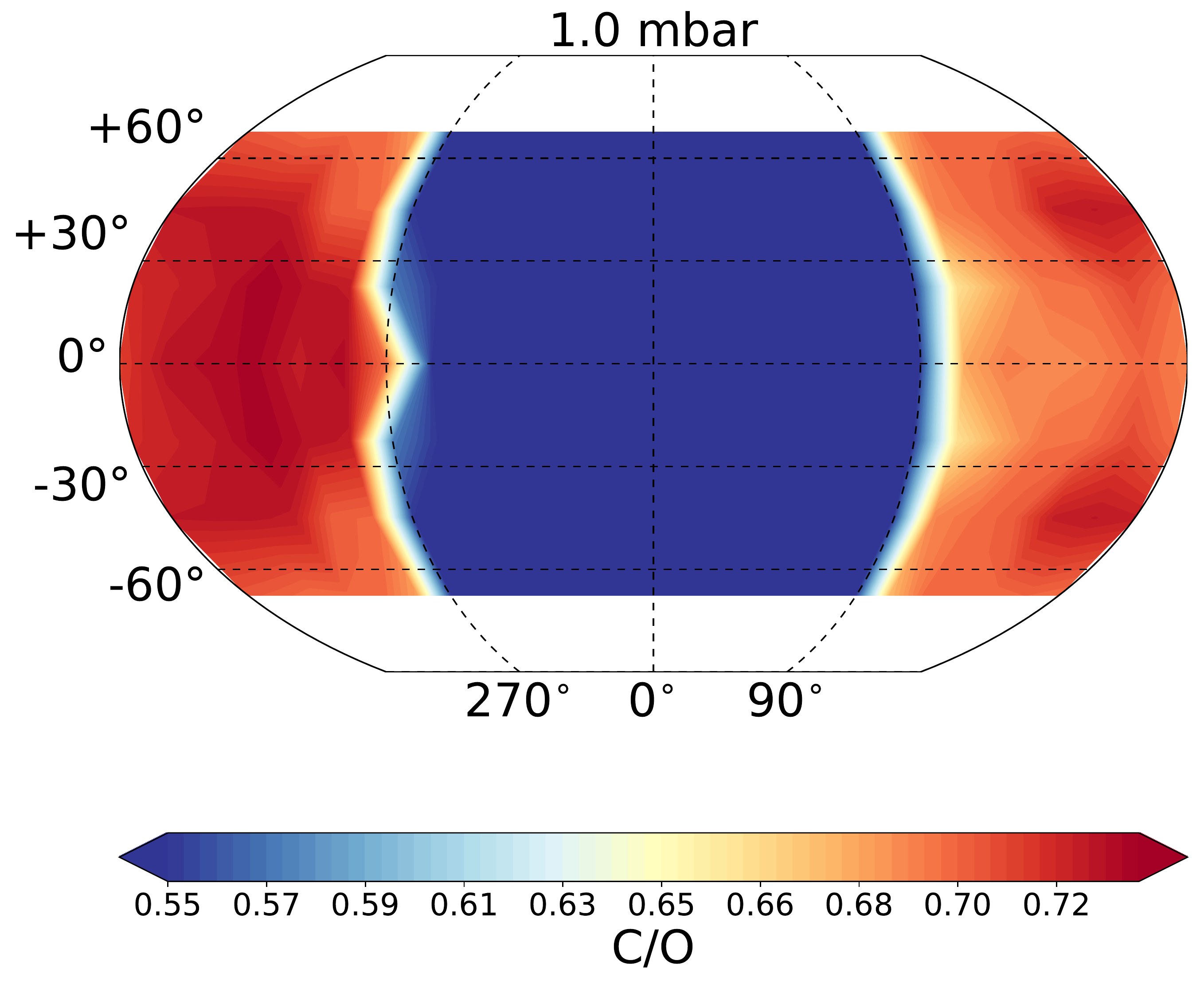}\\
      \includegraphics[width=\columnwidth]{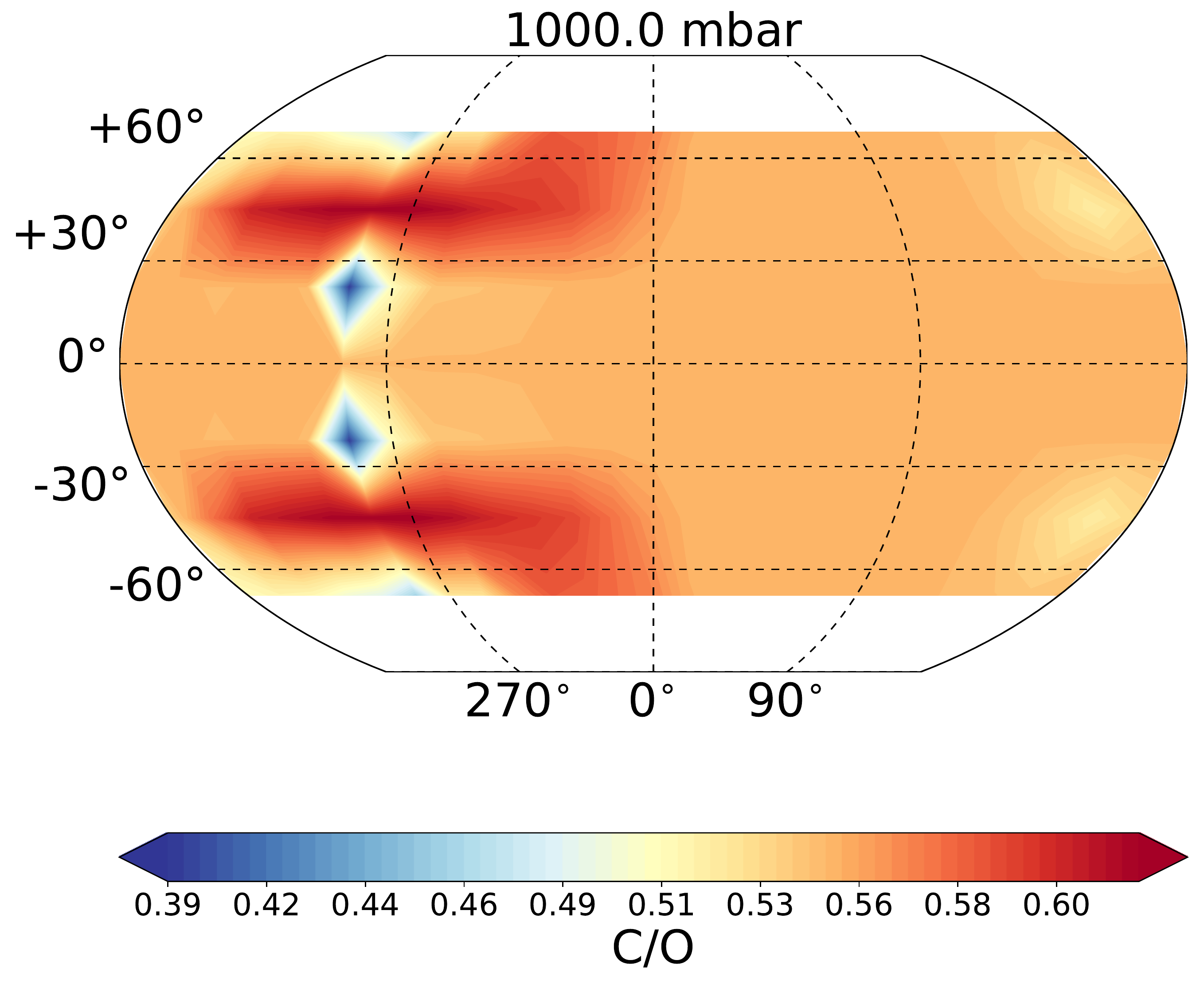}
    \caption{Global gas-phase C/O  after cloud formation. {\bf Top:} All 1D profiles (line color codes as in Fig.~\ref{fig:1DTp_1}). The black horizontal dash line represents the solar value  C/O$\sim$0.54. {\bf Middle:} 2D map at 1\,bar,
    {\bf Bottom:}  2D map at 1\,mbar. The dayside C/O is consistent with solar, whereas the nightside shows an enhancement of C/O ($\sim 0.7$) due to oxygen depletion by cloud formation. Deep atmospheric layers may have C/O$\sim 0.25$ due to oxygen enrichment from evaporating cloud particles. }
    \label{fig:CO_level}
\end{figure}

\begin{figure}[h!]
 \includegraphics[width=\columnwidth]{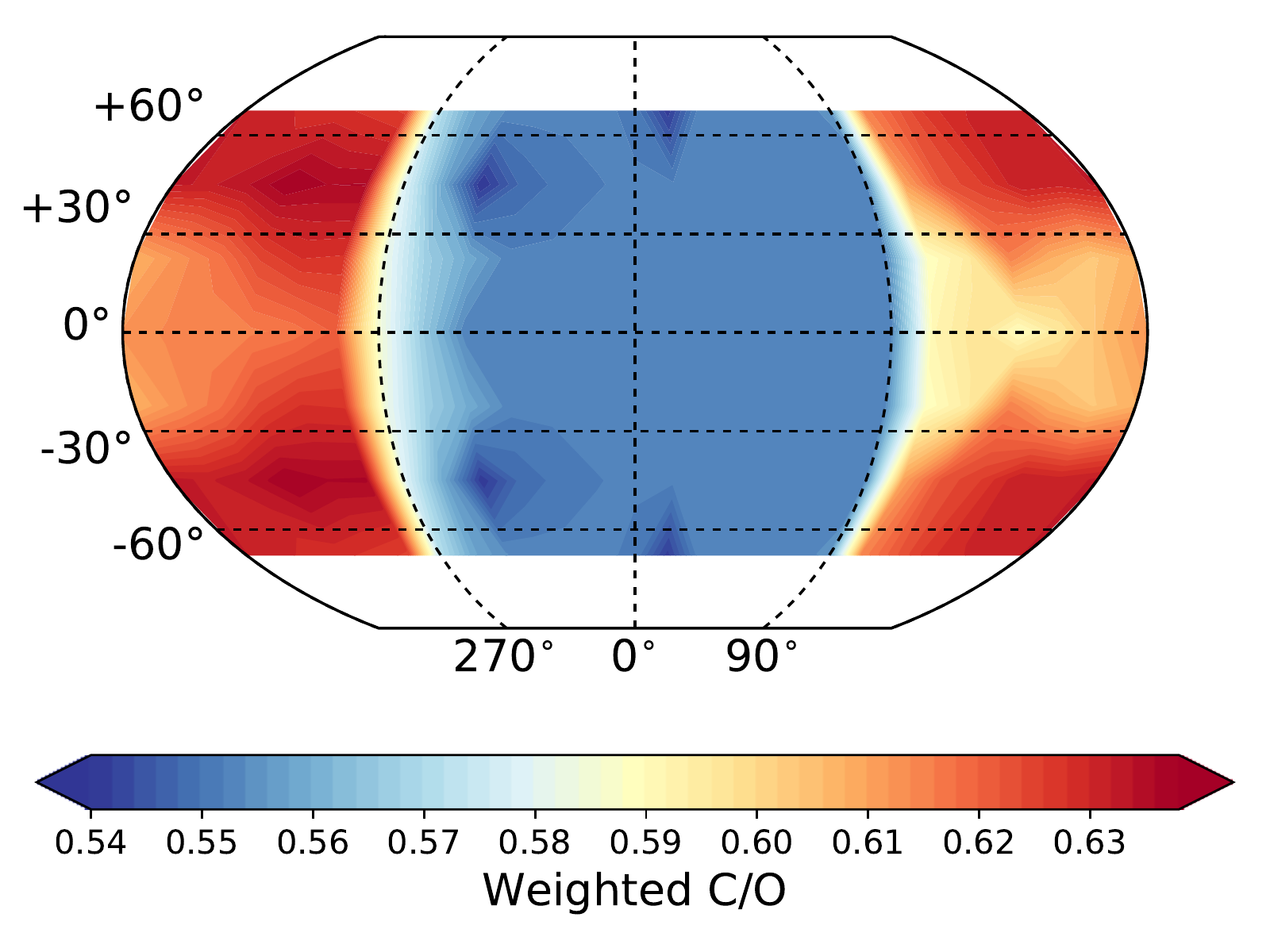}
    \caption{C/O integrated over atmospheric height and normalised to the local height of each integrated column.}
    \label{fig:CO_integrated}
\end{figure}

\section{The gas-phase composition on HAT-P-7b}
\label{s:maps}

The composition of the atmospheric gas in a planetary atmosphere is affected by the local temperature and pressure, the local element abundances in addition to an external radiation field and material transport. Here, we study the effect of the gas-temperature and pressure, and that of the element abundances which are affected by cloud formation and destruction processes. We assume that the gas is collisionaly dominated. The effect of irradiation is taken into account by radiative heating of the atmospheric gas as part of the 3D GCM model. We do not discuss ionisations processes and refer to \cite{2019arXiv190304565H} for a recent summary.

The gas-phase composition of HAT-P-7b's atmosphere is inhomogeneous as result of the different thermodynamic day- and nightside regimes (Sect.~\ref{ss:ggchem_dn}). 
Section~\ref{ss:tau1mol_abovecloud} investigates how the amount of gas that is optically thin and that is accessible by observations changes with latitude and longitude across the globe. We therefore discuss \ce{H2}, \ce{H2O}, CO, \ce{CH4}, 
SiO, TiO, \ce{Na} and \ce{Na+}  in more detail here. Mineral ratios, such as the carbon-to-oxygen ratio (C/O), are helpful properties to possibly link the atmosphere chemistry to planet formation and/or evolution processes, and an analysis of the global gas-phase C/O is presented in Sect.~\ref{ss:CtoO}. We demonstrate that C/O is hugely affected by cloud formation such that complex physical models are required if a link to planet formation and/or evolution is desired. Paper II will address quenching and photoionisation to provide a first study of how much hydrodynamical mixing processes may alter the gas-phase composition (quenching) and what we may expect from host-star driven photochemistry on HAT-P-7b. 

\subsection{Changing chemical regimes from day to nightside}\label{ss:ggchem_dn}

The temperature difference of $\sim 2500$\;K causes the atmosphere of HAT-P-7b to expose a molecule-dominated low-temperature chemistry on the nightside (Fig.~\ref{fig:Molecs}, anti-stellar point  ($\theta,\phi)=(0\degree, 180\degree)$) and an atom/ion-dominated high temperature chemical composition on the dayside (Fig.~\ref{fig:Molecs}, substellar point  ($\theta,\phi)=(0\degree, 0\degree)$). The dayside gas composition is that of a typical, solar element abundance gas, except for those altitudes that {may be} affected by quenching and/or photoionisation (see Paper II) and/or those latitudes affected by cloud formation.  This allows us to understand which species might be affected given the modelling framework available.  The thermal gas composition is dominated by atomic hydrogen (HI) in the upper layers and by molecular hydrogen (\ce{H2}) at p$_{\rm gas}< 0.15$\;bar.
The next most abundant molecules are CO, \ce{H2O} and SiO in the higher atmosphere which changes with increasing atmospheric depth, hence increasing temperature and pressure. CO and \ce{H2O} appear with similar number densities for p$_{\rm gas}< 0.15$\;bar. Atomic species like Fe, Na and S, but also ions like \ce{Al+} and \ce{Ca+} are the next most numerous. Atoms and ions are less abundant at the evening terminator   $(\theta,\phi)=(0\degree, 90\degree)$ and the gas pressure where H/\ce{H2} and CO/(CO, \ce{H2O}) domination switches over is lower than the sub-stellar point, hence moving into higher atmospheric altitudes. The gas becomes gradually more dominated by molecules when moving to the nightside, with more SiS and \ce{H2S}.

The nightside is affected by cloud formation and the resulting chemical composition of the atmospheric gas is therefore expected to be different for every exoplanet with different system parameters (Table~\ref{tab:params}). The nightside (anti-stellar point, $(\theta,\phi)=(0\degree, 180\degree)$) resembles a H$_2$-dominated, cool atmosphere with CO, H$_2$O, H$_2$S, and SiS, being particularly abundant. The nightside is affected by element depletion due to cloud formation which manifests via a decreased abundance of species containing Mg, Si, O, Fe, Al, or S (e.g. SiS, SiO, Fe). The equatorial morning terminator shown in Fig.~\ref{fig:Molecs} demonstrates the combined effect of cloud formation and a temperature inversion on the gas phase composition.

 Figure~\ref{fig:mixing_ratios} presents the mixing ratios (concentration $q(x)=n_{\rm x}/n_{\rm <H>}$, with $n_{\rm x}$ [cm$^{-3}$]) for CO, CH$_4$, TiO and Na  for all investigated 1D profiles colour coded for day- and nightside according to Fig.~\ref{fig:1DTp_1}. CO is one of the most abundant molecules on HAT-P-7b on the day- and the nightside. CH$_4$ is discussed as one of the small hydrocarbon molecules, while TiO is an important opacity carrier in a cool atmosphere that is affected by cloud formation through depletion of Ti. Na is a tracer for the thermal ionisation of gas and is also used for element abundance determination through transmission spectroscopy \citep{2018Natur.557..526N}. The CO concentration on the dayside does not change, but drops substantially on the nightside. In contrast, the CH$_4$ concentration is gradually decreasing from the night- to the dayside which is a purely thermodynamic effect due to the systematically increasing temperatures in the dayside. The TiO concentration is affected by the changes of the local gas temperature and cloud formation. While the dayside temperatures do not allow for TiO to be present in thermal equilibrium, the nightside shows even lower TiO concentrations because of an substantial element depletion. We note an increase of the TiO concentration for many profiles for higher pressures. At the dayside, thermal stability does increase with increasing pressure (10-100\;bar), but the increase in TiO at the nightside occurs at lower temperatures and is linked to the enrichment of the gas phase with Ti due to the evaporation of the cloud particles at these temperatures. The change of the local element abundances is not shown, but has been demonstrated for other planets (see e.g., Fig.~7 of \citealt{2019arXiv190108640H}). Figure~\ref{fig:mixing_ratios} shows that CO, \ce{CH4}, TiO and Na do substantially change horizontally or vertically throughout the atmosphere of HAT-P-7b in the case of thermochemical equilibrium.

\paragraph{Water in the atmosphere of HAT-P-7b:}
 \cite{2018AJ....156...10M} suggest that water dissociation plays a crucial role in shaping the dayside emission spectrum of HAT-P-7b. We demonstrated in Fig.~\ref{fig:Molecs} that molecular hydrogen \ce{H2} is dissociated on the dayside, such that the atomic hydrogen H is the dominant hydrogen-bearing species. With respect to water, we further show that the H$_2$O abundance depends on longitude and latitude due to the thermal structure of the atmosphere and cloud formation (Fig.~\ref{fig:Molecs}). Figure~\ref{fig:H2O} (top left) demonstrates how the \ce{H2O} concentration,  $q(x)=n_{\rm H2O}/n_{\rm <H>}$  ($n_{\rm x}$ [cm$^{-3}$]), the  mixing ratio,  varies for all considered 1D trajectories in HAT-P-7b's atmosphere. In particular, the \ce{H2O} mixing ratio is roughly vertically homogeneous (it varies only within one order of magnitude) on the nightside (blue curves) and across the terminator (light blue and light orange curves). 
 Many of the dayside trajectories (red/orange curves) experience a rapid decrease in the \ce{H2O} mixing ratio by orders-of-magnitude
 at p$_{\rm gas} \lesssim 10^{-2}$\;bar, in agreement with the \cite{2018AJ....156...10M} result. For higher pressures (lower altitudes), the \ce{H2O} mixing ratio for all trajectories tends toward a solar value of $\sim 10^{-4}$ {i.e. solar element abundance results}. The implications of this variable \ce{H2O} mixing ratio on its contribution to the opacity  is considered in Sect.~\ref{s:opacity}.
 

 The H$_2$O abundance plays a crucial role in shaping exoplanet spectra. \cite{2016Natur.529...59S} demonstrated that \ce{H2O} absorption is the dominant near-infrared feature seen in the observed transmission spectra of relatively cloud-free atmospheres. For atmospheres with detected H$_2$O absorption features, atmospheric retrieval algorithms can be applied to empirically derive the range of \ce{H2O} abundances consistent with an observed spectrum. Recently, \cite{2019MNRAS.482.1485P} applied such an algorithm to derive \ce{H2O} abundances in the optically thin (i.e., above any cloud deck) upper atmospheres of the ten giant planet transmission spectra in \cite{2016Natur.529...59S}. However, a common assumption of such retrieval approaches is that the \ce{H2O} abundance is uniform across the terminator and as a function of altitude \citep[e.g., see][for a discussion of such 1D assumptions]{MacDonald2017}, which will not hold generally across all global locations \citep{2016MNRAS.460..855H}. As Fig.~\ref{fig:H2O} demonstrates, the \ce{H2O} abundance at the terminator is relatively constant from the morning- to the evening side and as a function of altitude, with variations around a factor of $\sim2$. This suggests that the (1D averaged) \ce{H2O}\  abundances retrieved from transmission spectra are likely a good proxy for the deep H$_2$O abundance of ultra-hot Jupiters such as HAT-P-7b. Indeed, our predicted terminator H$_2$O abundances are consistent with those found by \cite{2019MNRAS.482.1485P} for planets with equilibrium temperatures $\gtrsim 1600$\;K. Further implications for observations of HAT-P-7b will be explored Paper III.

 

\paragraph{Global molecular ratios CO/\ce{H2O}, TiO/\ce{TiO2}, SiO/\ce{SiO2}:}
Figure~\ref{fig:CoH2O} (top) shows that CO is more abundant than \ce{H2O} on the dayside of HAT-P-7b because all of the hydrogen is atomic in the upper atmospheric regions, as Fig.~\ref{fig:Molecs} shows for the substellar point and evening- and the morning terminators.
CO and \ce{H2O} are of comparable abundance in deeper atmospheric regions with p$_{\rm gas}>10^{-3}\,\ldots\,10^{-2}$\;bar. This is also the case on the nightside, except for two profiles which correspond to the local temperature minima at the nightside (see Fig.~\ref{fig:1DTp_1}). The differences between the CO and \ce{H2O} increase with decreasing pressures in the higher atmospheric regions. This is emphasised by the 2D maps in Fig.~\ref{fig:CoH2O}. The 2D maps further demonstrate the difference of the CO and \ce{H2O} in the terminator regions, and that the offset anti-stellar point shows the highest abundance of \ce{H2O}  ($n_{\rm CO}/n_{\rm \ce{H2O}}$ smallest). This is linked to the lowest temperatures in our 2D map (compared Fig.~\ref{fig:1DTp_1}).

Figure~\ref{fig:TiOTiO2} demonstrates the day/night effect on the abundance of TiO and \ce{TiO2}. TiO is more abundant than \ce{TiO2} on the dayside where the atmosphere is the warmest. The differences become less pronounced in the terminator regions. \ce{TiO2} is substantially more abundant that TiO on the nightside. The differences between TiO and \ce{TiO2} also increase with decreasing pressure. Figure~\ref{fig:SiOSiO2} depicts the day/night effect on the abundance of SiO and \ce{SiO2}. In contrast to TiO/\ce{TiO2}, SiO appears always more abundant than \ce{SiO2} in thermochemical equilibrium in the atmospheric temperature regimes of HAT-P-7b. The lowest SiO/\ce{SiO2} ratio appears in the terminator regions.

\paragraph{Global ion/atom and atom/molecules ratios Ti/TiO, Si/SiO, \ce{Na+}/Na:} 
In order to capture the changing chemical regimes on the day- and nightside, we now discuss the ratios Ti/TiO, Si/SiO and \ce{Na+}/Na. Ti/TiO and Si/SiO consider the effect of thermal dissociation of species that play an important role in cloud formation. Al, Ti, TiO, Si and SiO can be affected by element depletion through the formation of clouds. Ti and TiO provide surface reaction channels for the growth of \ce{TiO2}[s] material and its depletion affects the nucleation process, which only proceeds through \ce{TiO2} (Sect.~\ref{s:ap}; Tables B.1 and B.2 in \citealt{2019arXiv190108640H}). SiO is a key molecule for various surface reaction channels leading to the formation of \ce{SiO}[s], \ce{SiO2}[s], \ce{MgSiO3}[s],  \ce{Mg2SiO4}[s], \ce{CaSiO3}[s], and \ce{Fe2SiO4}[s] of the 12 growth materials considered here. 

The ratio \ce{Na+}/Na is of particular interest as it traces the level of thermal ionisation across the globe in a planetary atmosphere. Na is a good tracer of thermal ionisation because it is one of the dominating electron donors in cool atmospheres, particularly in low density regions  (e.g., Fig.~6 of \citealt{2015MNRAS.454.3977R}). Other atoms that ionise easily include K, Ca, Mg, as well as Al and Ti. In fact, \ce{Na+}, \ce{Ca+} and \ce{Al+} reach very similar  abundances at the substellar point in the low-pressure atmosphere where p$_{\rm gas}<10^{-4}\,$bar (0.1\,mbar), as shown in Fig.~\ref{fig:Molecs} (top right). \ce{K+} is about 2 orders of magnitude less abundant. We note, however, that atomic S and Fe remain more abundant than these positive ions.

Figures~\ref{fig:TiTiO} and~\ref{fig:SiSiO} demonstrate  that the atomic species Si and Ti dominate over their molecular counterparts at the dayside at 0.1\,mbar and 1\,bar level.  A similar conclusion can be drawn for S/\ce{H2S} based on Fig.~\ref{fig:Molecs}. However,  Si is only slightly more abundant than  SiO  on the dayside and  n(Si)<n(SiO) already occurs well before the terminator regions for a given pressure level.  Figure~\ref{fig:rATIOS_MIXING} shows that easy-to-ionise metals (like Na, K, Ca) appear in the second ionisation state on the dayside.  Figure~\ref{fig:Molecs} demonstrates that this is also the case for \ce{Al+}.  Therefore, the {\it opacity of the dayside} will be affected by Ti rather than TiO, H rather than \ce{H2} (or \ce{H2O}), S  rather than \ce{H2S}, by \ce{Na+}, \ce{Ca+}, \ce{Al+} and \ce{K+} rather than Na, Ca, Al, and K. SiO should still be an appreciable opacity source amongst the Si-binding gas species.

\subsection{Carbon-to-oxygen ratio (C/O) in HAT-P-7b}\label{ss:CtoO}

Planetary carbon-to-oxygen (C/O) ratios are largely influenced by atmospheric chemistry 
and may provide key insights into the formation histories of gas giants. Protoplanetary disk models predict that C/O for planets vary depending on distance from the host star, since water vapour and carbon monoxide have different condensation temperatures \citep{Oberg11,2014Life....4..142H,2018A&A...613A..14E}. We note that our initial C/O has a solar value and the resulting ``local'' C/O is the result of our cloud and chemistry simulation  rather than a parameter of it. To explore the effect of cloud formation on the atmospheric composition of this ultra-hot Jupiter, we provide C/O for all profiles tested and map the gas-phase C/O globally at specific pressure levels and integrated through the atmosphere in Fig.~\ref{fig:CO_level}. We construct these plots using the gas-phase element abundances of carbon and oxygen for HAT-P-7b from the cloud formation models described in Sect.~\ref{ss:cm}. We demonstrate that C/O appears to be a well-suited tracer of cloud formation regions in HAT-P-7b, and may even be a useful probe in all cloud-forming atmospheres.

In our models, the cloud-free C/O (prior to cloud formation) is uniformly solar ($\sim$ 0.54). Figure~\ref{fig:CO_level} shows C/O maps for HAT-P-7b after cloud formation at p$_{\rm gas}=10^{-4}$\;bar and p$_{\rm gas}=0.1$\;bar. These level pressure maps indicate dayside C/O values consistent with solar, i.e. with the deep atmosphere C/O. The nightside shows an enhancement of C/O after cloud formation, with values of $\sim$ 0.7 at these pressure levels which are accessible by observations. This is consistent with \citet{molliere2015model} and \citet{molaverdikhani2019cold} findings, where they report that colder Class-II planets (global temperature ranges from $\sim$1000\;K to $\sim$1650\;K) result in enhanced condensation of oxygen-bearing condensates, which in turn result in a higher C/O ratio at the photospheric levels. Both approaches are utilizing phase-equilibrium, and they do not report any condensation in the atmosphere of planets hotter than T\textsubscript{global}$\gtrsim$1650\;K(\footnote{We note, that the values of these global temperatures can not be compared between 1D atmosphere models and 3D GCM models since a 1D atmosphere would mimic a homogeneous 3D atmosphere and the 3D average will not produce the 1D value. This emphasises the necessity of a 3D treatment of hot/ultra-hot Jupiters (where the day/night temperature difference becomes more significant) for connecting the estimated observable (local) C/O ratios to the bulk (global) C/O ratio which is as of great interest for planet formation.}).
If considering phase-equilibrium (in contrast to a kinetic non-phase equilibrium approach as utilized in this paper), the derived value of the maximum C/O depends on the completeness of the set of elements and materials considered (see Sect. 7 in  \cite{2018A&A...614A...1W}). In particular, if models neglect Mg, Fe, Si as part of a retrieval approach, C/O will be overestimated.

We find that C/O $<<$ 0.54 in the deep atmosphere where the cloud particles evaporation causes the local oxygen abundance to increase. Values of C/O $\sim$ 0.25 are reached for dayside profiles where (few but) the largest cloud particles formed and which, hence, fall deepest into the atmosphere. Such vertical transport of elements has been pointed out previously  \citep{2017A&A...603A.123H,2019arXiv190108640H}. Figure~\ref{fig:CO_level} (top) emphasises the large range of  C/O values in the atmosphere of HAT-P-7b. It is reasonable to expect that this is typical for ultra-hot giant gas planets.  These changes in the C/O  as a result of cloud formation affirms our previous notions that there is no single value of C/O to describe a cloud-forming exoplanet \citep{2018arXiv181203793H}.

We also integrate the C/O ratio along atmospheric height and map the global C/O structure in Fig.~\ref{fig:CO_integrated}. The normalised, height-integrated C/O takes into account that the dayside atmosphere has a larger geometrical extension compared to the nightside (see Sect.~\ref{s:cloud}). 
The steep temperature gradient appears in the maps as a clearly defined shape corresponding to the hottest part of the HAT-P-7b's atmosphere where clouds cannot form. {\it C/O traces out the cloud-affected area in a planetary atmosphere very well.}

We find that the nightside value of C/O  $\sim$ 0.7 appears in agreement with planet formation models, which predict that water ice in planetismals results in C/O values less than unity in gas giant atmospheres (\citealt{Mordasini2016,Espinoza2017}). This coincidence shows that C/O values  for planets affected by cloud formation can masque formation scenario and create a false-positive. Therefore, any C/O ratio retrieval linking to  planetary formation interpretation should be accompanied with a detailed modeling of cloud formation, to properly connect the {\it observed local} C/O to the {\it global} (bulk) C/O value deduced from formation models.

\begin{figure*}[t!]
    \vspace*{-1.5cm}
    \includegraphics[width=\textwidth]{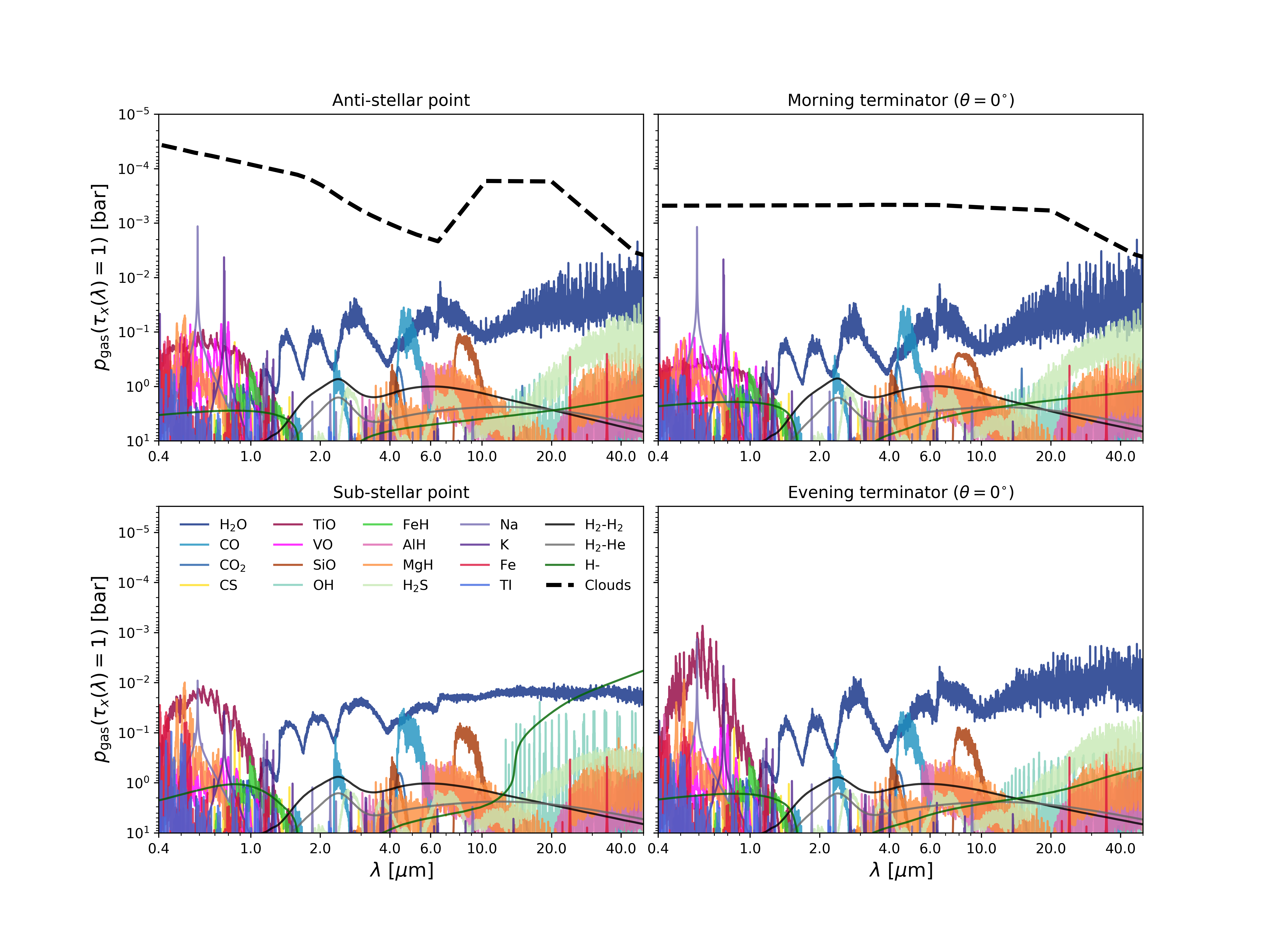}
    \caption{Prominent sources of gas-phase opacity. The atmospheric gas pressure, $p_{\rm gas}$ [bar], at which the vertical wavelength-dependent optical depth, $\tau_{ x}(\lambda)$, of species $x$ reaches unity is shown for $\lambda= 0.4\,\ldots\, 50\mu$m. All gas phase opacities are plotted at a spectral resolution of $R = 1000$. Four equatorial regions are considered: (i) the anti-stellar point $(\phi, \theta) = (180\degree, 0\degree)$, (ii) the morning terminator $(\phi, \theta) = (270\degree, 0\degree)$, (iii) the sub-stellar point $(\phi, \theta) = (0\degree, 0\degree)$, and (iv) the evening terminator $(\phi, \theta) = (90\degree, 0\degree)$. In regions containing clouds, we also show the atmospheric pressure where the cloud becomes optically thick, i.e. $\tau_{\rm cloud}(\lambda)=1$, for comparison. The gas pressure at which the atmosphere of HAT-P-7b becomes optically thick varies by orders of magnitude between clear and cloudy regions.}
    \label{fig:gasopacdepth}
\end{figure*}
\begin{figure*}
    \centering
    \includegraphics[width=\hsize]{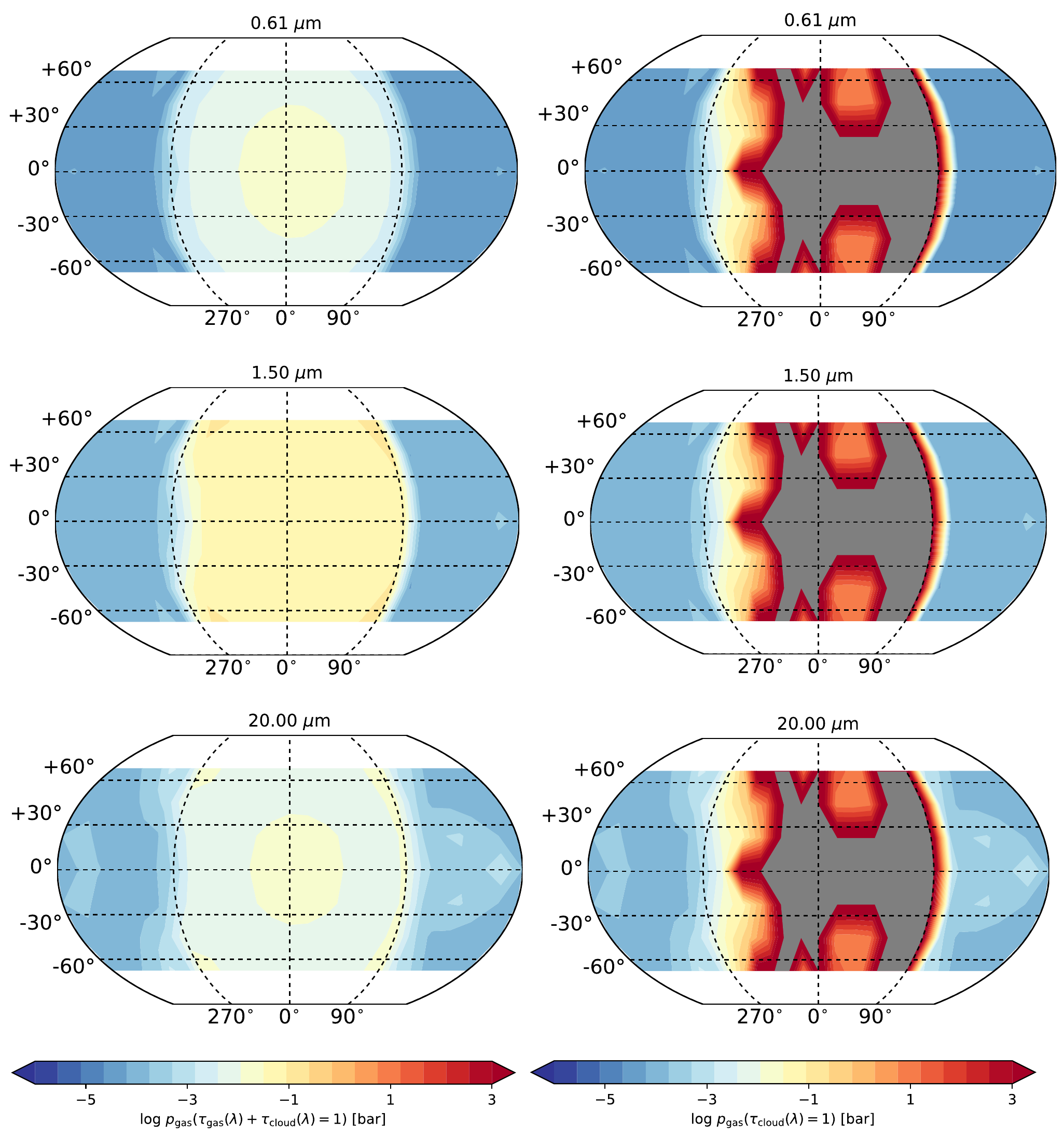}
    \caption{The pressure levels, log(p$_{\rm gas}$ / [bar]), where the atmosphere becomes optically thick at several distinct wavelengths: the optical (0.6\;$\mu$m, top), infrared (1.5\;$\mu$m, middle), and mid-infrared (20\;$\mu$m, bottom). The left side shows the gas pressure level at which  $\tau_{\rm cloud}(\lambda)+\tau_{\rm gas}(\lambda)=1$. The right side shows the same but for $\tau_{\rm cloud}(\lambda)=1$ only.  Because there are no clouds near the equator on the dayside, the map shows the highest pressure layer used in the simulation (100\;bar).}
    \label{fig:cloudtau}
\end{figure*}

\begin{figure*}
    \centering
    \includegraphics[width=\hsize]{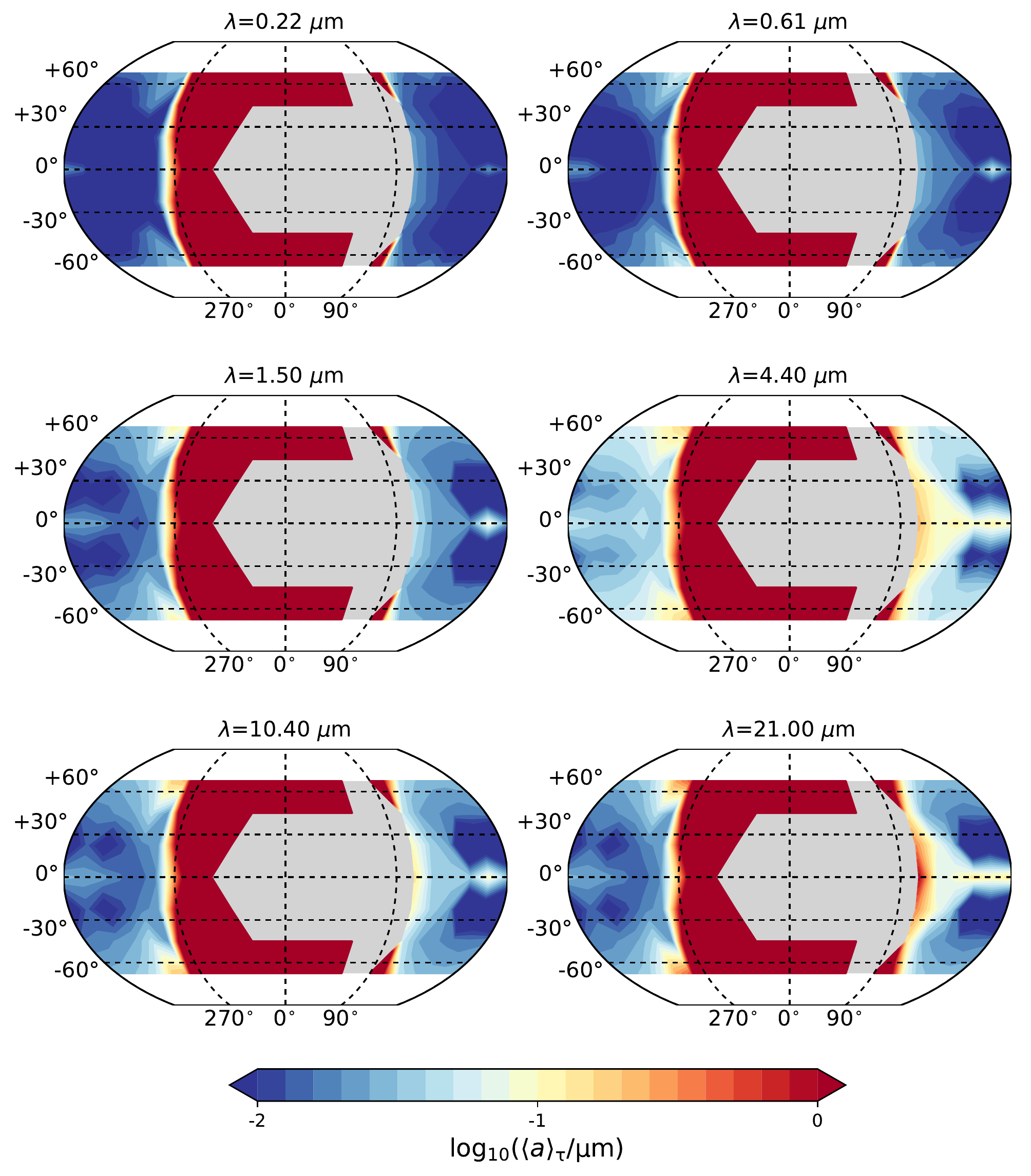}
    \caption{The 2D distribution of the wavelength-dependent optical-depth weighted mean particle radius, $\langle a\rangle_{\tau}$ [$\mu$m].  Note that cloud-free region in the dayside hemisphere is colored by light grey. The atmosphere of HAT-P-7b becomes optically thick due to small cloud particles on the nightside where $\langle a\rangle <10^{-1}\,\ldots\,10^{-2}\mu$m and due to large particles of $\langle a\rangle \approx 1\mu$m on the morning terminator and in the higher latitudes.}
    \label{fig:atau}
\end{figure*}

\section{Atmospheric opacity from gas and clouds} \label{s:opacity}

Having considered the formation of clouds and the equilibrium gas phase chemistry in HAT-P-7b's atmosphere, we now assess their relative importance as sources of atmospheric opacity. Due to the fact that an abundant species with a negligible absorption cross section can matter far less for radiative transfer, atmospheric heating, etc., than a trace species with a sizable cross section, it is the \textit{abundance-weighted} cross sections (\textit{opacity}) that are the key quantities of interest.

For the purposes of this section, we imagine a vertically downwards line-of-sight commencing at the top of the atmosphere for four locations of interest: the sub-stellar point $(\phi, \theta) = (180\degree, 0\degree)$, the anti-stellar point $(\phi, \theta) = (0\degree, 0\degree)$, the equatorial region at the morning terminator $(\phi, \theta) = (270\degree, 0\degree)$, and at the evening terminator $(\phi, \theta) = (90\degree, 0\degree)$. By integrating the wavelength-dependant opacities of the gas phase species and clouds along each line-of-sight (Eq.~\ref{eq:taud}), we obtain an initial picture of the dominant opacity source at each wavelength on the dayside, nightside, and on each side of the terminator. Our results enable the identification of the regions in HAT-P-7b's atmosphere where clouds will influence the radiative transfer, heating, and cooling. A full radiative transfer calculation capable of simulating spectral observations will be considered in Paper III.

\subsection{Preeminent opacity sources: p$_{\rm gas}(\tau(\lambda)=1)$}
\label{ss:tau1mol_abovecloud}

Led by our results from Sect.~\ref{ss:ggchem_dn}, we consider the influence of various gas phase species, $x=\,$ \ce{H2O, C2H2, CH4, CO2}, CO, CH, OH, HCN, NH, NO, \ce{NH3, O3}, SiH, SiO, AlH, AlO, MgH, Cs, FeH, \ce{H2S, SO2, H2, N2, O2}, TiO, CaO, CaH, TiH, LiH, VO, Na, K, Li, Fe, Ti, and H-, along with \ce{H2-H2 and H2-He} CIA, on the vertical optical depth of HAT-P-7b's atmosphere. For each species, we determine the gas pressure where the atmosphere becomes optically thick at a wavelength $\lambda$, i.e. p$_{\rm gas}(\tau_x(\lambda)=1)$ according to Sect.~\ref{s:ap}.

In Fig.~\ref{fig:gasopacdepth}, we show the atmospheric pressure level where $\tau_x(\lambda) = 1$ (integrated downwards from the top of the atmosphere), calculated at the four profiles of interest. At optical wavelengths ($0.3-0.8~\mu$m), the most prominent gas phase opacity sources are Na and K at the anti-stellar point and morning terminator, whereas TiO and MgH become more important at the sub-stellar point and evening terminator. At longer infrared, wavelengths ($\lambda > 1~\mu$m), CO and \ce{H2O} become the dominant gas phase opacity sources. The only exception is for mid-infrared wavelengths $> 10~\mu$m, where OH opacity competes with that of \ce{H2O} at the sub-stellar point. \cite{2007ApJS..168..140S} demonstrate that the HI continuum opacity remains small compared to the \ce{H2O} line opacity.

However, cloud particles have large absorption and scattering cross-sections, and hence can completely obscure gas phase properties of an atmosphere. We compare the relative strengths of gas phase and cloud opacities to assess their anticipated influence on the optical depth --- and by proxy, spectral observations --- in each atmospheric region. Figure~\ref{fig:gasopacdepth} overplots the pressure where the atmosphere becomes optically thick based on the cloud opacity alone (p$_{\rm gas}(\tau_{\rm cloud}(\lambda)=1)$, thick black dashed line). This is only relevant for cloud forming regions, i.e., for the anti-stellar point and the morning terminator in Fig.~\ref{fig:gasopacdepth}. 

The cloud opacity on the morning terminator appears wavelength independent (`grey') for $\lambda < 20\mu$m, due to the presence of larger cloud particles ($a \sim 0.5\mu$m, described in Sect.~\ref{ss:OpacityMaps}). Smaller cloud particles can survive on the night-side, as seen at the anti-stellar point, causing extinction that is more wavelength dependent. We also see hints of an extinction bump at $\lambda \sim 10\mu$m, which is caused by silicate absorption and similar what has been shown for brown dwarf atmospheres (see Fig. 9 in \citealt{2008A&A...485..547H}). {\cite{2018MNRAS.481..194L} demonstrate that observing the mineral absorption features in a transmission spectrum  maybe a non-trivial task.}

Figure~\ref{fig:cloudtau} (right) shows the 2D map of the atmospheric pressure at which the vertical optical depth due to cloud opacity alone reaches $\tau_{\mathrm{cloud}}(\lambda)= 1$ for some selected wavelengths. Across the optical and near-infrared, HAT-P-7b's nightside is completely obscured by clouds, such that only the uppermost layers of the atmosphere (p$_{\rm gas} < 10^{-3}$\,bar) are spectroscopically accessible. Therefore, the nightside emission spectrum should be dominated by clouds. The cloud opacity across the terminators, however, exhibits a high degree of spatial variability. Figure~\ref{fig:cloudtau} shows that, due to global circulation {affecting the temperature structure}, clouds obscure the atmosphere from $10^{-2}-10^{-1}$\,bar on the morning terminator, while the evening terminator exhibits clear skies (i.e., p$_{\rm gas}(\tau_{\rm cloud}(\lambda)=1) > 1$\,bar). The absence of clouds in the upper atmosphere at the evening terminator allows the gas-phase opacity to dominate from the optical to mid-infrared (Fig.~\ref{fig:gasopacdepth}, left). This result leads us to expect a roughly 50\% cloud coverage for the terminator of HAT-P-7b -- similar to that inferred from the transmission spectrum of the somewhat cooler hot Jupiter HD 209458b \citep{MacDonald2017}. This `patchy cloud' scenario is known to strongly influence transmission spectra \citep{Line2016}. We shall examine predictions for the transmission spectrum of HAT-P-7b, along with observational implications, in Paper III.


\citet{2016NatAs...1E...4A} reported that the emission and reflection phase curve of HAT-P-7b exhibits temporal variability. Specifically, the phase offset between secondary eclipse and when the peak planet flux is observed oscillates between times before and after secondary eclipse. This has been  interpreted as reflected light variations due to time-variable cloud coverage on the dayside \citep{Oreschenko16_phase-curve,2016ApJ...828...22P,2019arXiv190304565H}.
However, our Figure \ref{fig:cloudtau} shows that the dayside opacity is overwhelmed by gas-phase contributions and that high-temperature cloud species formed on the dayside are likely not responsible for the phase curve observations. 
It was argued that the temporal increase of winds might transport cloud from nightside to dayside \citet{2016NatAs...1E...4A}. However, Sect.~\ref{s:ap} showed that the evaporation timescale is orders of magnitude shorter than the horizontal transport timescale; hence, any cloud particles that are transported to the dayside will evaporate quickly.
Our results suggest that the time variability of the phase curve of HAT-P-7b originate from alternative mechanisms, such as the interaction between ionized atmosphere and magnetic fields \citep{Rogers17_magnetic,Hindel19_magnetic,2019arXiv190304565H}. Future 3D modeling coupling atmospheric dynamics and cloud microphysics is needed to justify this hypothesis.
At the very least, our result that the gas-phase dominates the dayside opacity is consistent with the portion of the results by \citet{2016NatAs...1E...4A} that show the light curve usually exhibits the flux peak before the secondary eclipse, which is expected for cloud-free tidally locked planets \citep[e.g.,][]{Showman2009}.

\begin{figure*}
    \centering
    \includegraphics[width=\hsize]{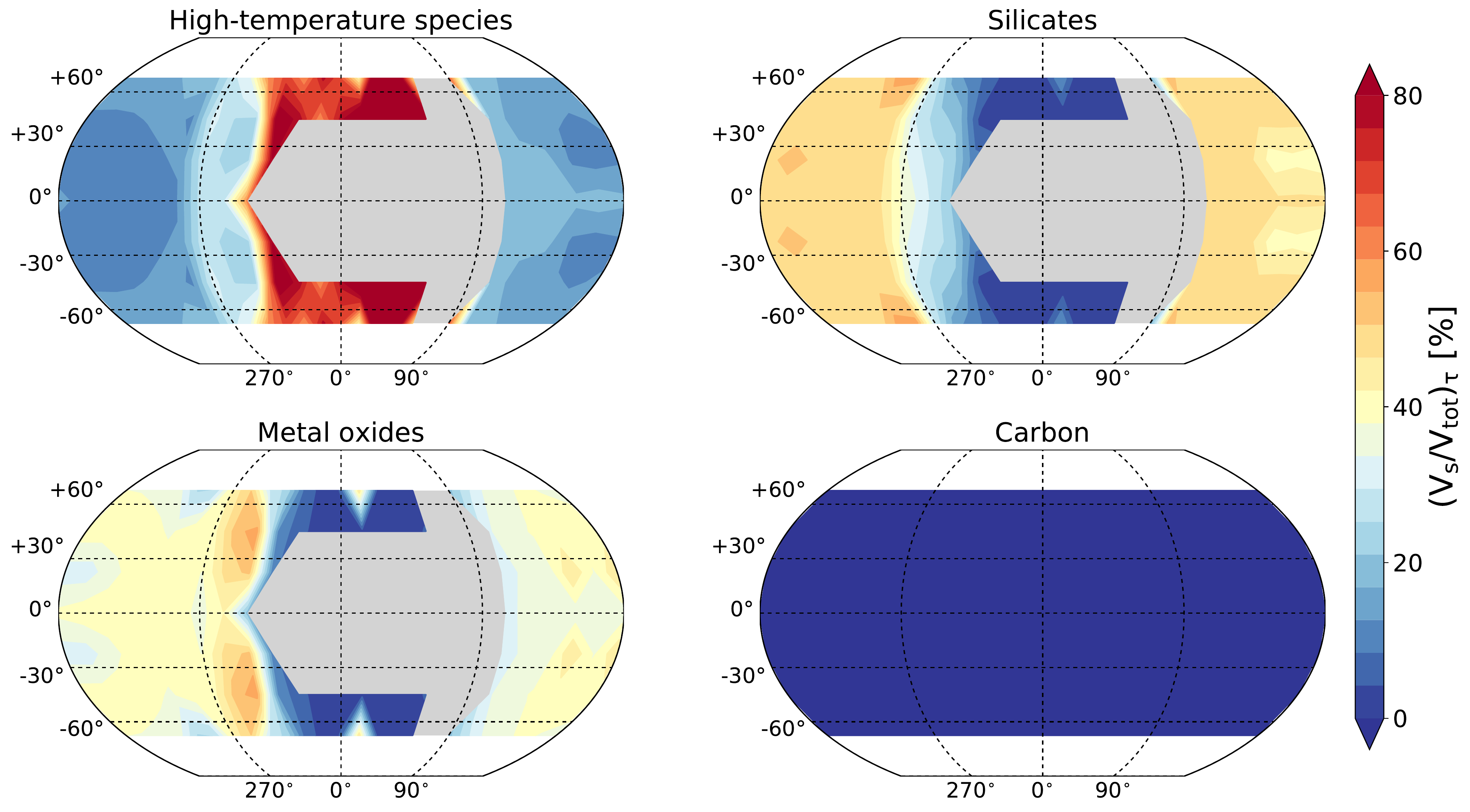}
    \caption{The 2D distribution of the wavelength-dependent optical-depth weighted volume fraction of condensate materials  of clouds ($\langle V_s/V_{\rm tot}\rangle_{\tau}$, color scale) for $\lambda=1.5$\;$\mu$m. Each material is categorized into the same groups as in Figure \ref{fig:cloud_map_1bar_2}. The maximum volume contribution to $\tau(\lambda)\approx 1$ of any of the species is 20\%.}
    \label{fig:map_comp1}
\end{figure*}
\begin{figure*}
    \centering
    \includegraphics[width=\hsize]{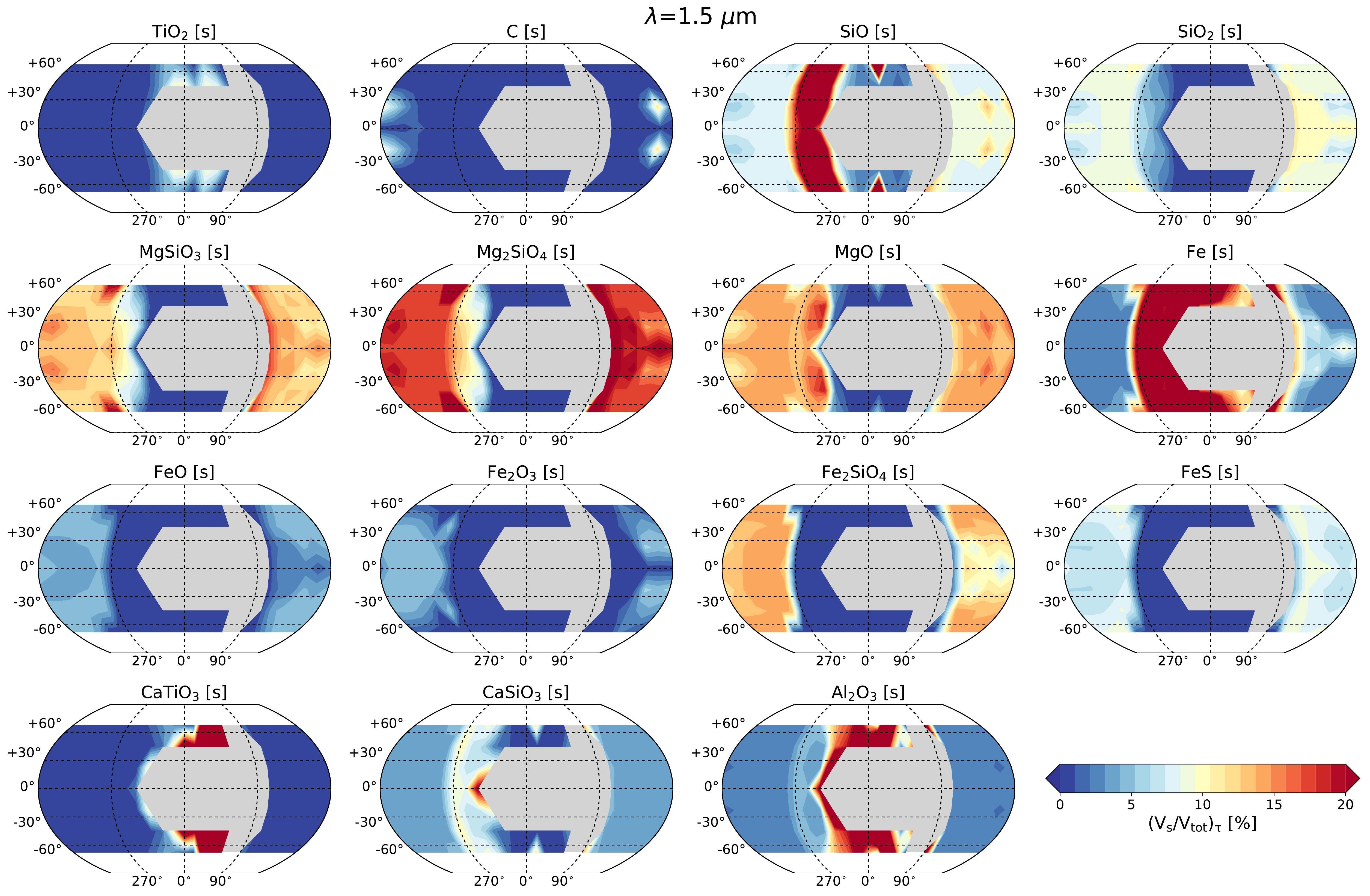}
    \caption{The 2D distribution of the wavelength-dependent optical-depth weighted volume fraction of condensate materials  of clouds ($\langle V_s/V_{\rm tot}\rangle_{\tau}$, color scale) for $\lambda=1.5$\;$\mu$m. The maximum volume contribution to $\tau(\lambda)\approx 1$ of any of the species is 20\%. }
    \label{fig:map_comp1}
\end{figure*}

\subsection{Optical-depth weighted cloud properties}
\label{ss:OpacityMaps}

To better understand which aspect of the cloud particles dominate the cloud contribution to the atmospheric, we calculate optical-depth weighted cloud properties. For a given physical quantity $f$, such as the mean particle size $\langle a\rangle$, the weighted quantity $f_{\rm \tau}$ is defined as
\begin{equation}
 f_{\rm \tau}\equiv \frac
 {\int_{\rm 0}^{\rm \infty} f\exp{(-\tau_{\rm cloud}(\lambda))}\,d\tau_{\rm cloud}(\lambda)}
 {\int_{\rm 0}^{\rm \infty} \exp{(-\tau_{\rm cloud}(\lambda))}\,d\tau_{\rm cloud}(\lambda)},
 \label{eq:ftau}
\end{equation}
where $\tau_{\rm cloud}(\lambda)$ is the wavelength-dependent vertical
optical depth of the cloud. Eq.~\ref{eq:ftau} was also used by
\citet{Ohno&Okuzumi17} to compare their cloud profiles with
observations of Jovian clouds. The weighting factor $e^{-\tau_{\rm
    cloud}(\lambda)}$ accounts for the fact that the cloud particles
will only be observable if their absorption remains optically thin at
this wavelength. In Eq.~\ref{eq:ftau}, $e^{-\tau_{\rm
    cloud}(\lambda)}$ decreases inwards (with increasing $\tau_{\rm
  cloud}$) while $f$ may increase. Hence $f_{\rm \tau}$ is the maximum
of the envelope of the product of the two functions ($e^{-\tau_{\rm
    cloud}(\lambda)}$ and $f$), i.e. the maximum of the convolution.
The weighted quantity $f_{\rm \tau}$ traces the properties of the
cloud particles with the largest contribution on the cloud opacity
above the cloud deck, and thereby, the particles that dominate
spectral observations. Effectively, $f_{\rm \tau}$ traces the cloud
properties on the $\tau_{\rm cloud}(\lambda)\sim 1$ plane.

First, we examine the optical-depth weighted mean particle size, $\langle a \rangle_{\tau}$ [$\mu$m], according to Eq.~\ref{eq:ftau} for different wavelengths. Figure~\ref{fig:atau} shows the 2D distribution in latitude and longitude of $\langle a \rangle_{\tau}$  for a few selected wavelengths. On the dayside, when present, the cloud opacity is dominated by particles with sizes of $\langle a \rangle_{\tau}\approx 1~\mu$m. 
Our simulation results in Fig.~\ref{fig:1Damean_1} (top left) demonstrate that cloud particles can grow to cm-sized particles on the dayside of HAT-P-7b, which will affect the remaining element budget of the local atmospheric gas. However, the day-side atmosphere is dominated by gas-phase opacity, as seen in Fig.~\ref{fig:gasopacdepth}, so these large cloud particles will not be responsible for observations such as emission and transmission spectra.
On the night-side, in contrast, small particles $\langle a \rangle \approx 10^{-1}-10^{-2}~\mu$m dominate the cloud opacity at $\lambda=0.22~\mu$m and $\lambda=0.61~\mu$m. The optical-depth weighted mean size slightly increases with increasing wavelength, hence, longer wavelengths can probe deeper regions where larger particles are present. On the entire night-side, $\langle a \rangle_{\tau} \la 0.1~\mu$m for all wavelengths examined here. This indicates that cloud particles on the nightside fall into the Rayleigh regime in which the opacity is wavelength-dependent. This explains why the opaque cloud level for the anti-stellar point decreases with increasing wavelength in Fig.~\ref{fig:gasopacdepth} (black dashed line).

The cloud particles that form in the atmosphere of HAT-P-7b are made of a mix of materials that changes throughout the cloud deck, as demonstrated in Sect.~\ref{ss:how}, which is a finding similar to all other giant gas planets that have been addressed by detailed cloud modelling   \citep[HD\,189733\,b, HD\,209458\,b and WASP-18b,][]{Lee2015,2018A&A...615A..97L,2019arXiv190108640H}. In Sect.~\ref{sec:map_comp} we have used 2D maps to trace the changing material mixes globally at two different pressure levels. It is therefore worth investigating the relative importance of each material in terms of cloud opacity. Figure~\ref{fig:map_comp1} shows the optical-depth weighted volume fraction of each condensate material, $\langle V_{\rm s}/V_{\rm tot}\rangle_{\tau}$, diagnosing the relative contribution of each material on total cloud opacity at $\lambda=1.5\mu$m only. The high-temperature species (TiO$_2$[s], Fe[s], CaTiO$_3$[s],  Al$_2$O$_3$[s]) dominate the cloud opacity on the dayside. On the other hand, the cloud opacity on the nightside is largely dominated by silicates (MgSiO$_3$[s], Mg$_2$SiO$_4$[s],  Al$_2$O$_3$[s], Fe$_2$SiO$_4$[s]).  The metal oxides have a relatively minor effect on the cloud opacity except for SiO[s] and MgO[s]. The impacts of silicate materials on cloud opacity is also seen in Fig.~\ref{fig:gasopacdepth}, in which the opaque cloud level steeply increases around $10~\mu$m. This is caused by the absorption signature of silicate species.

%

\section{Summary}

{We investigated cloud formation and its effects on the gas-phase composition for 1D atmosphere profiles extracted from a 3D GCM solution for the ultra-hot Jupiter HAT-P-7b. We  provide a global overview of clouds on HAT-P-7b, including details on particle sizes, material compositions, cloud particle load, C/O, $p_{\rm gas}(\tau_{\lambda}=1)$.  Our detailed account of the global cloud properties and associated element depletion effects on the gas phase enables us to build  a comparative understanding for these class of easy-to-access JWST targets. }

\begin{itemize}
\item The high radiation input from the host star HAT-P-7 causes substantial thermodynamic day/night differences of the ultra-hot Jupiter HAT-P-7b. A maximum day/night temperature difference of 2500\;K occurs, where the dayside appears geometrically more extended than the nightside.

\item  Ultra-hot Jupiters cannot be expected to have an average isothermal upper atmosphere on the dayside due to the strong variation of the local temperature throughout the atmosphere. 

\item  The global wind dynamics in the atmosphere of HAT-P-7b in the equator regions create a chemical asymmetry between the terminator regions that emerges as a typical feature for tidally locked giant gas planets. 
 
\item  The global wind dynamics also result in the temperature inversion, which is more profound at the morning terminator.
 
\item Asymmetric terminator properties, in particular a combination of cloud-forming (morning) and cloud-free (evening) regions result. Observing such asymmetries may be possible even with photometric surveys (e.g. CHEOPS. PLATO) without recourse to spectroscopy, however, more detailed studies are required. Ultra-hot Jupiters like HAT-P-7b may this be the best candidates to study terminator asymmetries.  
 
\item Clouds form predominantly on the nightside of HAT-P-7b. The dayside features an almost completely cloud-free equatorial region where clouds can extend into the dayside from the morning terminator. Clouds also form at higher latitude at the sub-stellar point on the dayside due to rapid formation and growth of high temperature condensates (TiO$_2$[s]) at high pressures (1~bar), forming a thin layer of cm-sized particles.

 \item The cloud opacity on the nightside is dominated by $\sim0.01$--$0.1~{\rm \mu m}$ particles, made mainly of silicates and metal oxides (\ce{MgSiO3}[s],\ce{Mg2SiO4}[s],\ce{Fe2SiO4}[s], MgO[s]). The cloud opacity on the dayside is, if present, dominated by large particles with $\sim1~{\rm \mu m}$ composed of a mix of high-temperature condensates and metal oxides (\ce{TiO2}[s], SiO[s], Fe[s], \ce{Al2O3}s]). These large $\mu$m-sized cloud particles  will not be responsible for the observations because gas species overwhelm the opacity on the dayside. 
 
 \item The gas pressure level that traces where the atmosphere becomes optically thick is highly wavelength dependent, particularly in the cloud-free parts of the atmosphere. This includes the evening terminator.  $p_{\rm gas}(\tau(\lambda)=1)\not=$\,const is reinforced on the nightside because of the cloud characteristics, whereas the clouds enforce a $p_{\rm gas}(\tau(\lambda)=1)=$\,const on the morning terminator except where Na and K absorb in the optical and CO in the near-IR.
 
\item C/O reaches super-solar values of $\sim 0.7$ in the cloud forming parts of the atmosphere of HAT-P-7b. C/O appears therefore well-suited to trace cloud forming regions in exoplanet atmospheres. Critically, our finding that C/O is influenced by cloud formation may hinder efforts to relate observed C/O ratios to those attained following planetary formation and evolution. Therefore, any C/O ratio derived from a retrieval analysis should be accompanied with a detailed modelling of clouds, to properly connect the {\it observed local} C/O to the {\it bulk/global} C/O value deduced from formation models.

\item The day/night gas chemistry differs strongly due to substantial day/night temperature differences. The nightside is a \ce{H2}/He/CO/\ce{H2O}-dominated gas and the dayside is dominated by HI/He/CO/\ce{H2O}. The nightside exhibits an element-depleted (due to cloud formation) molecular gas and the dayside has somewhat depleted atomic and partially ionised atmosphere. Ions like \ce{Na+} and \ce{Al+} appear considerably more abundant than on the nightside and in the terminator regions.

\item The cloud particle mass distribution is measured through the dust-to-gas ratio, $\rho_{\rm d}/\rho_{\rm gas}$, which reaches a maximum of $4.3\times 10^{-3}$ in the center regions of the clouds that form on the nightside. This values suggests an almost complete condensation of the gas in the mid-cloud regions. However,  $\rho_{\rm d}/\rho_{\rm gas}$ is not constant in the cloud forming regions of the atmosphere of HAT-P-7b.

\item {We find HAT-P-7b to be comparable to WASP-18b in that both planets form clouds predominantly on the nightside, have  a higher number of ionised species on the dayside than on the nightside, and their nightside is \ce{H2}/He dominated but the dayside is dominated by HI/He.}

\end{itemize}

\label{s:conclusions}


\begin{acknowledgements}
    We acknowledge {\it Cloud Academy I} 2018 at the Les Houches
    School of Physics, France, during which this project was kicked off. M.S. was supported by NASA Headquarters under the NASA Earth and Space Science Fellowship Program - Grant 80NSSC18K1248. M.K.A. acknowledges support by the National Science Foundation through a Graduate Research Fellowship. R.M. acknowledges support from a Royal Astronomical Society (RAS) travel grant and financial support from the Science and Technology Facilities Council (STFC), UK, towards his doctoral program. K.O. acknowledges support from JSPS KAKENHI Grant Number JP18J14557. B.L acknowledges support for this work provided by NASA through grant number HST-GO-14241.001-A from the Space Telescope Science Institute, which is operated by AURA, Inc., under NASA contract NAS 5-26555.
\end{acknowledgements}

\appendix
\section{Supplemental Tables and Figures}
\label{s:appendix}

\begin{figure*}
        \includegraphics[width=8.5cm]{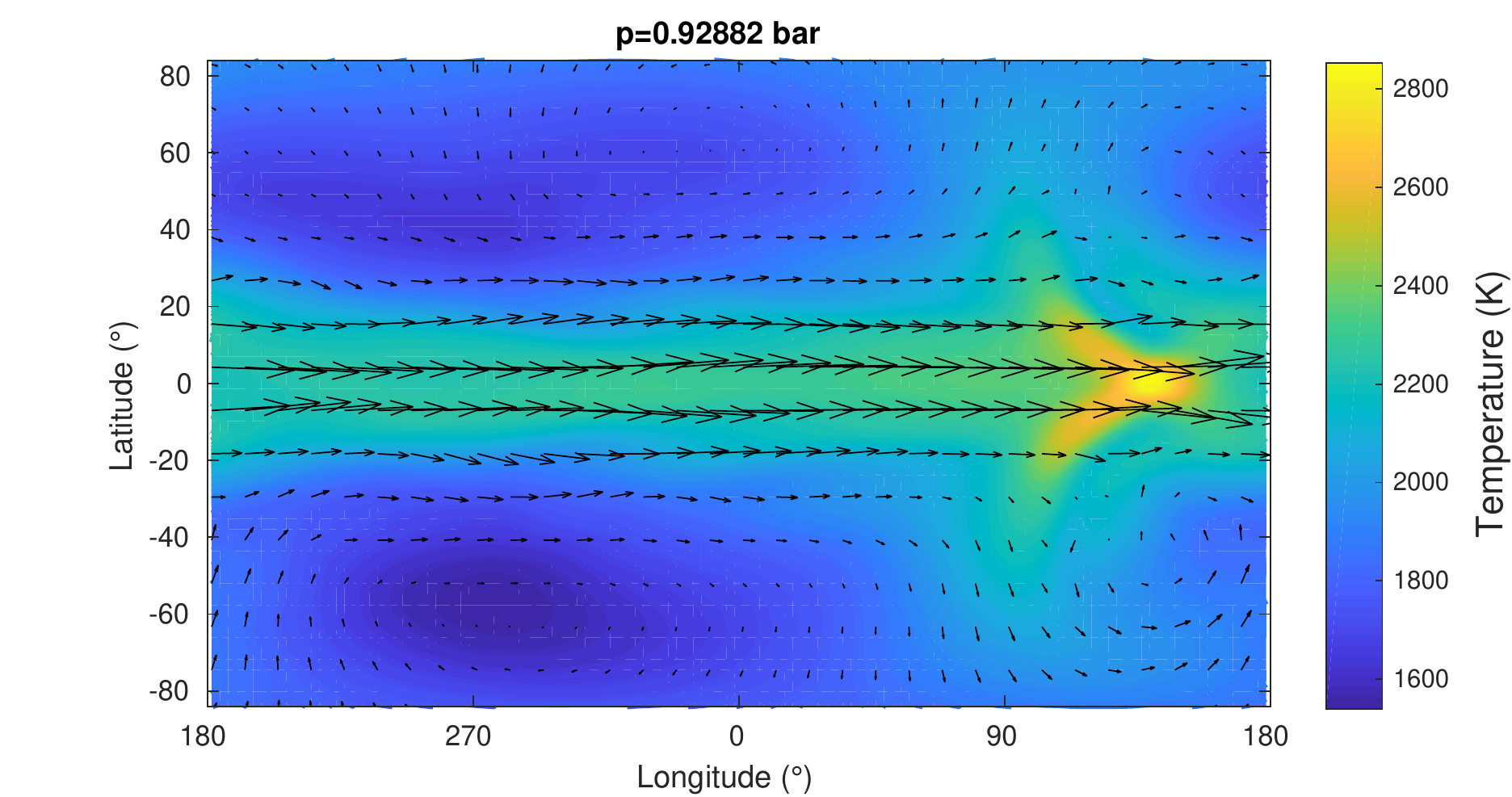}
      \includegraphics[width=8.5cm]{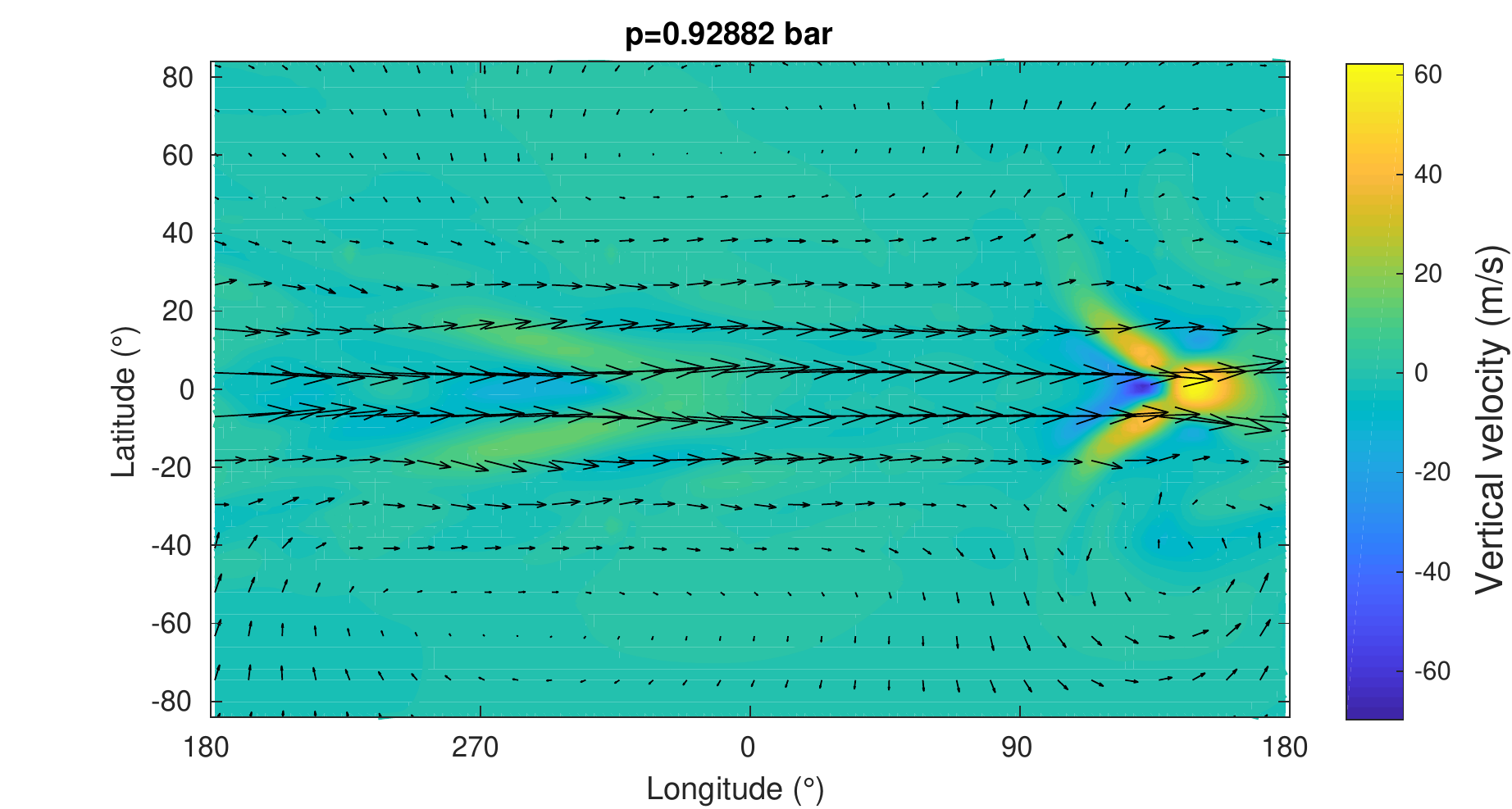}\\
      \includegraphics[width=8.5cm]{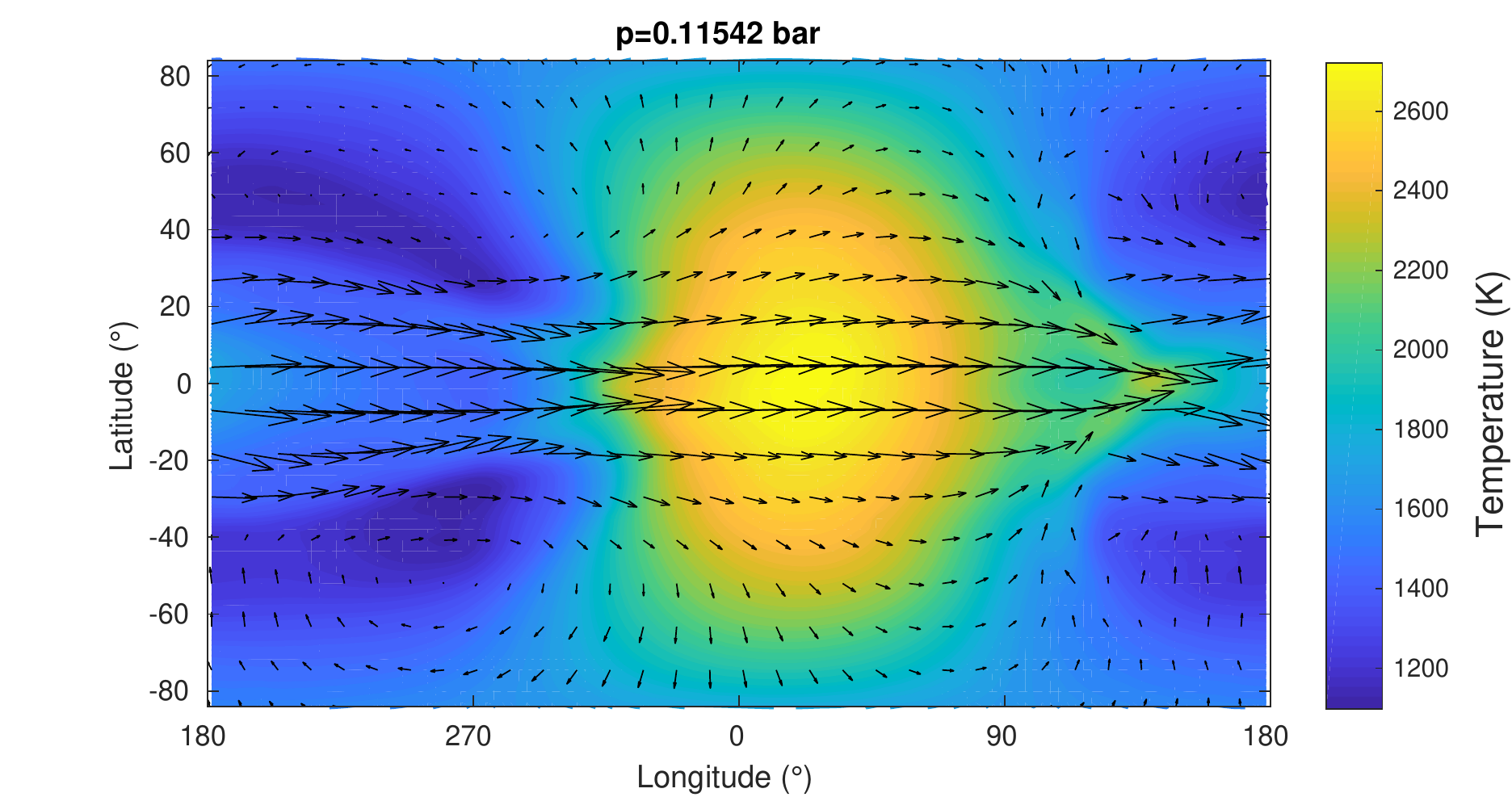}
      \includegraphics[width=8.5cm]{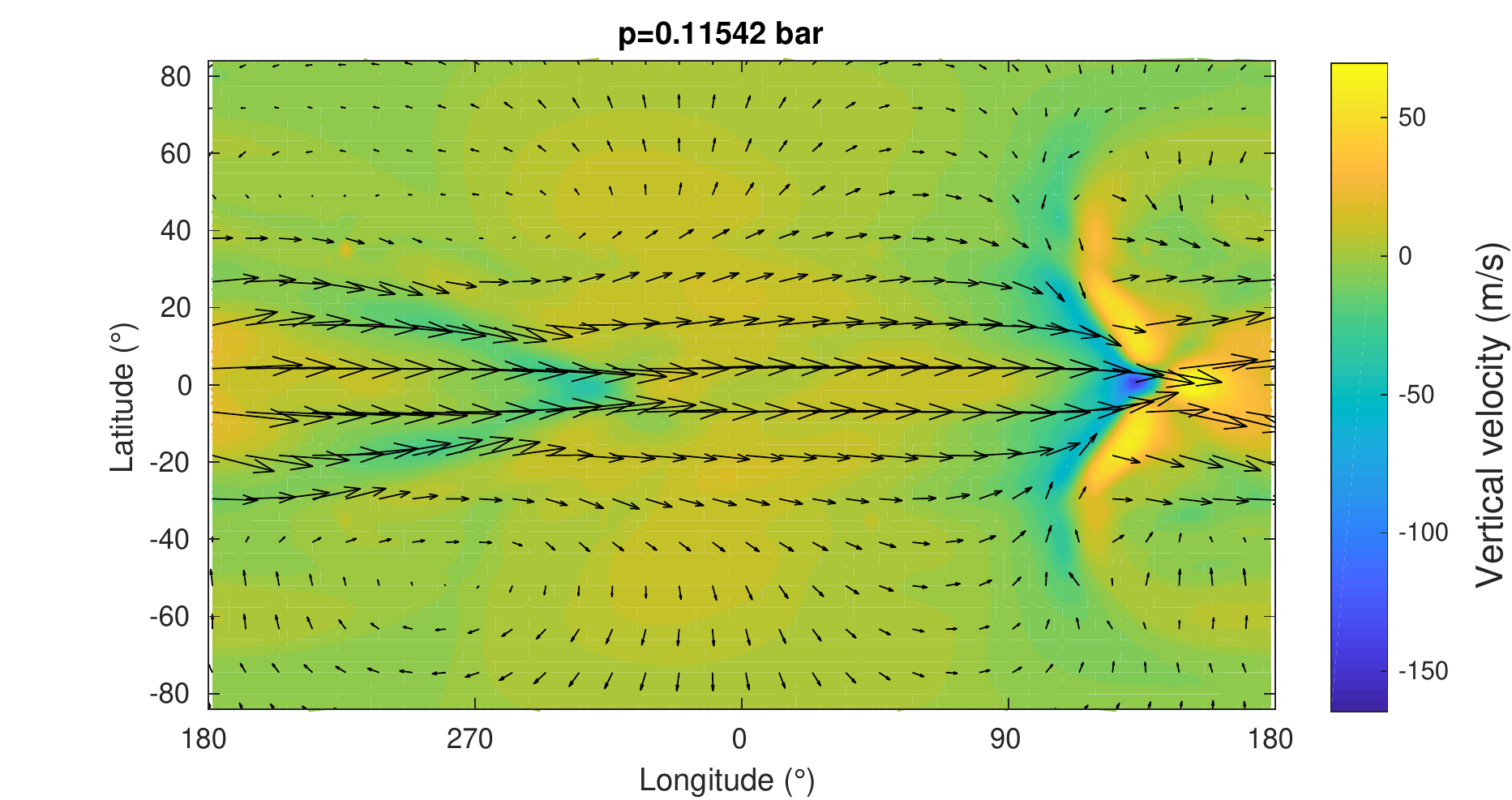}\\
      \includegraphics[width=8.5cm]{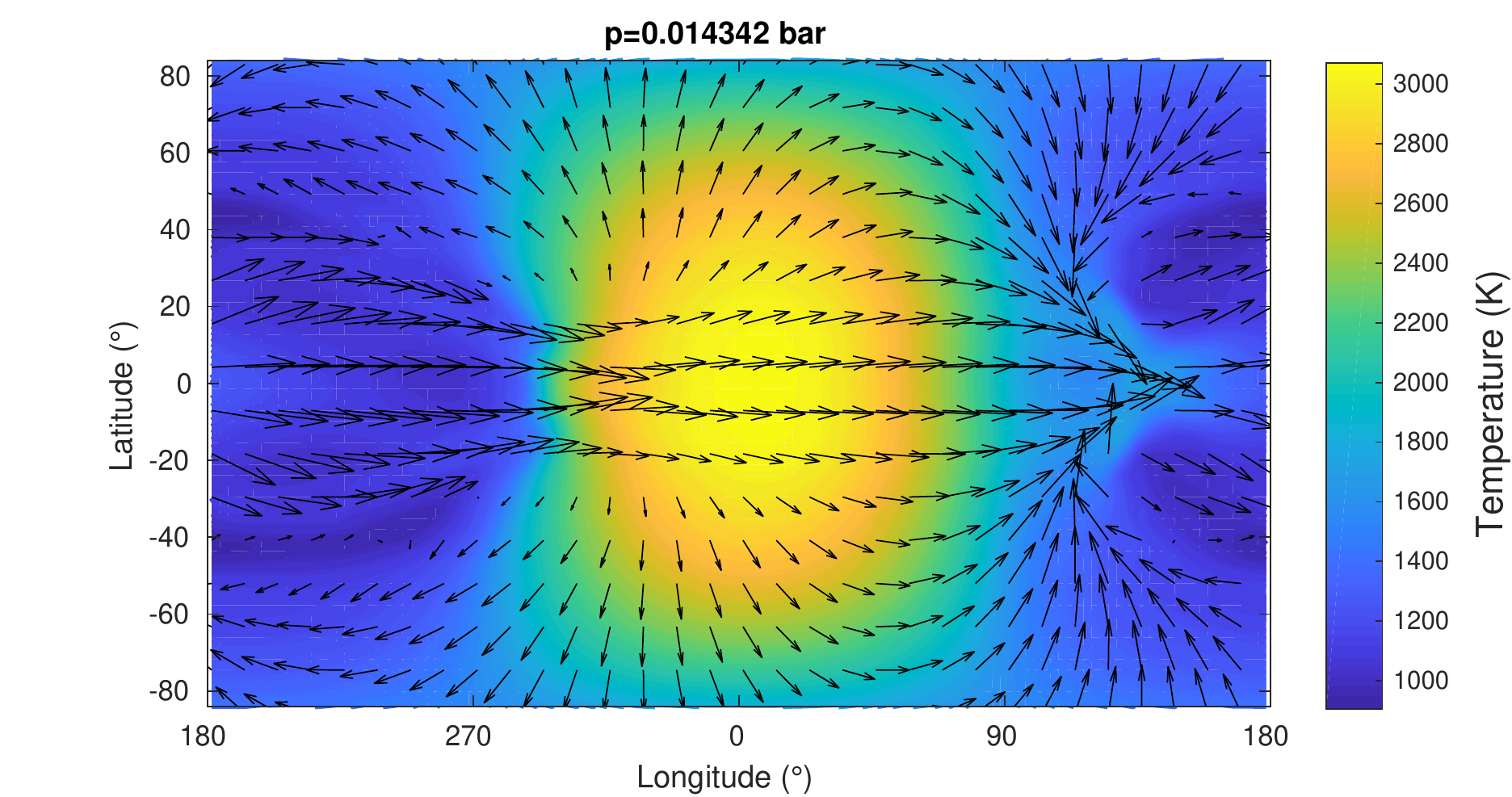}
      \includegraphics[width=8.5cm]{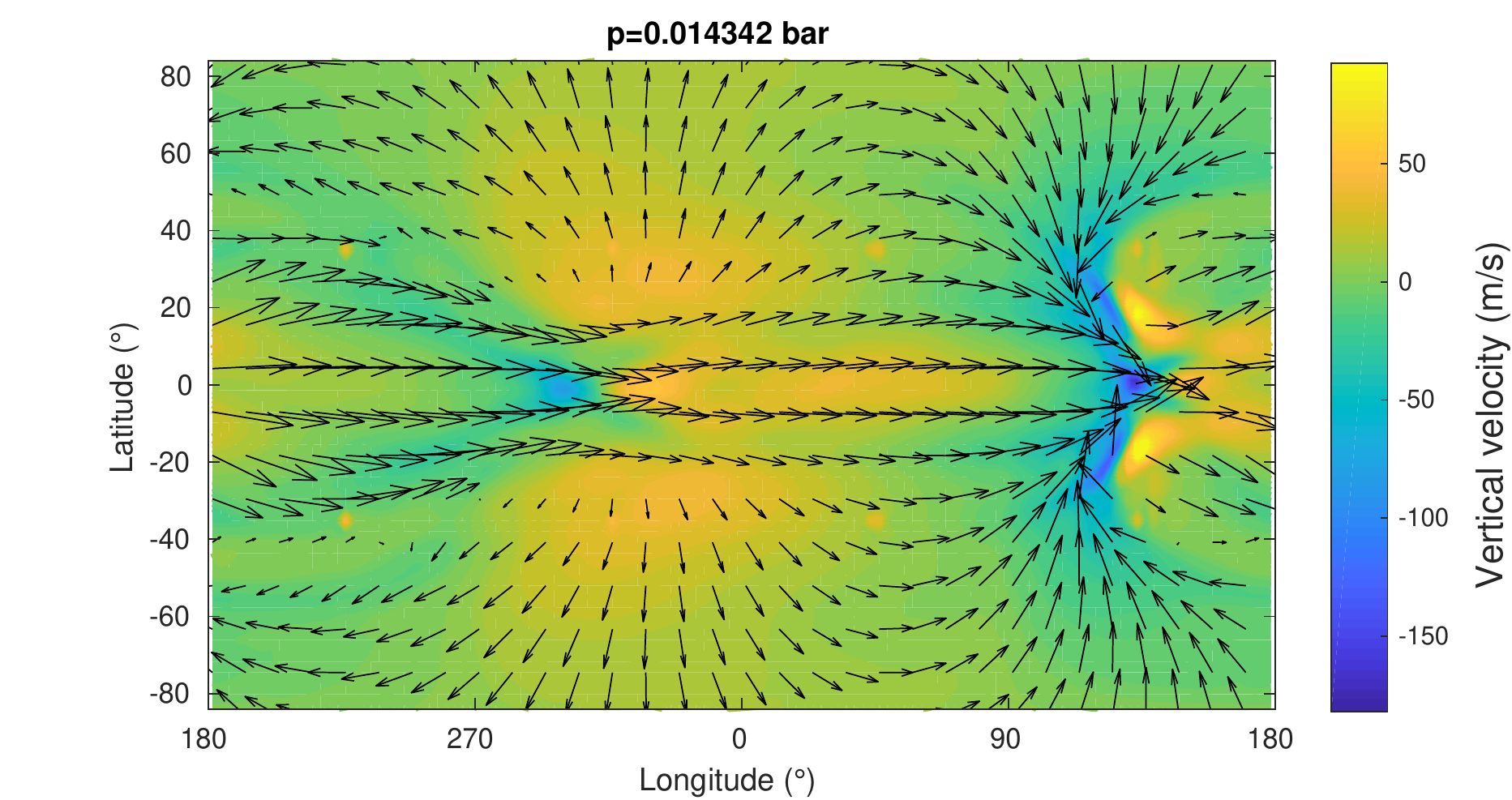}\\
    \caption{The numerical data of the 3D GCM atmosphere simulations for HAT-P-7b according to Sect.~\ref{ss:ib} at $\approx 1\,$bar, 0.1\,bar and 0.01\,bar. {\bf Left:} the gas temperature (colour coded) and horizontal velocity (arrows).  {\bf Right:}
    the vertical (colour coded)  and the horizontal (arrows) velocity components.   These are the original numerical data from which we extract the 97 1D profiles shown Fig.~\ref{fig:1DTp_1}. The velocity field (right) shows the equatorial jet,  the global-scale overturning circulation with upwelling on the dayside and downwelling on the nightside, i.e. a global
    Hadley-cell like large-scale vertical velocity components. }
    \label{fig:orgData}
\end{figure*}

\begin{figure}
        \includegraphics[width=8.5cm]{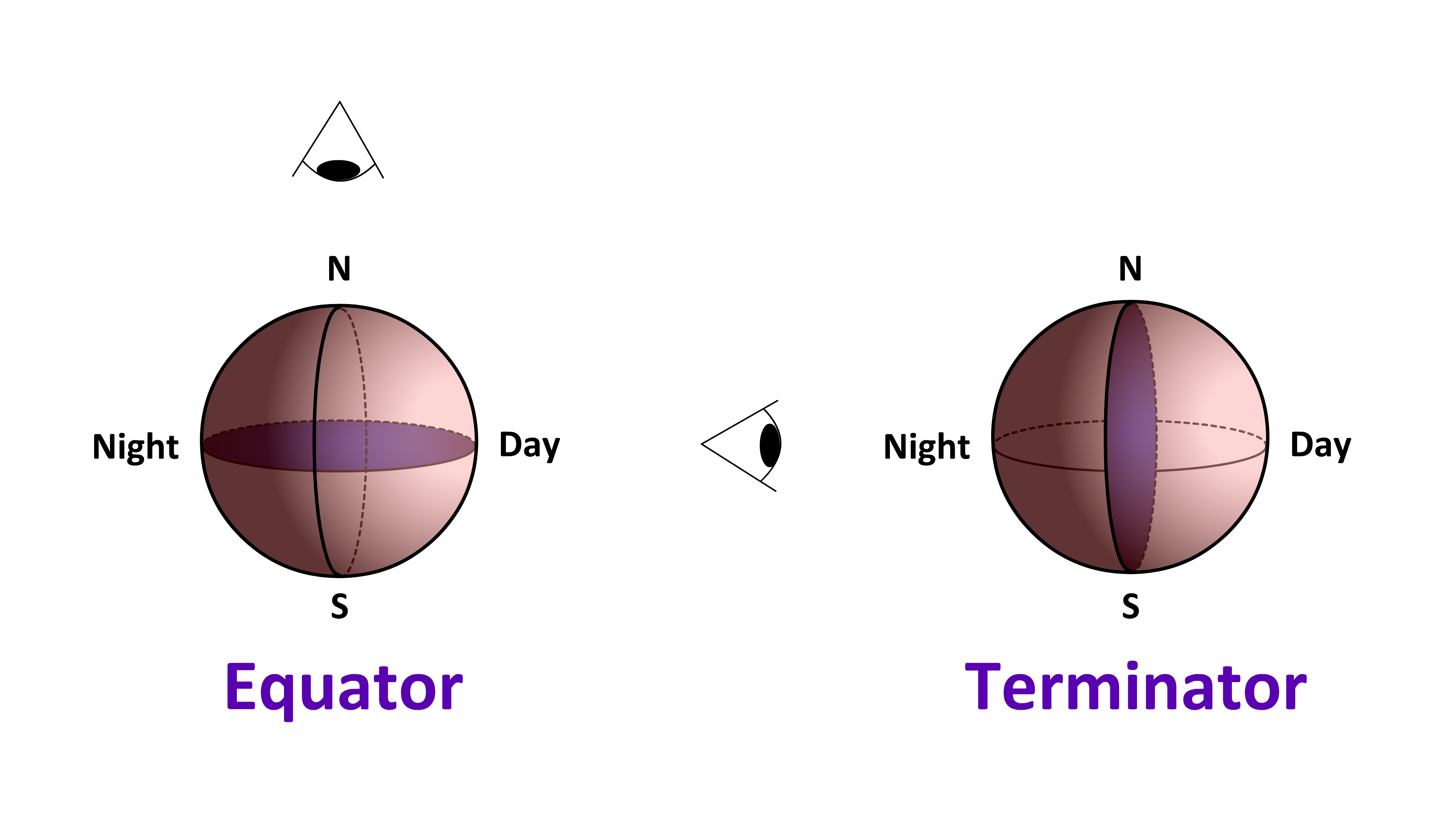}
    \caption{Viewing geometry for slice-plots in Figs.~\ref{fig:1DTp_1},~\ref{fig:1DJ*} etc. The purple shaded disk indicates the location of the slices we refer to by `equator' (left) and `terminator' (right).}
    \label{fig:vgeo}
\end{figure}

\begin{figure}
     \includegraphics[width=\columnwidth]{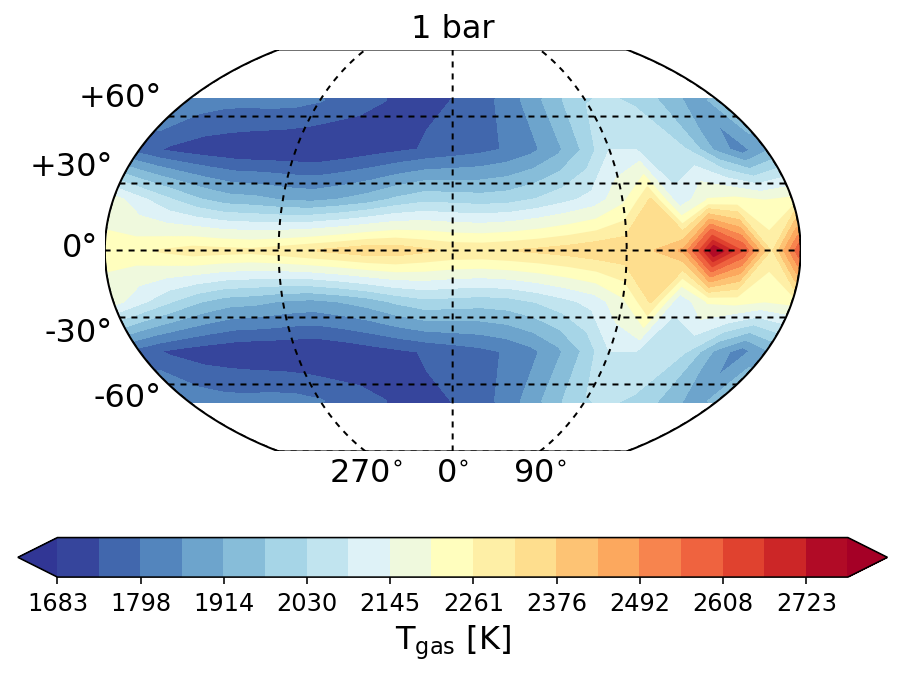}
   \includegraphics[width=\columnwidth]{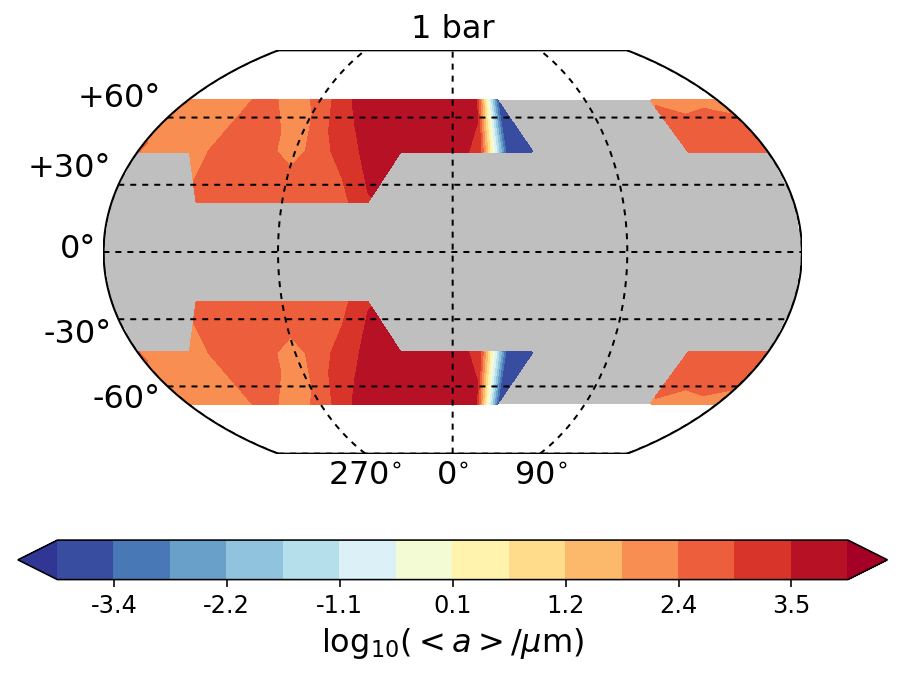}
        \includegraphics[width=\columnwidth]{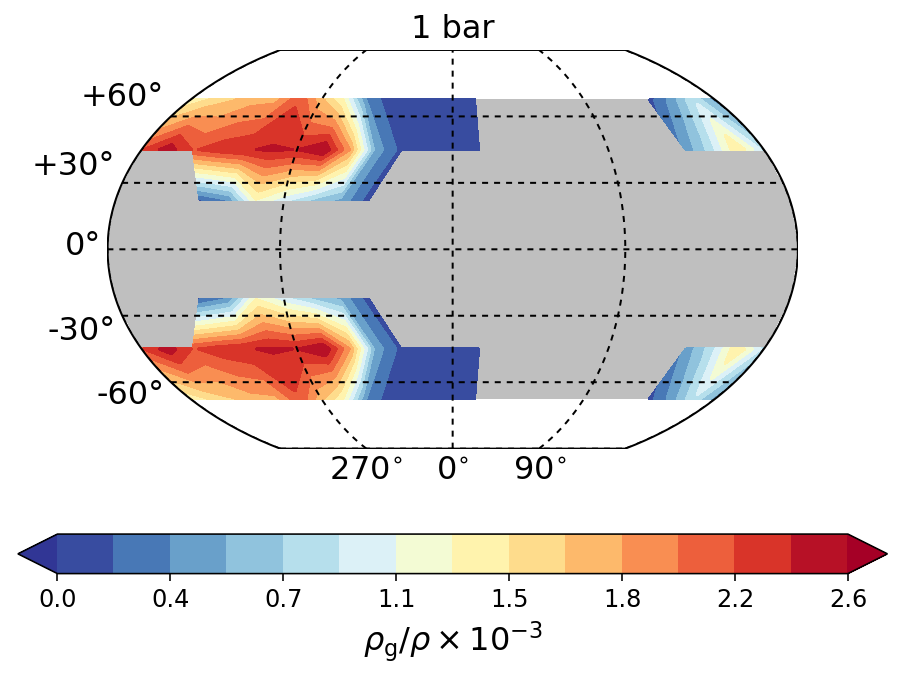}
  \caption{Cloud properties mapped in 2D at the 1bar-level:
  {\bf Top:} gas temperature T$_{\rm gas}$ [K], {\bf Middle} mean cloud particle sizes $\langle a \rangle$ [$\mu$m], {\bf Bottom:} the cloud particle mass load, the dust-to-gas ratio $\rho_{\rm d}/\rho_{\rm gas}$.}
      \label{fig:cloud_map_1bar}
\end{figure}

We provide additional figures, information about material data and visualisations to help the reader through the amount of data that we have been evaluating here.
Figure~\ref{fig:orgData} shows the original numerical data of the 3D GCM results for HAT-P-7b according to Sect.~\ref{ss:ib} at $\approx 10$bar, 1 bar and 0.1 bar for the local gas temperature, the vertical and horizontal velocity components. Figure~\ref{fig:vgeo} visualises the viewing geometry for the slice-plots we introduced in Fig.~\ref{fig:1DTp_1}. Figure~\ref{fig:cloud_map_1bar} provides supplementary 2D cloud maps for the 1bar-level.

\begin{table*}[t]
  \centering
   \caption{References of Optical Constants for Cloud Materials}
  \begin{tabular}{ccc}
     \hline\hline 
     Material species & Reference & Wavelength range $\rm (\mu m)$\\ \hline \hline
    \ce{TiO2}[s] (rutile) & \citet{TiO2_opt} & $0.47$--$36$\\
    \ce{SiO2}[s] (alpha-Quartz) & \citet{1985hocs.book.....P}, \citet{SiO2_opt} & $0.00012$--$10000$\\
    \ce{SiO}[s] (polycrystalline)& Philipp in \citet{1985hocs.book.....P} & $0.0015$--$14$\\
    \ce{MgSiO3}[s] (grass)& \citet{MgSiO3_opt} & $0.20$--$500$\\
    \ce{Mg2SiO4}[s] (crystalline)& \citet{2006MNRAS.370.1599S} & $0.10$--$1000$\\
    \ce{MgO}[s] (cubic)& \citet{1985hocs.book.....P} & $0.017$--$625$\\
    \ce{Fe}[s] (metallic)& \citet{1985hocs.book.....P} & $0.00012$--$285$\\
    \ce{FeO}[s] (amorphous)& \citet{FeO_opt} & $0.20$--$500$\\
    \ce{Fe2O3}[s] (amorphous)& Amaury H.M.J. Triaud (unpublished) & $0.10$--$1000$\\
    \ce{Fe2SiO4}[s] (amorphous)& \citet{MgSiO3_opt} & $0.20$--$500$ \\
    \ce{FeS}[s] (amorphous)& Henning (unpublished) & $0.10$--$100000$\\
    \ce{CaTiO3}[s] (amorphous)& \citet{2003ApJS..149..437P} & $2$--$5843$\\
    \ce{Al2O3}[s] (grass)& \citet{1997ApJ...476..199B} & $0.10$--$200$ \\
    \ce{C}[s] (graphite)& \citet{1985hocs.book.....P} & $0.20$--$794$\\\hline\hline
  \end{tabular}
  \label{table:opt}
\end{table*}


\bibliographystyle{aa}
\bibliography{reference.bib}

\begin{thebibliography}{102}
\expandafter\ifx\csname natexlab\endcsname\relax\def\natexlab#1{#1}\fi

\bibitem[{{Ackerman} \& {Marley}(2001)}]{2001ApJ...556..872A}
{Ackerman}, A.~S. \& {Marley}, M.~S. 2001, \apj, 556, 872

\bibitem[{{Allard} {et~al.}(2001){Allard}, {Hauschildt}, {Alexander},
  {Tamanai}, \& {Schweitzer}}]{2001ApJ...556..357A}
{Allard}, F., {Hauschildt}, P.~H., {Alexander}, D.~R., {Tamanai}, A., \&
  {Schweitzer}, A. 2001, \apj, 556, 357

\bibitem[{{Apai} {et~al.}(2013){Apai}, {Radigan}, {Buenzli}, {Burrows}, {Reid},
  \& {Jayawardhana}}]{2013ApJ...768..121A}
{Apai}, D., {Radigan}, J., {Buenzli}, E., {et~al.} 2013, \apj, 768, 121

\bibitem[{{Arcangeli} {et~al.}(2018){Arcangeli}, {D{\'e}sert}, {Line}, {Bean},
  {Parmentier}, {Stevenson}, {Kreidberg}, {Fortney}, {Mansfield}, \&
  {Showman}}]{Arcangeli18}
{Arcangeli}, J., {D{\'e}sert}, J.-M., {Line}, M.~R., {et~al.} 2018, \apjl, 855,
  L30

\bibitem[{{Arcangeli} {et~al.}(2019){Arcangeli}, {D{\'e}sert}, {Parmentier},
  {Stevenson}, {Bean}, {Line}, {Kreidberg}, {Fortney}, \&
  {Showman}}]{2019A&A...625A.136A}
{Arcangeli}, J., {D{\'e}sert}, J.-M., {Parmentier}, V., {et~al.} 2019, \aap,
  625, A136

\bibitem[{{Armstrong} {et~al.}(2016){Armstrong}, {de Mooij}, {Barstow},
  {Osborn}, {Blake}, \& {Saniee}}]{2016NatAs...1E...4A}
{Armstrong}, D.~J., {de Mooij}, E., {Barstow}, J., {et~al.} 2016, Nature
  Astronomy, 1, 0004

\bibitem[{{Asplund} {et~al.}(2009){Asplund}, {Grevesse}, {Sauval}, \&
  {Scott}}]{2009ARA&A..47..481A}
{Asplund}, M., {Grevesse}, N., {Sauval}, A.~J., \& {Scott}, P. 2009, \araa, 47,
  481

\bibitem[{{Begemann} {et~al.}(1997){Begemann}, {Dorschner}, {Henning},
  {Mutschke}, {G{\"u}rtler}, {K{\"o}mpe}, \& {Nass}}]{1997ApJ...476..199B}
{Begemann}, B., {Dorschner}, J., {Henning}, T., {et~al.} 1997, \apj, 476, 199

\bibitem[{{Bell} \& {Cowan}(2018)}]{2018ApJ...857L..20B}
{Bell}, T.~J. \& {Cowan}, N.~B. 2018, \apj, 857, L20

\bibitem[{{Bohren} \& {Huffman}(1983)}]{1983asls.book.....B}
{Bohren}, C.~F. \& {Huffman}, D.~R. 1983, {Absorption and scattering of light
  by small particles} (Wiley)

\bibitem[{{Bruggeman}(1935)}]{1935AnP...416..636B}
{Bruggeman}, D.~A.~G. 1935, Annalen der Physik, 416, 636

\bibitem[{{Buenzli} {et~al.}(2015){Buenzli}, {Saumon}, {Marley}, {Apai},
  {Radigan}, {Bedin}, {Reid}, \& {Morley}}]{2015ApJ...798..127B}
{Buenzli}, E., {Saumon}, D., {Marley}, M.~S., {et~al.} 2015, \apj, 798, 127

\bibitem[{{Caldas} {et~al.}(2019){Caldas}, {Leconte}, {Selsis}, {Waldmann},
  {Bord{\'e}}, {Rocchetto}, \& {Charnay}}]{Caldas2019}
{Caldas}, A., {Leconte}, J., {Selsis}, F., {et~al.} 2019, arXiv e-prints,
  arXiv:1901.09932

\bibitem[{{Charbonneau} {et~al.}(2009){Charbonneau}, {Berta}, {Irwin}, {Burke},
  {Nutzman}, {Buchhave}, {Lovis}, {Bonfils}, {Latham}, {Udry}, {Murray-Clay},
  {Holman}, {Falco}, {Winn}, {Queloz}, {Pepe}, {Mayor}, {Delfosse}, \&
  {Forveille}}]{Charbonneau09}
{Charbonneau}, D., {Berta}, Z.~K., {Irwin}, J., {et~al.} 2009, \nat, 462, 891

\bibitem[{{Dorschner} {et~al.}(1995){Dorschner}, {Begemann}, {Henning},
  {Jaeger}, \& {Mutschke}}]{MgSiO3_opt}
{Dorschner}, J., {Begemann}, B., {Henning}, T., {Jaeger}, C., \& {Mutschke}, H.
  1995, \aap, 300, 503

\bibitem[{{Eistrup} {et~al.}(2018){Eistrup}, {Walsh}, \& {van
  Dishoeck}}]{2018A&A...613A..14E}
{Eistrup}, C., {Walsh}, C., \& {van Dishoeck}, E.~F. 2018, \aap, 613, A14

\bibitem[{{Espinoza} {et~al.}(2017){Espinoza}, {Fortney}, {Miguel},
  {Thorngren}, \& {Murray-Clay}}]{Espinoza2017}
{Espinoza}, N., {Fortney}, J.~J., {Miguel}, Y., {Thorngren}, D., \&
  {Murray-Clay}, R. 2017, \apj, 838, L9

\bibitem[{{Evans} {et~al.}(2017){Evans}, {Sing}, {Kataria}, {Goyal}, {Nikolov},
  {Wakeford}, {Deming}, {Marley}, {Amundsen}, {Ballester}, {Barstow},
  {Ben-Jaffel}, {Bourrier}, {Buchhave}, {Cohen}, {Ehrenreich}, {Garc{\'{\i}}a
  Mu{\~n}oz}, {Henry}, {Knutson}, {Lavvas}, {Lecavelier Des Etangs}, {Lewis},
  {L{\'o}pez-Morales}, {Mandell}, {Sanz-Forcada}, {Tremblin}, \&
  {Lupu}}]{Evans17}
{Evans}, T.~M., {Sing}, D.~K., {Kataria}, T., {et~al.} 2017, \nat, 548, 58

\bibitem[{{Freytag} {et~al.}(2008){Freytag}, {Allard}, {Ludwig}, {Homeier}, \&
  {Steffen}}]{2008PhST..133a4005F}
{Freytag}, B., {Allard}, F., {Ludwig}, H.-G., {Homeier}, D., \& {Steffen}, M.
  2008, Physica Scripta Volume T, 133, 014005

\bibitem[{{Gray}(2008)}]{Gray2008}
{Gray}, D.~F. 2008, {The Observation and Analysis of Stellar Photospheres}

\bibitem[{{Haynes} {et~al.}(2015){Haynes}, {Mandell}, {Madhusudhan}, {Deming},
  \& {Knutson}}]{Haynes15}
{Haynes}, K., {Mandell}, A.~M., {Madhusudhan}, N., {Deming}, D., \& {Knutson},
  H. 2015, \apj, 806, 146

\bibitem[{{Heinze} {et~al.}(2013){Heinze}, {Metchev}, {Apai}, {Flateau},
  {Kurtev}, {Marley}, {Radigan}, {Burgasser}, {Artigau}, \&
  {Plavchan}}]{2013ApJ...767..173H}
{Heinze}, A.~N., {Metchev}, S., {Apai}, D., {et~al.} 2013, \apj, 767, 173

\bibitem[{{Helling}(2018)}]{2018arXiv181203793H}
{Helling}, C. 2018, arXiv e-prints [\eprint[arXiv]{1812.03793}]

\bibitem[{{Helling} \& {Fomins}(2013)}]{helling2013RSPTA}
{Helling}, C. \& {Fomins}, A. 2013, Philosophical Transactions of the Royal
  Society of London Series A, 371, 10581

\bibitem[{{Helling} {et~al.}(2019){Helling}, {Gourbin}, {Woitke}, \&
  {Parmentier}}]{2019arXiv190108640H}
{Helling}, C., {Gourbin}, P., {Woitke}, P., \& {Parmentier}, V. 2019, arXiv
  e-prints [\eprint[arXiv]{1901.08640}]

\bibitem[{{Helling} {et~al.}(2004){Helling}, {Klein}, {Woitke}, {Nowak}, \&
  {Sedlmayr}}]{2004A&A...423..657H}
{Helling}, C., {Klein}, R., {Woitke}, P., {Nowak}, U., \& {Sedlmayr}, E. 2004,
  \aap, 423, 657

\bibitem[{{Helling} {et~al.}(2016){Helling}, {Lee}, {Dobbs-Dixon}, {Mayne},
  {Amundsen}, {Khaimova}, {Unger}, {Manners}, {Acreman}, \&
  {Smith}}]{2016MNRAS.460..855H}
{Helling}, C., {Lee}, G., {Dobbs-Dixon}, I., {et~al.} 2016, \mnras, 460, 855

\bibitem[{{Helling} {et~al.}(2001){Helling}, {Oevermann}, {L{\"u}ttke},
  {Klein}, \& {Sedlmayr}}]{2001A&A...376..194H}
{Helling}, C., {Oevermann}, M., {L{\"u}ttke}, M.~J.~H., {Klein}, R., \&
  {Sedlmayr}, E. 2001, \aap, 376, 194

\bibitem[{{Helling} \& {Rimmer}(2019)}]{2019arXiv190304565H}
{Helling}, C. \& {Rimmer}, P.~B. 2019, arXiv e-prints
  [\eprint[arXiv]{1903.04565}]

\bibitem[{{Helling} {et~al.}(2017){Helling}, {Tootill}, {Woitke}, \&
  {Lee}}]{2017A&A...603A.123H}
{Helling}, C., {Tootill}, D., {Woitke}, P., \& {Lee}, G. 2017, \aap, 603, A123

\bibitem[{{Helling} \& {Woitke}(2006)}]{Helling2006}
{Helling}, C. \& {Woitke}, P. 2006, \aap, 455, 325

\bibitem[{{Helling} {et~al.}(2014){Helling}, {Woitke}, {Rimmer}, {Kamp}, {Thi},
  \& {Meijerink}}]{2014Life....4..142H}
{Helling}, C., {Woitke}, P., {Rimmer}, P.~B., {et~al.} 2014, Life, 4
  [\eprint[arXiv]{1403.4420}]

\bibitem[{{Helling} {et~al.}(2008){Helling}, {Woitke}, \&
  {Thi}}]{2008A&A...485..547H}
{Helling}, C., {Woitke}, P., \& {Thi}, W.-F. 2008, \aap, 485, 547

\bibitem[{{Heng} \& {Demory}(2013)}]{heng13}
{Heng}, K. \& {Demory}, B.-O. 2013, \apj, 777, 100

\bibitem[{{Henning} {et~al.}(1995){Henning}, {Begemann}, {Mutschke}, \&
  {Dorschner}}]{FeO_opt}
{Henning}, T., {Begemann}, B., {Mutschke}, H., \& {Dorschner}, J. 1995, \aaps,
  112, 143

\bibitem[{{Hindle} {et~al.}(2019){Hindle}, {Bushby}, \&
  {Rogers}}]{Hindel19_magnetic}
{Hindle}, A.~W., {Bushby}, P.~J., \& {Rogers}, T.~M. 2019, \apjl, 872, L27

\bibitem[{{Hou} {et~al.}(2019){Hou}, {Aoyama}, {Hirashita}, {Nagamine}, \&
  {Shimizu}}]{2019MNRAS.485.1727H}
{Hou}, K.-C., {Aoyama}, S., {Hirashita}, H., {Nagamine}, K., \& {Shimizu}, I.
  2019, \mnras, 485, 1727

\bibitem[{{Juncher} {et~al.}(2017){Juncher}, {J{\o}rgensen}, \&
  {Helling}}]{2017A&A...608A..70J}
{Juncher}, D., {J{\o}rgensen}, U.~G., \& {Helling}, C. 2017, \aap, 608, A70

\bibitem[{{Kopparapu} {et~al.}(2012){Kopparapu}, {Kasting}, \&
  {Zahnle}}]{2012ApJ...745...77K}
{Kopparapu}, R.~k., {Kasting}, J.~F., \& {Zahnle}, K.~J. 2012, \apj, 745, 77

\bibitem[{{Kreidberg} {et~al.}(2018){Kreidberg}, {Line}, {Parmentier},
  {Stevenson}, {Louden}, {Bonnefoy}, {Faherty}, {Henry}, {Williamson},
  {Stassun}, {Beatty}, {Bean}, {Fortney}, {Showman}, {D{\'e}sert}, \&
  {Arcangeli}}]{2018AJ....156...17K}
{Kreidberg}, L., {Line}, M.~R., {Parmentier}, V., {et~al.} 2018, \aj, 156, 17

\bibitem[{{Lee} {et~al.}(2016){Lee}, {Dobbs-Dixon}, {Helling}, {Bognar}, \&
  {Woitke}}]{Lee2016}
{Lee}, G., {Dobbs-Dixon}, I., {Helling}, C., {Bognar}, K., \& {Woitke}, P.
  2016, \aap, 594, A48

\bibitem[{{Lee} {et~al.}(2015{\natexlab{a}}){Lee}, {Helling}, {Dobbs-Dixon}, \&
  {Juncher}}]{Lee2015}
{Lee}, G., {Helling}, C., {Dobbs-Dixon}, I., \& {Juncher}, D.
  2015{\natexlab{a}}, ArXiv e-prints [\eprint[arXiv]{1505.06576}]

\bibitem[{{Lee} {et~al.}(2015{\natexlab{b}}){Lee}, {Helling}, {Giles}, \&
  {Bromley}}]{2015A&A...575A..11L}
{Lee}, G., {Helling}, C., {Giles}, H., \& {Bromley}, S.~T. 2015{\natexlab{b}},
  \aap, 575, A11

\bibitem[{{Lee} {et~al.}(2018){Lee}, {Blecic}, \&
  {Helling}}]{2018A&A...614A.126L}
{Lee}, G.~K.~H., {Blecic}, J., \& {Helling}, C. 2018, \aap, 614, A126

\bibitem[{{Lew} {et~al.}(2016){Lew}, {Apai}, {Zhou}, {Schneider}, {Burgasser},
  {Karalidi}, {Yang}, {Marley}, {Cowan}, {Bedin}, {Metchev}, {Radigan}, \&
  {Lowrance}}]{2016ApJ...829L..32L}
{Lew}, B.~W.~P., {Apai}, D., {Zhou}, Y., {et~al.} 2016, \apjl, 829, L32

\bibitem[{Line {et~al.}(2010)Line, Liang, \& Yung}]{line2010high}
Line, M.~R., Liang, M.-C., \& Yung, Y.~L. 2010, The Astrophysical Journal, 717,
  496

\bibitem[{{Line} \& {Parmentier}(2016)}]{Line2016}
{Line}, M.~R. \& {Parmentier}, V. 2016, \apj, 820, 78

\bibitem[{{Lines} {et~al.}(2018{\natexlab{a}}){Lines}, {Manners}, {Mayne},
  {Goyal}, {Carter}, {Boutle}, {Lee}, {Helling}, {Drummond}, {Acreman}, \&
  {Sing}}]{2018MNRAS.481..194L}
{Lines}, S., {Manners}, J., {Mayne}, N.~J., {et~al.} 2018{\natexlab{a}},
  \mnras, 481, 194

\bibitem[{{Lines} {et~al.}(2018{\natexlab{b}}){Lines}, {Mayne}, {Boutle},
  {Manners}, {Lee}, {Helling}, {Drummond}, {Amundsen}, {Goyal}, {Acreman},
  {Tremblin}, \& {Kerslake}}]{2018A&A...615A..97L}
{Lines}, S., {Mayne}, N.~J., {Boutle}, I.~A., {et~al.} 2018{\natexlab{b}},
  \aap, 615, A97

\bibitem[{{Lothringer} {et~al.}(2018){Lothringer}, {Barman}, \&
  {Koskinen}}]{2018ApJ...866...27L}
{Lothringer}, J.~D., {Barman}, T., \& {Koskinen}, T. 2018, \apj, 866, 27

\bibitem[{Lothringer {et~al.}(2018)Lothringer, Barman, \&
  Koskinen}]{lothringer2018extremely}
Lothringer, J.~D., Barman, T., \& Koskinen, T. 2018, The Astrophysical Journal,
  866, 27

\bibitem[{{Lunine} {et~al.}(1986){Lunine}, {Hubbard}, \&
  {Marley}}]{1986ApJ...310..238L}
{Lunine}, J.~I., {Hubbard}, W.~B., \& {Marley}, M.~S. 1986, \apj, 310, 238

\bibitem[{{MacDonald} \& {Madhusudhan}(2017)}]{MacDonald2017}
{MacDonald}, R.~J. \& {Madhusudhan}, N. 2017, \mnras, 469, 1979

\bibitem[{{MacDonald} \& {Madhusudhan}(2019)}]{MacDonald2019}
{MacDonald}, R.~J. \& {Madhusudhan}, N. 2019, arXiv e-prints, arXiv:1903.09151

\bibitem[{{Mallonn} {et~al.}(2019){Mallonn}, {K{\"o}hler}, {Alexoudi}, {von
  Essen}, {Granzer}, {Poppenhaeger}, \& {Strassmeier}}]{2019arXiv190207944M}
{Mallonn}, M., {K{\"o}hler}, J., {Alexoudi}, X., {et~al.} 2019, arXiv e-prints
  [\eprint[arXiv]{1902.07944}]

\bibitem[{{Mansfield} {et~al.}(2018){Mansfield}, {Bean}, {Line}, {Parmentier},
  {Kreidberg}, {D{\'e}sert}, {Fortney}, {Stevenson}, {Arcangeli}, \&
  {Dragomir}}]{2018AJ....156...10M}
{Mansfield}, M., {Bean}, J.~L., {Line}, M.~R., {et~al.} 2018, \aj, 156, 10

\bibitem[{{Marley} \& {McKay}(1999)}]{Marley1999}
{Marley}, M.~S. \& {McKay}, C.~P. 1999, \icarus, 138, 268

\bibitem[{{Masuda}(2015)}]{2015ApJ...805...28M}
{Masuda}, K. 2015, \apj, 805, 28

\bibitem[{Molaverdikhani {et~al.}(2019)Molaverdikhani, Henning, \&
  Molli{\`e}re}]{molaverdikhani2019cold}
Molaverdikhani, K., Henning, T., \& Molli{\`e}re, P. 2019, The Astrophysical
  Journal, 873, 32

\bibitem[{Molli{\`e}re {et~al.}(2015)Molli{\`e}re, van Boekel, Dullemond,
  Henning, \& Mordasini}]{molliere2015model}
Molli{\`e}re, P., van Boekel, R., Dullemond, C., Henning, T., \& Mordasini, C.
  2015, The Astrophysical Journal, 813, 47

\bibitem[{{Mordasini} {et~al.}(2016){Mordasini}, {van Boekel}, {Molli{\`e}re},
  {Henning}, \& {Benneke}}]{Mordasini2016}
{Mordasini}, C., {van Boekel}, R., {Molli{\`e}re}, P., {Henning}, T., \&
  {Benneke}, B. 2016, \apj, 832, 41

\bibitem[{{Moses} {et~al.}(2011){Moses}, {Visscher}, {Fortney}, {Showman},
  {Lewis}, {Griffith}, {Klippenstein}, {Shabram}, {Friedson}, {Marley}, \&
  {Freedman}}]{2011ApJ...737...15M}
{Moses}, J.~I., {Visscher}, C., {Fortney}, J.~J., {et~al.} 2011, \apj, 737, 15

\bibitem[{{Nikolov} {et~al.}(2018){Nikolov}, {Sing}, {Fortney}, {Goyal},
  {Drummond}, {Evans}, {Gibson}, {De Mooij}, {Rustamkulov}, {Wakeford},
  {Smalley}, {Burgasser}, {Hellier}, {Helling}, {Mayne}, {Madhusudhan},
  {Kataria}, {Baines}, {Carter}, {Ballester}, {Barstow}, {McCleery}, \&
  {Spake}}]{2018Natur.557..526N}
{Nikolov}, N., {Sing}, D.~K., {Fortney}, J.~J., {et~al.} 2018, \nat, 557, 526

\bibitem[{{{\"O}berg} {et~al.}(2011){{\"O}berg}, {Murray-Clay}, \&
  {Bergin}}]{Oberg11}
{{\"O}berg}, K.~I., {Murray-Clay}, R., \& {Bergin}, E.~A. 2011, \apj, 743, L16

\bibitem[{{Ohno} \& {Okuzumi}(2017)}]{Ohno&Okuzumi17}
{Ohno}, K. \& {Okuzumi}, S. 2017, \apj, 835, 261

\bibitem[{{Oreshenko} {et~al.}(2016){Oreshenko}, {Heng}, \&
  {Demory}}]{Oreschenko16_phase-curve}
{Oreshenko}, M., {Heng}, K., \& {Demory}, B.-O. 2016, \mnras, 457, 3420

\bibitem[{{P{\'a}l} {et~al.}(2008){P{\'a}l}, {Bakos}, {Torres}, {Noyes},
  {Latham}, {Kov{\'a}cs}, {Marcy}, {Fischer}, {Butler}, {Sasselov}, {Sip{\H
  o}cz}, {Esquerdo}, {Kov{\'a}cs}, {Stefanik}, {L{\'a}z{\'a}r}, {Papp}, \&
  {S{\'a}ri}}]{2008ApJ...680.1450P}
{P{\'a}l}, A., {Bakos}, G.~{\'A}., {Torres}, G., {et~al.} 2008, \apj, 680, 1450

\bibitem[{{Palik}(1985)}]{1985hocs.book.....P}
{Palik}, E.~D. 1985, {Handbook of optical constants of solids} (Academic Press)

\bibitem[{{Parmentier} {et~al.}(2016{\natexlab{a}}){Parmentier}, {Fortney},
  {Showman}, {Morley}, \& {Marley}}]{2016ApJ...828...22P}
{Parmentier}, V., {Fortney}, J.~J., {Showman}, A.~P., {Morley}, C., \&
  {Marley}, M.~S. 2016{\natexlab{a}}, \apj, 828, 22

\bibitem[{{Parmentier} {et~al.}(2016{\natexlab{b}}){Parmentier}, {Fortney},
  {Showman}, {Morley}, \& {Marley}}]{parmentier16_phase-curve}
{Parmentier}, V., {Fortney}, J.~J., {Showman}, A.~P., {Morley}, C., \&
  {Marley}, M.~S. 2016{\natexlab{b}}, \apj, 828, 22

\bibitem[{{Parmentier} {et~al.}(2018{\natexlab{a}}){Parmentier}, {Line},
  {Bean}, {Mansfield}, {Kreidberg}, {Lupu}, {Visscher}, {D{\'e}sert},
  {Fortney}, {Deleuil}, {Arcangeli}, {Showman}, \& {Marley}}]{Parmentier18}
{Parmentier}, V., {Line}, M.~R., {Bean}, J.~L., {et~al.} 2018{\natexlab{a}},
  \aap, 617, A110

\bibitem[{{Parmentier} {et~al.}(2018{\natexlab{b}}){Parmentier}, {Line},
  {Bean}, {Mansfield}, {Kreidberg}, {Lupu}, {Visscher}, {D{\'e}sert},
  {Fortney}, {Deleuil}, {Arcangeli}, {Showman}, \& {Marley}}]{Parmentier2018}
{Parmentier}, V., {Line}, M.~R., {Bean}, J.~L., {et~al.} 2018{\natexlab{b}},
  \aap, 617, A110

\bibitem[{{Pinhas} {et~al.}(2019){Pinhas}, {Madhusudhan}, {Gandhi}, \&
  {MacDonald}}]{2019MNRAS.482.1485P}
{Pinhas}, A., {Madhusudhan}, N., {Gandhi}, S., \& {MacDonald}, R. 2019, \mnras,
  482, 1485

\bibitem[{{Posch} {et~al.}(2003){Posch}, {Kerschbaum}, {Fabian}, {Mutschke},
  {Dorschner}, {Tamanai}, \& {Henning}}]{2003ApJS..149..437P}
{Posch}, T., {Kerschbaum}, F., {Fabian}, D., {et~al.} 2003, \apjs, 149, 437

\bibitem[{{Priestley} {et~al.}(2019){Priestley}, {Barlow}, \& {De
  Looze}}]{2019MNRAS.485..440P}
{Priestley}, F.~D., {Barlow}, M.~J., \& {De Looze}, I. 2019, \mnras, 485, 440

\bibitem[{{Radigan} {et~al.}(2012){Radigan}, {Jayawardhana}, {Lafreni{\`e}re},
  {Artigau}, {Marley}, \& {Saumon}}]{2012ApJ...750..105R}
{Radigan}, J., {Jayawardhana}, R., {Lafreni{\`e}re}, D., {et~al.} 2012, \apj,
  750, 105

\bibitem[{{Richard} {et~al.}(2012){Richard}, {Gordon}, {Rothman}, {Abel},
  {Frommhold}, {Gustafsson}, {Hartmann}, {Hermans}, {Lafferty}, {Orton},
  {Smith}, \& {Tran}}]{Richad2012}
{Richard}, C., {Gordon}, I.~E., {Rothman}, L.~S., {et~al.} 2012, Journal of
  Quantitative Spectroscopy and Radiative Transfer, 113, 1276

\bibitem[{{Rodr{\'{\i}}guez-Barrera} {et~al.}(2015){Rodr{\'{\i}}guez-Barrera},
  {Helling}, {Stark}, \& {Rice}}]{2015MNRAS.454.3977R}
{Rodr{\'{\i}}guez-Barrera}, M.~I., {Helling}, C., {Stark}, C.~R., \& {Rice},
  A.~M. 2015, \mnras, 454, 3977

\bibitem[{{Rogers}(2017)}]{Rogers17_magnetic}
{Rogers}, T.~M. 2017, Nature Astronomy, 1, 0131

\bibitem[{{Rogers} \& {Komacek}(2014)}]{Rogers2014}
{Rogers}, T.~M. \& {Komacek}, T.~D. 2014, \apj, 794, 132

\bibitem[{{Sharp} \& {Burrows}(2007)}]{2007ApJS..168..140S}
{Sharp}, C.~M. \& {Burrows}, A. 2007, \apjs, 168, 140

\bibitem[{{Showman} {et~al.}(2009){Showman}, {Fortney}, {Lian}, {Marley},
  {Freedman}, {Knutson}, \& {Charbonneau}}]{Showman2009}
{Showman}, A.~P., {Fortney}, J.~J., {Lian}, Y., {et~al.} 2009, \apj, 699, 564

\bibitem[{{Showman} \& {Polvani}(2011)}]{ShowmanPolvani2011}
{Showman}, A.~P. \& {Polvani}, L.~M. 2011, \apj, 738, 71

\bibitem[{{Sing} {et~al.}(2016){Sing}, {Fortney}, {Nikolov}, {Wakeford},
  {Kataria}, {Evans}, {Aigrain}, {Ballester}, {Burrows}, {Deming},
  {D{\'e}sert}, {Gibson}, {Henry}, {Huitson}, {Knutson}, {Lecavelier Des
  Etangs}, {Pont}, {Showman}, {Vidal-Madjar}, {Williamson}, \&
  {Wilson}}]{2016Natur.529...59S}
{Sing}, D.~K., {Fortney}, J.~J., {Nikolov}, N., {et~al.} 2016, \nat, 529, 59

\bibitem[{{Spiegel} \& {Burrows}(2010)}]{2010ApJ...722..871S}
{Spiegel}, D.~S. \& {Burrows}, A. 2010, \apj, 722, 871

\bibitem[{{Street} {et~al.}(2015){Street}, {Fulton}, {Scholz}, {Horne},
  {Helling}, {Juncher}, {Lee}, \& {Valenti}}]{2015ApJ...812..161S}
{Street}, R.~A., {Fulton}, B.~J., {Scholz}, A., {et~al.} 2015, \apj, 812, 161

\bibitem[{{Suto} {et~al.}(2006){Suto}, {Sogawa}, {Tachibana}, {Koike},
  {Karoji}, {Tsuchiyama}, {Chihara}, {Mizutani}, {Akedo}, {Ogiso}, {Fukui}, \&
  {Ohara}}]{2006MNRAS.370.1599S}
{Suto}, H., {Sogawa}, H., {Tachibana}, S., {et~al.} 2006, \mnras, 370, 1599

\bibitem[{{Tsuji} {et~al.}(1996){Tsuji}, {Ohnaka}, \&
  {Aoki}}]{1996A&A...305L...1T}
{Tsuji}, T., {Ohnaka}, K., \& {Aoki}, W. 1996, \aap, 305, L1

\bibitem[{{Van Eylen} {et~al.}(2012){Van Eylen}, {Kjeldsen},
  {Christensen-Dalsgaard}, \& {Aerts}}]{vaneylen13}
{Van Eylen}, V., {Kjeldsen}, H., {Christensen-Dalsgaard}, J., \& {Aerts}, C.
  2012, Astronomische Nachrichten, 333, 1088

\bibitem[{Venot {et~al.}(2018)Venot, Drummond, Miguel, Waldmann, Pascale, \&
  Zingales}]{venot2018better}
Venot, O., Drummond, B., Miguel, Y., {et~al.} 2018, Experimental Astronomy, 46,
  101

\bibitem[{{Visscher} {et~al.}(2006){Visscher}, {Lodders}, \&
  {Fegley}}]{2006ApJ...648.1181V}
{Visscher}, C., {Lodders}, K., \& {Fegley}, Jr., B. 2006, \apj, 648, 1181

\bibitem[{{Visscher} {et~al.}(2010){Visscher}, {Lodders}, \&
  {Fegley}}]{2010ApJ...716.1060V}
{Visscher}, C., {Lodders}, K., \& {Fegley}, Jr., B. 2010, \apj, 716, 1060

\bibitem[{{Vos} {et~al.}(2019){Vos}, {Biller}, {Bonavita}, {Eriksson}, {Liu},
  {Best}, {Metchev}, {Radigan}, {Allers}, {Janson}, {Buenzli}, {Dupuy},
  {Bonnefoy}, {Manjavacas}, {Brandner}, {Crossfield}, {Deacon}, {Henning},
  {Homeier}, {Kopytova}, \& {Schlieder}}]{2019MNRAS.483..480V}
{Vos}, J.~M., {Biller}, B.~A., {Bonavita}, M., {et~al.} 2019, \mnras, 483, 480

\bibitem[{{Winters} {et~al.}(2000){Winters}, {Le Bertre}, {Jeong}, {Helling},
  \& {Sedlmayr}}]{2000A&A...361..641W}
{Winters}, J.~M., {Le Bertre}, T., {Jeong}, K.~S., {Helling}, C., \&
  {Sedlmayr}, E. 2000, \aap, 361, 641

\bibitem[{{Witte} {et~al.}(2009){Witte}, {Helling}, \&
  {Hauschildt}}]{2009A&A...506.1367W}
{Witte}, S., {Helling}, C., \& {Hauschildt}, P.~H. 2009, \aap, 506, 1367

\bibitem[{{Woitke} \& {Helling}(2003)}]{Woitke2003}
{Woitke}, P. \& {Helling}, C. 2003, \aap, 399, 297

\bibitem[{{Woitke} {et~al.}(2018){Woitke}, {Helling}, {Hunter}, {Millard},
  {Turner}, {Worters}, {Blecic}, \& {Stock}}]{2018A&A...614A...1W}
{Woitke}, P., {Helling}, C., {Hunter}, G.~H., {et~al.} 2018, \aap, 614, A1

\bibitem[{{Wong} {et~al.}(2016){Wong}, {Knutson}, {Kataria}, {Lewis},
  {Burrows}, {Fortney}, {Schwartz}, {Shporer}, {Agol}, {Cowan}, {Deming},
  {D{\'e}sert}, {Fulton}, {Howard}, {Langton}, {Laughlin}, {Showman}, \&
  {Todorov}}]{2016ApJ...823..122W}
{Wong}, I., {Knutson}, H.~A., {Kataria}, T., {et~al.} 2016, \apj, 823, 122

\bibitem[{{Wright} {et~al.}(2012){Wright}, {Marcy}, {Howard}, {Johnson},
  {Morton}, \& {Fischer}}]{Wright2012}
{Wright}, J.~T., {Marcy}, G.~W., {Howard}, A.~W., {et~al.} 2012, \apj, 753, 160

\bibitem[{{Zahnle} {et~al.}(2009){Zahnle}, {Marley}, {Freedman}, {Lodders}, \&
  {Fortney}}]{2009ApJ...701L..20Z}
{Zahnle}, K., {Marley}, M.~S., {Freedman}, R.~S., {Lodders}, K., \& {Fortney},
  J.~J. 2009, \apjl, 701, L20

\bibitem[{{Zeidler} {et~al.}(2013){Zeidler}, {Posch}, \& {Mutschke}}]{SiO2_opt}
{Zeidler}, S., {Posch}, T., \& {Mutschke}, H. 2013, \aap, 553, A81

\bibitem[{{Zeidler} {et~al.}(2011){Zeidler}, {Posch}, {Mutschke}, {Richter}, \&
  {Wehrhan}}]{TiO2_opt}
{Zeidler}, S., {Posch}, T., {Mutschke}, H., {Richter}, H., \& {Wehrhan}, O.
  2011, \aap, 526, A68

\end{thebibliography}

\end{document}